\documentclass[12pt,a4paper]{scrartcl}
\pdfoutput=1

\usepackage[utf8]{inputenc}
\usepackage[T1]{fontenc}
\usepackage{amsmath,amssymb,mathtools}
\usepackage[separate-uncertainty=true,retain-unity-mantissa=false]{siunitx}
\usepackage{graphicx}
\usepackage{booktabs,multicol,multirow}
\usepackage{enumitem}
\usepackage{hyperref}
\usepackage[capitalize]{cleveref}
\usepackage{xspace}
\usepackage[noblocks]{authblk}
\usepackage[dvipsnames]{xcolor}
\usepackage{soul}
\usepackage[textsize=tiny]{todonotes}
\usepackage{subcaption}
\usepackage[sorting=none,%
  citestyle=numeric-comp,%
  bibstyle=mynumeric,%
  giveninits=true]{biblatex}

\NewBibliographyString{refname}
\NewBibliographyString{refsname}
\DefineBibliographyStrings{english}{%
  refname = {Ref\adddot},
  refsname = {Refs\adddot}
}
\DeclareCiteCommand{\ccite}
  {%
  \ifnum\thecitetotal=1
    \bibstring{refname}%
  \else%
    \bibstring{refsname}%
  \fi%
  \addnbspace\bibopenbracket%
  \usebibmacro{cite:init}%
   \usebibmacro{prenote}}
  {\usebibmacro{citeindex}%
   \usebibmacro{cite:comp}}
  {}
  {\usebibmacro{cite:dump}%
   \usebibmacro{postnote}%
   \bibclosebracket}
\newrobustcmd*{\Ccite}{\bibsentence\ccite}

\DeclareSIUnit{\pb}{\pico\barn}
\DeclareSIUnit{\fb}{\femto\barn}
\DeclareSIUnit{\ab}{\atto\barn}

\newcommand{\cp}{\ensuremath{\mathcal{CP}}\xspace}

\newcommand{\HB}{\texttt{HiggsBounds}\xspace}

\newcommand{\HBv}[1]{\texttt{HiggsBounds-#1}\xspace}
\newcommand{\HSv}[1]{\texttt{HiggsSignals-#1}\xspace}

\newcommand{\sw}{s_{W}}
\newcommand{\cw}{c_{W}}
\newcommand{\cba}{c_{\beta-\alpha}}

\newcommand{\mbarsq}{\overline{m}^2}
\newcommand{\eqcomma}{\,,}
\newcommand{\eqdot}{\,.}
\newcommand{\SU}[1]{\mathrm{SU}(#1)}
\newcommand{\SM}{{\text{SM}}}
\newcommand{\BSM}{{\text{BSM}}}

\newcommand{\order}[1]{\mathcal{O}(#1)}

\newcommand{\Pdd}{\Phi_1^\dagger\Phi_1}
\newcommand{\Puu}{\Phi_2^\dagger\Phi_2}
\newcommand{\Pdu}{\Phi_1^\dagger\Phi_2}
\newcommand{\Pud}{\Phi_2^\dagger\Phi_1}

\newcommand{\BPh}{\texorpdfstring{$\text{cH}(Wh_\BSM)$}{cH(WhBSM)}\xspace}
\newcommand{\BPA}{\texorpdfstring{$\text{cH}(WA)$}{cH(WA)}\xspace}
\newcommand{\BPff}{\texorpdfstring{$\text{cH}(Wh_\BSM^\text{fphob})$}{cH(WhBSM fphob)}\xspace}
\newcommand{\BPlight}{\texorpdfstring{$\text{cH}(Wh_\BSM^\text{light})$}{cH(WhBSM light)}\xspace}
\newcommand{\BPlightLS}{\texorpdfstring{$\text{cH}(Wh_\BSM^{\ell\text{phil}})$}{cH(WhBSM l-phil)}\xspace}

\graphicspath{{./figures/}}
\addtokomafont{disposition}{\rmfamily}

\title{\Large The forgotten channels: charged Higgs boson decays to a $W^\pm$ and a non-SM-like Higgs boson}
\author{Henning Bahl\footnote{\href{mailto:henning.bahl@desy.de}{henning.bahl@desy.de}}}
\author{Tim Stefaniak\footnote{\href{mailto:tim.stefaniak@desy.de}{tim.stefaniak@desy.de}}}
\affil{Deutsches Elektronen-Synchrotron DESY,
       Notkestraße~85,
       D-22607 Hamburg, Germany}
\author{Jonas Wittbrodt\footnote{\href{mailto:jonas.wittbrodt@thep.lu.se}{jonas.wittbrodt@thep.lu.se}}}
\affil{Department of Astronomy and Theoretical Physics,
       Lund University,
       Sölvegatan~14A, 223~62~Lund,
       Sweden}
\date{}
\titlehead{\hfill \parbox{3cm}{\vspace{-2cm} \texttt{DESY 21-035\\LU TP 21-09}}}

\begin{document}
\maketitle

\begin{abstract}\noindent
    The presence of charged Higgs bosons is a generic prediction of multiplet extensions of the Standard Model (SM) Higgs sector. Focusing on the Two-Higgs-Doublet-Model (2HDM) with type I and lepton-specific Yukawa sectors, we discuss the charged Higgs boson collider phenomenology in the theoretically and experimentally viable parameter space. While almost all existing experimental searches at the LHC target the fermionic decays of charged Higgs bosons, we point out that the bosonic decay channels --- especially the decay into a non-SM-like Higgs boson and a $W$ boson --- often dominate over the fermionic channels. Moreover, we revisit two genuine BSM effects on the properties of the discovered Higgs boson --- the charged Higgs contribution to the diphoton rate and the Higgs decay to two light Higgs bosons --- and their implication for the charged Higgs boson phenomenology. As main result of the present paper, we propose five two-dimensional benchmark scenarios with distinct phenomenological features in order to facilitate the design of dedicated LHC searches for charged Higgs bosons decaying into a $W$ boson and a light, non-SM-like Higgs boson.
\end{abstract}
\setcounter{footnote}{0}

\newpage

\tableofcontents

\newpage

\section{Introduction}

The discovery of a Higgs boson at the LHC in 2012~\cite{Chatrchyan:2012xdj,Aad:2012tfa} has initiated the survey of the scalar sector and the concomitant mechanism of spontaneous electroweak (EW) symmetry breaking. The Higgs boson --- predicted in the Standard Model (SM) to be a fundamental scalar boson $h$ originating from a complex scalar $\SU{2}_L$ doublet field with hypercharge $Y = +\tfrac{1}{2}$ and  vacuum expectation value (vev) $v \approx \SI{246}{\GeV}$ --- is the first of its kind, and the thorough investigation of its properties is of paramount importance for the understanding of particle physics. To this end, the LHC collaborations perform detailed measurements of the signal rates of the discovered Higgs boson in various production and decay modes, from which (under certain model assumptions) the strengths of the Higgs-boson couplings to the $W$ and $Z$-bosons and to the third generation fermions can be inferred.\footnote{The precise determination of the Higgs boson trilinear and quartic self-couplings as well as the much weaker couplings to first and second generation fermions is challenging at the LHC~\cite{Cepeda:2019klc}.} By the end of Run-2 of the LHC, with $\sim \SI{140}{\per\fb}$ of data each collected by the ATLAS and CMS experiment, the Higgs-boson rate measurements agree remarkably well with the predictions in the Standard Model (SM). Yet, with the currently achieved precision of $\gtrsim \mathcal{O}(\SI{10}{\%})$ in the coupling determination~\cite{Khachatryan:2016vau,Aad:2019mbh,Sirunyan:2018koj,CMS:2020gsy}, it is far from certain that the scalar sector as predicted in the SM is realized in nature.

Indeed, there are many reasons to anticipate effects from beyond the SM (BSM) physics in the scalar sector. The Higgs field may interact with the dark matter (DM) sector through so-called Higgs portal interactions, or DM itself may be composed of a stable scalar particle that originates from an extension of the SM Higgs sector (see Ref.~\cite{Arcadi:2019lka} for a recent review). Furthermore, a BSM scalar sector may lead to a strong first-order EW phase transition and feature new sources of \cp-violation --- these may enable the successful generation of the baryon asymmetry observed in the Universe (see e.g.~\cite{Cline:2011mm,Shu:2013uua,Blinov:2015vma,Dorsch:2016nrg,Basler:2019iuu,Fabian:2020hny} for recent works). BSM theories addressing the hierarchy problem, e.g.~Supersymmetry (SUSY)~\cite{Haber:1984rc,Gunion:1984yn,Gunion:1986nh,Ellis:1988er}, often modify or extend the scalar sector. Lastly, one may wonder why the scalar sector should be \emph{minimal}, while we clearly have a \emph{non-minimal} matter sector with three generations of fermions.

In many BSM extensions of the scalar sector, the discovered Higgs state $h_{125}$ can acquire tree-level couplings to fermions and gauge bosons identical to those predicted in the SM in the so-called \emph{alignment limit}~\cite{Gunion:2002zf,Craig:2013hca,Bernon:2015qea, Bernon:2015wef}. In this limit the physical Higgs state $h_{125}$ is \emph{aligned} in field space with the direction of the vacuum expectation value~$v$. The current LHC Higgs signal rate measurements imply that this alignment limit is at least approximately realized. However, the \emph{origin} of this alignment --- whether it is dynamical, by symmetry, or accidental --- is unknown, and is obviously model-dependent.\footnote{For instance, in supersymmetric models, alignment is automatically realized if the second Higgs doublet is decoupled, i.e.~if all other Higgs bosons of the model are very heavy~\cite{Gunion:2002zf}. However, SUSY scenarios of accidental alignment \emph{without} decoupling exist as well, see~\ccite{Carena:2013ooa,Carena:2014nza,Bechtle:2016kui,Haber:2017erd,Bahl:2018zmf,Hollik:2018wrr,Carena:2015moc,Coyle:2019exn}.} Given the experimental observations in the Higgs signal rates, any phenomenologically viable BSM model therefore has to contain a Higgs boson with a mass around \SI{125}{\GeV} that is approximately ``SM-like'' in its coupling properties, as achieved near the alignment limit.

Extending the scalar sector also introduces additional scalar particles that could be either neutral or carry electromagnetic charge. Detecting these new states complements the precision studies of the discovered Higgs state in the quest of unraveling the details of the electroweak symmetry breaking mechanism. To this end, the ATLAS and CMS collaborations have performed searches for the direct production and decay of additional (electrically neutral and charged) Higgs bosons. The targeted collider signatures are often guided by popular BSM theories, e.g.~the Minimal Supersymmetric Standard Model (MSSM), and predicted to be experimentally accessible with current detector capabilities and the current amount of accumulated data. Other experimental BSM searches look for decays of known particles (e.g.~the discovered Higgs boson $h_{125}$) into BSM particles. A prominent example are searches for an invisibly decaying Higgs boson (see e.g.~\ccite{Djouadi:2012zc} and references therein).

However, all these searches have not found any convincing hints for the existence of new particles yet, and have thus only produced upper limits on their possible signal cross section. Both experimental results --- the Higgs-boson signal rate and mass\footnote{The Higgs boson mass has been determined to $125.09\pm 0.21 (\text{stat}) \pm 0.11 (\text{syst})\si{\GeV}$ in the combined ATLAS and CMS analysis of LHC Run-1 data~\cite{Aad:2015zhl}.} measurements and the upper limits from searches for additional Higgs bosons --- give rise to important constraints on the parameters of BSM models. At the same time, they challenge the theoretical community to provide reasonable explanations for the absence of hints for BSM physics  at the LHC, and to give guidance for future strategies to remedy this situation. Given the prospect of an LHC upgrade to the high-luminosity (HL) phase, with an anticipated \SI{3}{\per\ab} of data per experiment, as well as recent and upcoming advances in data analysis techniques (see e.g.~Refs.~\cite{Baldi:2014kfa,Kasieczka:2017nvn,Larkoski:2017jix,Louppe:2017ipp,Macaluso:2018tck,Carleo:2019ptp,Araz:2021wqm}), so-far disregarded collider signatures of additional Higgs bosons (re)gain attention (see e.g.~\ccite{Gao:2016ued,Bahl:2018zmf,Gori:2018pmk,Bahl:2019ago,Robens:2019kga,Adhikary:2020ujn,Liu:2020muv,Aiko:2019mww,Cheung:2020xij,Arganda:2018hdn,Arganda:2021qgi} for recent proposals).

In this work we focus on collider signatures that arise from the production of a charged Higgs boson $H^\pm$ and its successive decay to a lighter neutral Higgs boson and a $W$ boson.  While this decay mode has been investigated in several phenomenological works~\cite{Akeroyd:1998dt,Basso:2012st,Coleppa:2014cca,Kling:2015uba,Haber:2015pua,Akeroyd:2016ymd,Arhrib:2016wpw,Bechtle:2016kui,Arhrib:2017wmo,Alves:2017snd,Chun:2018vsn,Bahl:2018zmf,Coleppa:2019cul}, up to now the signatures arising from this process have not been actively searched for by the LHC experiments\footnote{A possible explanation for this omission is that LHC searches for charged Higgs bosons were ever-so-often guided and motivated by the expectations within the MSSM Higgs sector. While MSSM parameter regions exist where the charged Higgs boson dominantly decays to a neutral Higgs and a $W$ boson --- e.g.~the $H^\pm \to W^\pm h$ decays in the $M_H^{125}$ scenario of \ccite{Bahl:2018zmf} --- these decays are absent in most of the parameter space due to an approximate mass degeneracy of the BSM Higgs bosons~\cite{Gunion:1989we,Carena:1999py,Gunion:2002zf,Djouadi:2005gj,Carena:2005ek}.}--- with the notable exception of \ccite{Sirunyan:2019zdq}.\footnote{The CMS collaboration has searched for the process $pp\to H^\pm t b$, $H^\pm \to W^\pm A$, with the light pseudoscalar boson $A$ decaying to $\mu^+\mu^-$, using $35.9~\text{fb}^{-1}$ of Run-2 data~\cite{Sirunyan:2019zdq}.} Confirming earlier works we shall show in the present paper that these decays occur at sizable rates quite naturally already in minimal models that contain a charged Higgs boson, namely Two-Higgs-Doublet Models (2HDM), and, therefore, need to be explored experimentally. The purpose of this work is to motivate and initiate dedicated experimental searches by exploring the possible signal rates and providing suitable benchmark models for these studies based on the 2HDM of type-I and the lepton-specific 2HDM\@. We base our benchmark model definitions on the latest experimental constraints and state-of-the-art model predictions.

This paper is organized as follows. We first review the coupling properties of charged Higgs bosons in multi-Higgs-doublet models in \cref{sec:chargedH_pheno}, and then discuss the charged Higgs boson phenomenology in the Two-Higgs-Doublet-Model (2HDM) in light of current experimental and theoretical constraints. In \cref{sec:BSMeffects} we revisit two important BSM effects on the discovered Higgs boson $h_{125}$ that subsist even in case of exact alignment --- the charged Higgs contribution to the Higgs-to-diphoton rate and the neutral Higgs ($h_{125}$) decay to two lighter neutral Higgs bosons --- and their implications on the charged Higgs boson phenomenology. We discuss the most relevant charged Higgs boson production and decay modes, as well as the current LHC searches in \cref{sec:chargedH_at_LHC}. We then elaborate on the experimentally unexplored charged Higgs boson signatures in \cref{sec:benchmarks} and present several benchmark scenarios for future searches for these signatures. We conclude in \cref{sec:conclusions}.

\section{Charged Higgs boson phenomenology in doublet extensions of the SM}
\label{sec:chargedH_pheno}

In this section we will first review the coupling structure of charged Higgs bosons in the general $N$-Higgs-Doublet model.\footnote{Charged Higgs bosons can also appear in higher-multiplet extensions of the SM-Higgs sector. We will not discuss those models (see~\ccite{Ivanov:2017dad} for a review).} Afterwards, we will focus on the 2HDM.

\subsection{Charged Higgs couplings to bosons in \texorpdfstring{$N$}{N}-Higgs-Doublet models}
\label{sec:nhdm}

A genuine prediction of Lorentz- and gauge-invariant BSM theories with additional scalar $SU(2)_L$ doublet fields is the existence of one pair of electrically charged Higgs bosons per doublet added to the SM\@.

The couplings of the charged scalars to vector bosons are
\begin{align}
    H^\pm_i H^\mp_j \gamma                  & :\quad g(H^\pm_i H^\mp_j \gamma) (p^\mu_{H^\pm_i}-p^\mu_{H^{\mp}_j}) = \frac{1}{2}\delta_{ij}e(p^\mu_{H^\pm_i}-p^\mu_{H^{\mp}_j})\eqcomma            \\
    H^\pm_i H^\mp_j Z                       & : \quad g(H^\pm_i H^\mp_j Z) (p^\mu_{H^\pm_i}-p^\mu_{H^{\mp}_j})    = \frac{\delta_{ij}}{2}(g\cw - g'\sw)(p^\mu_{H^\pm_i}-p^\mu_{H^{\mp}_j})\eqcomma \\
    H^\pm_i W^\mp Z,\; H^\pm_i W^\mp \gamma & : \quad 0\eqdot
\end{align}
Here, $H_i^\pm$ denotes the charged Higgs boson mass eigenstates, $g$ and $g'$ are the $SU(2)_L$ and $U(1)_Y$ gauge coupling, respectively, and $e$ is the electric charge. We use the short-hand notation $c_W \equiv \cos\theta_W$ and $s_W \equiv \sin\theta_W$, with the weak mixing angle $\theta_W$. All momenta $p^\mu$ are considered as incoming.

The couplings of the neutral Higgs boson $h_i$ to gauge bosons are defined by
\begin{align}
    h_i Z Z & : \quad g(h_i Z Z) g_{\mu\nu}\eqcomma \\
    h_i W W & : \quad g(h_i W W) g_{\mu\nu}\eqcomma
\end{align}
and fulfill the sum rules (see e.g.~\cite{Bento:2017eti})
\begin{align}
    \sum_i g(h_i Z Z)^2 = g(h_\SM ZZ)^2 = \frac{g^2}{\cw^2}  M_Z^2\eqcomma \label{eq:hzzsum} \\
    \sum_i g(h_i W W)^2 = g(h_\SM WW)^2 = g^2  M_W^2\eqcomma \label{eq:hwwsum}
\end{align}
where the sum runs over all neutral Higgs bosons and $h_\SM$ is a state with exactly SM-like couplings.\footnote{If the SM-Higgs sector is extended only by $SU(2)_L$ doublets and singlets, it is always possible to construct such a SM-like state. In general, this state is, however, not a mass eigenstate (see e.g.~\ccite{Bento:2017eti}).} Furthermore, gauge invariance requires
\begin{equation}
    \frac{g(h_iZZ)}{g(h_\SM ZZ)} =\frac{g(h_iWW)}{g(h_\SM WW)} \equiv c(h_iVV)
\end{equation}
such that
\begin{equation}
    \sum_i c(h_i VV)^2 = 1\eqcomma\label{eq:hVVsum}
\end{equation}
where $V = W, Z$.

We define the charged Higgs boson couplings to a neutral \cp-even or \cp-odd Higgs boson --- $h_j$ and $a_j$, respectively --- and the $W$ boson via
\begin{align}
    H_i^\pm W^\mp h_j & : \quad g(H_i^\pm W^\mp h_j)(p^\mu_{h_i}-p^\mu_{H^{\pm}})\eqcomma \\
    H_i^\pm W^\mp a_j & : \quad g(H_i^\pm W^\mp a_j)(p^\mu_{a_i}-p^\mu_{H^{\pm}})\eqdot
\end{align}
They obey sum rules for any $j$~\cite{Bento:2017eti},\footnote{In the presence of \cp violation, the sum rule holds separately for the real and imaginary components of the \cp-admixed neutral Higgs bosons.}
\begin{align}
    \sum_i |g(H_i^\pm W^\mp h_j)|^2 & = \frac{g^2}{4}\left(1 - c(h_j VV)^2\right)\eqcomma \label{eq:sumrule_HpmWmh} \\
    \sum_i |g(H_i^\pm W^\mp a_j)|^2 & = \frac{g^2}{4}\eqcomma \label{eq:sumrule_HpmWma}
\end{align}
where the sum runs over all charged Higgs bosons (excluding the charged Goldstone boson).

This sum-rule structure leads to important correlations between constraints on the couplings $c(h_j VV)$ and the couplings $g(H^\pm_i W^\mp h_j)$. In particular, in the alignment limit where one of the neutral $\cp$-even Higgs bosons is SM-like, $h_j = h_\SM$, these correlations imply $\sum_i g(H^\pm_i W^\mp h_j) = 0$. In turn, the remaining neutral Higgs bosons $h_i$ ($i\ne j$) will have rather large couplings to the charged Higgs bosons and the $W$ boson. In the special case of the \cp-conserving Two-Higgs-Doublet Model (2HDM), which contains only two \cp-even neutral Higgs bosons and a single pair of charged Higgs bosons, this implies that the coupling between the charged Higgs boson, the non-SM-like \cp-even Higgs boson and the $W$ boson is maximal in the alignment limit. We will discuss this case in more detail in \cref{sec:2hdm}.

Phenomenologically, these correlations have very important implications for collider searches for charged and neutral Higgs bosons. In particular, if there is a sizable difference between a charged Higgs boson mass, $m_{H_i^\pm}$, and the mass of a neutral, non-SM-like Higgs bosons, $m_{h_j}$, the decay modes
\begin{equation}
    H_i^\pm \to W^\pm h_j  \qquad \mbox{if} \quad m_{H_i^\pm} > m_{h_j} \label{eq:Hpm_to_Wh_i}
\end{equation}
and
\begin{equation}
    h_i \to  H^\pm_jW^\mp \qquad \mbox{if} \quad m_{h_i} > m_{H_j^\pm} \label{eq:h_i_to_WHpm}
\end{equation}
can have sizable rates that potentially dominate and, in turn, suppress the Higgs decay modes to SM fermions and gauge bosons. We shall focus on the first case, \cref{eq:Hpm_to_Wh_i}, in this work.

\subsection{The Two Higgs Doublet Model}
\label{sec:2hdm}

The two Higgs doublet model (2HDM) (see~\ccite{Gunion:1989we, Branco:2011iw} for reviews) is the simplest extension of the SM containing a charged Higgs state $H^\pm$, as it adds one additional Higgs doublet to the SM\@. In the present work, we focus on the most commonly studied version: the \cp-conserving 2HDM with a softly broken $\mathbb{Z}_2$ symmetry. Its scalar potential is given by
\begin{align}
    V_{\text{2HDM}}(\Phi_1,\Phi_2) ={} & m_{11}^2\,\Pdd + m_{22}^2\,\Puu - m_{12}^2\left(\Pdu + \Pud\right) \nonumber                               \\
                                       & + \frac{1}{2}\lambda_1 (\Pdd)^2 + \frac{1}{2}\lambda_2 (\Puu)^2  + \lambda_3 (\Pdd)(\Puu) \nonumber        \\
                                       & + \lambda_4 (\Pdu)(\Pud) + \frac{1}{2}\lambda_5 \left((\Pdu)^2 + (\Pud)^2\right) \label{eq:HiggsPotential}
\end{align}
with the scalar doublets
\begin{align}
    \Phi_{1,2} =
    \begin{pmatrix}
        \phi_i^+ \\ \frac{1}{\sqrt{2}}(v_i + \phi_i + i\chi_i)
    \end{pmatrix}\eqdot\label{eq:doubletfields}
\end{align}
It is useful to rotate to the Higgs basis (see e.g.~\cite{Gunion:2002zf}),
\begin{align}
    \begin{pmatrix}
        H_1 \\ H_2
    \end{pmatrix}
    =
    \begin{pmatrix}
        c_\beta  & s_\beta \\
        -s_\beta & c_\beta
    \end{pmatrix}
    \begin{pmatrix}
        \Phi_1 \\ \Phi_2
    \end{pmatrix}\eqdot
\end{align}
where we introduced the abbreviations $s_\gamma \equiv \sin\gamma$ and $c_\gamma \equiv \cos\gamma$ for a generic angle $\gamma$. The angle $\beta$ is defined via the ratio of the vacuum expectation values (vevs), $t_\beta \equiv \tan\beta = v_2/v_1$. In this basis, only $H_1$ receives a vev.

In the Higgs basis the charged Higgs field $H^\pm$ and the \cp-odd scalar field $A$ are automatically mass eigenstates. To obtain the mass eigenstates $h_1$ and $h_2$ of the \cp-even scalars --- whose masses fulfill, by definition, $m_{h_1} \le m_{h_2}$ --- a further rotation is necessary,
\begin{align}
    \begin{pmatrix}
        h_1 \\ h_2
    \end{pmatrix}
    =
    R
    \begin{pmatrix}
        h_1^\text{HB} \\ h_2^\text{HB}
    \end{pmatrix}
    =
    \begin{pmatrix}
        s_{\beta - \alpha} & c_{\beta - \alpha}   \\
        c_{\beta - \alpha} & - s_{\beta - \alpha}
    \end{pmatrix}
    \begin{pmatrix}
        h_1^\text{HB} \\ h_2^\text{HB}
    \end{pmatrix}
    \eqcomma\label{eq:mixmat}
\end{align}
where $h_{1,2}^\text{HB}$ are the \cp-even scalars in the Higgs basis and $\alpha$ is the rotation angle relating the fields $\phi_i$ of \cref{eq:doubletfields} to the mass eigenstates (see e.g.~Ref.~\cite{Gunion:2002zf} for explicit formulas relating $\alpha$ to the potential parameters in \cref{eq:HiggsPotential}). In the following, we will denote the SM-like Higgs boson among $h_{1,2}$ by $h_{125}$ and the non-SM-like Higgs boson by $h_\BSM$.

The unitary matrix $R$ is crucial for the phenomenology of the 2HDM. For instance, the couplings of the \cp-even scalars to gauge bosons $V\in \{W^\pm,Z\}$ are given by
\begin{equation}
    c(h_i VV) = R_{i1}\eqcomma
\end{equation}
where $R_{i1}$ is the $(i1)$-entry of the mixing matrix $R$.

In the alignment limit, where $s_{\beta-\alpha}\simeq 1$ if $h_1\to h_\SM$ and $c_{\beta-\alpha}\simeq 1$ if $h_2\to h_\SM$, these couplings are maximized for the SM-like Higgs boson, i.e.~it acquires SM-like couplings to gauge bosons. At the same time, orthogonality of the mixing matrix, \cref{eq:mixmat}, or equivalently \cref{eq:hVVsum}, implies that the couplings of the other \cp-even neutral Higgs boson to gauge bosons have to vanish.

As already discussed in \cref{sec:nhdm}, this has important consequences for the charged Higgs boson $H^\pm$. Its coupling to a neutral Higgs boson $h_i$ and a $W^\pm$ is given by
\begin{align}
    g(H^\pm W^\mp h_i) & =-\frac{g}{2}R_{i2} (p^\mu_{h_i}-p^\mu_{H^{\pm}})\eqdot\label{eq:HpHWcoup}
\end{align}
In the alignment limit this coupling vanishes for $h_{125}$ and is maximized for $h_\BSM$ since $R$ is a $2\times2$ orthogonal matrix. The coupling of the charged Higgs boson to the $A$ boson and a $W^\pm$ boson,
\begin{align}
    g(H^\pm W^\mp A) & =-\frac{g}{2}\eqcomma\label{eq:HpAWcoup}
\end{align}
is not affected by the alignment of the SM-like Higgs boson.

The Yukawa sector of the 2HDM is determined by its symmetry structure. In order to avoid tree-level flavor changing neutral currents, the $\mathbb{Z}_2$ symmetry can be extended to the fermion sector resulting in four distinct types of Yukawa sectors (see \cref{tab:thdm_types}): type~I, type~II, flipped, and lepton specific. In this work we will focus on type-I and lepton-specific models since in type~II (and flipped) 2HDMs measurements of flavor observables constrain the charged Higgs boson to be very heavy~\cite{Misiak:2017bgg,Arbey:2017gmh}. In type-I Yukawa sectors, the SM-normalized couplings of each Higgs boson to all fermion types are equal.

\begin{table}
    \centering
    \small
    \begin{tabular}{ c  c  c  c  c  c  c }
        \toprule
        Type            & $u_R$ & $d_R$ & $l_R$ & $\lambda_{uu}$ & $\lambda_{dd}$ & $\lambda_{ll}$ \\
        \midrule
        I               & $+$   & $+$   & $+$   & $\cot\beta$    & $ \cot\beta$   & $ \cot\beta$   \\
        II              & $+$   & $-$   & $-$   & $\cot\beta$    & $-\tan\beta$   & $-\tan\beta$   \\
        Flipped         & $+$   & $-$   & $+$   & $\cot\beta$    & $-\tan\beta$   & $ \cot\beta$   \\
        Lepton-specific & $+$   & $+$   & $-$   & $\cot\beta$    & $ \cot\beta$   & $-\tan\beta$   \\
        \bottomrule
    \end{tabular}
    \caption{Assignment of the $\mathbb{Z}_2$ charges to the right-handed fermions for the different types of 2HDMs. The resulting coefficients of the charged Higgs--fermion interaction are also shown.}
    \label{tab:thdm_types}
\end{table}

Via this Yukawa interaction, also the charged Higgs boson interacts with fermions,
\begin{equation}
    \mathcal{L}_\text{Yuk} \supset -\sum_i H^+\left(
    \frac{\sqrt{2}V_{u_id_j}^\text{CKM}}{v}\bar{u_i}\left(m_{u_i}\lambda_{uu}P_L + m_{d_j}\lambda_{dd} P_R\right)d_j
    + \frac{\sqrt{2}}{v}m_{l_j}\lambda_{ll}\bar{\nu_i}_L {l_j}_R
    \right)\eqcomma\label{eq:fermcoup}
\end{equation}
where $V^\text{CKM}$ is the Cabibbo-Kobayashi-Maskawa matrix, and $P_{L,R}$ are the left- and right-handed chirality projection operators.

As a consequence of the coupling dependence on the quark masses ($m_{u_i}$ and $m_{d_i}$) and the lepton masses ($m_{l_j}$), the most important charged Higgs decay modes to fermions are $H^+\to c\bar{b}$ and $H^+\to \tau^+\nu$ for low masses below the top quark mass, and $H^+\to t\bar{b}$ for heavier masses.

For the rest of the paper we will work in the 2HDM framework introduced above. In many extended models, the charged Higgs boson phenomenology is very similar. Often, only the overall rates of the $H^\pm_i\rightarrow W^\pm h_j$ decays are expected to be lower due to the corresponding sum rules (see \cref{eq:sumrule_HpmWmh,eq:sumrule_HpmWma}).

\subsection{Phenomenological scan of the Two Higgs Doublet Model}%
\label{sec:2hdm_scan}

We base our phenomenological study on a large parameter scan of the 2HDM type-I parameter space using the code \texttt{ScannerS}~\cite{Coimbra:2013qq,Ferreira:2014dya,Costa:2015llh,Muhlleitner:2016mzt,Muhlleitner:2020wwk}. We require all parameter points to fulfill current theoretical and experimental constraints. These include
\begin{itemize}
    \item tree-level perturbative unitarity and boundedness from below (BfB)~\cite{Branco:2011iw},
    \item absolute stability\footnote{A sufficiently long-lived metastable EW vacuum would be acceptable as well. In the 2HDM, most parameter regions that could feature a metastable EW vacuum are, however, excluded by LHC measurements~\cite{Barroso:2013awa}. We, therefore, do not expect any different collider phenomenology by allowing metastable vacua, and exclude them for simplicity.} of the tree-level vacuum~\cite{Barroso:2013awa},
    \item electroweak precision constraints through the oblique parameters $S$, $T$ and $U$ using the prediction of Refs.~\cite{Grimus:2007if,Grimus:2008nb} and the fit result of Ref.~\cite{Haller:2018nnx},
    \item flavor constraints using the results of Ref.~\cite{Haller:2018nnx},
    \item bounds from searches for additional scalars using \HBv{5.9.0}~\cite{Bechtle:2008jh,Bechtle:2011sb,Bechtle:2013gu,Bechtle:2013wla,Bechtle:2015pma,Bechtle:2020pkv},
    \item and agreement with the Higgs signal measurements using \HSv{2.6.0}~\cite{Stal:2013hwa,Bechtle:2013xfa,Bechtle:2014ewa,Bechtle:2020uwn}, which incorporates the combined LHC Run-1 results~\cite{Khachatryan:2016vau} as well as the latest Run-2 Higgs measurements by the ATLAS~\cite{Aaboud:2018gay,Aaboud:2017rss,ATLAS:2019nvo,Aaboud:2018jqu,Aad:2020mkp,ATLAS:2019jst,Aaboud:2018pen,Aad:2019lpq,Aad:2020jym} and CMS~\cite{Sirunyan:2018hbu,Sirunyan:2017elk,Sirunyan:2017dgc,Sirunyan:2018mvw,CMS:2019lcn,Sirunyan:2018shy,CMS:2018dmv,Sirunyan:2020tzo,CMS:2019chr,CMS:1900lgv,Sirunyan:2020sum,CMS:2019pyn} collaborations.
\end{itemize}
The required branching ratios of the scalars are calculated using \texttt{HDECAY}~\cite{Djouadi:1997yw,Harlander:2013qxa,Djouadi:2018xqq} and the neutral scalar production cross sections using \texttt{SusHi}~\cite{Harlander:2012pb, Harlander:2016hcx}.\footnote{The used cross-section calculations for charged Higgs boson production are described in detail in \cref{sec:Hpm_prod}.} See \ccite{Muhlleitner:2020wwk} for details on the scanning procedure.

\begin{table}
    \centering
    \begin{tabular}{lS[table-format=4.0]S[table-format=4.0]S[table-format=4.0]S[table-format=2.1]S[table-format=+1.1]c}
        \toprule
            & {$m_{h_\BSM}$ [\si{\GeV}]} & {$m_A$ [\si{\GeV}]} & {$m_{H^\pm}$ [\si{\GeV}]} & {$\tan\beta$} & {$c(h_\BSM VV)$} & {$m_{12}^2$  [\si{\GeV^2}]} \\
        \midrule
        min & 30                         & 30                  & 50                        & 0.8           & -0.5             & 0                           \\
        max & 1000                       & 1000                & 1000                      & 25.0          & 0.5              & $10^6$                      \\
        \bottomrule
    \end{tabular}
    \caption{Input parameter ranges for the parameter scan in the type I 2HDM.}%
    \label{tab:scanpars}
\end{table}

We fix the mass of the $h_{125}$ boson to its observed value from the ATLAS and CMS LHC Run-1 combined analysis, $m_{h_{125}}=\SI{125.09}{\GeV}$~\cite{Aad:2015zhl}, and uniformly sample the remaining model parameters within the ranges given in \cref{tab:scanpars}. For convenience, we choose the coupling $c(h_\BSM VV)$ as input parameter in order to cover the two possible cases $h_1 \simeq h_\SM$ and $h_2 \simeq h_\SM$ together in one scan (see \ccite{Muhlleitner:2020wwk} for details). Valid parameter points with $\tan\beta$ larger than the chosen upper limit are possible. However, since we are interested in scenarios where both the fermionic channels --- suppressed by large $\tan\beta$ --- and the bosonic channels --- mostly independent of $\tan\beta$ --- are potentially observable at the LHC, we focus on the low and medium $\tan\beta$ region and use the arbitrary upper limit of $\tan\beta<25$. In the following, we show results for a sample of $10^6$ parameter points that fulfill all of the above constraints (at the $2\sigma$ level, where applicable).

\begin{figure}
    \centering
    \includegraphics[width=0.6\textwidth]{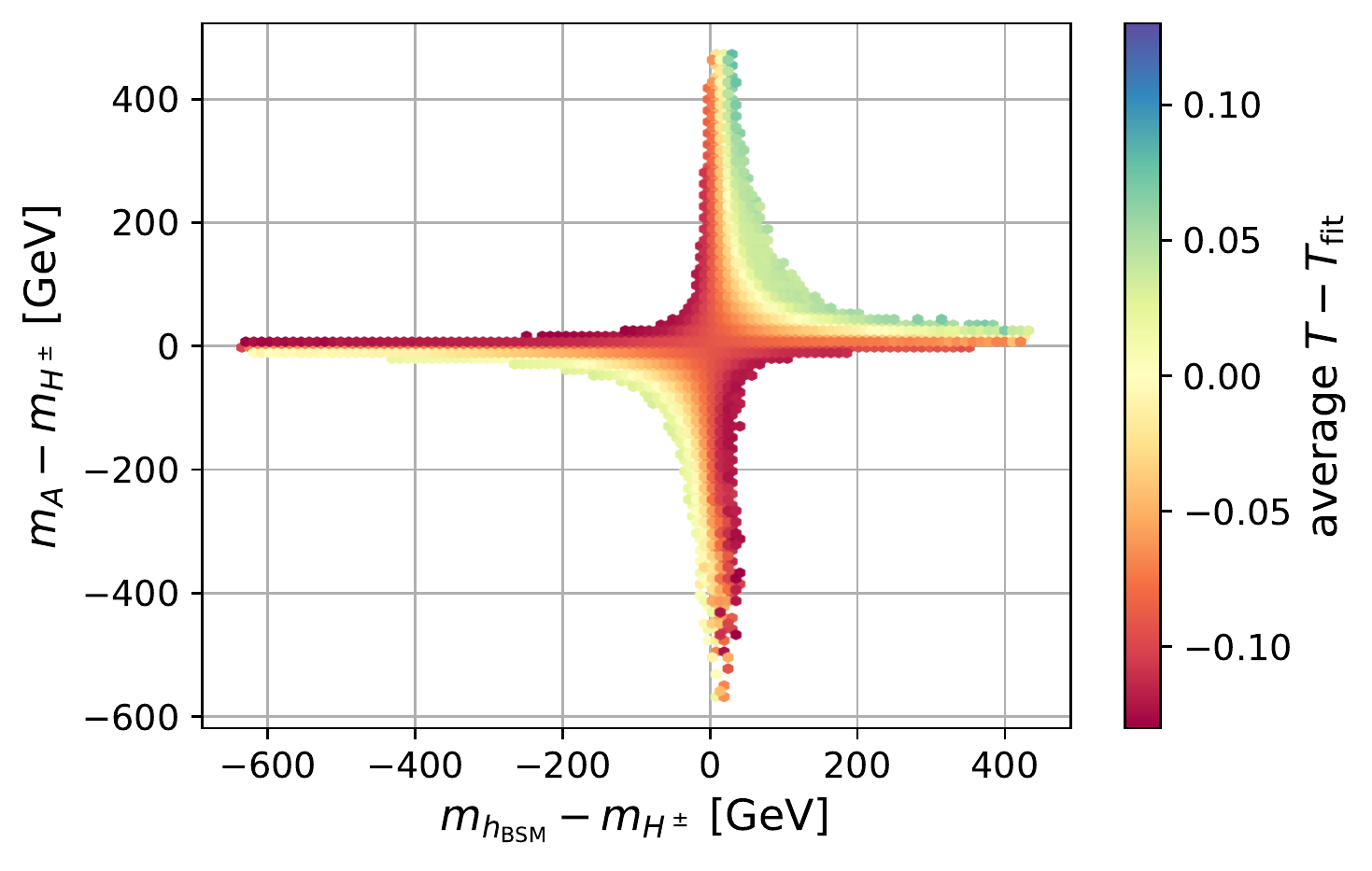}
    \caption{Possible mass separations between the charged Higgs boson and the non-$h_{125}$ neutral Higgs bosons. The color code indicates the (per-bin averaged) deviation of the  $T$ parameter from the central value of the fit from \ccite{Haller:2018nnx}.}%
    \label{fig:Tparam}
\end{figure}

As a first scan result, we investigate the well-known and important impact of the electroweak precision constraints on the Higgs mass spectrum (see e.g.~\ccite{Haber:2010bw}). Especially the constraint on the $T$ parameter forces $m_{H^\pm}$ to be always close to one of the neutral Higgs masses. This is illustrated in \cref{fig:Tparam} showing the deviation of the $T$ parameter from the central value of the fit in \ccite{Haller:2018nnx} in the $(m_{h_\BSM}-m_{H^\pm},m_{A}-m_{H^\pm})$ plane. As mentioned above, only parameter points with a deviation less than $2\sigma$ are shown. It is clearly visible that either $m_{h_\BSM} \sim m_{H^\pm}$ or $m_{A} \sim m_{H^\pm}$ needs to be fulfilled. In the context of the charged Higgs boson decay into a $W$ boson and a lighter Higgs boson, as discussed in \cref{sec:nhdm,sec:2hdm}, this constraint implies that either $h_\BSM$ or $A$ --- but not both --- can be significantly lighter than the charged Higgs boson. As a consequence, at least one of the channels $H^\pm \to h_\BSM W^\pm$ or $H^\pm\to A W^\pm$ can be kinematically accessible in large parts of the parameter space.

\begin{figure}
    \centering
    \includegraphics[width=0.6\textwidth]{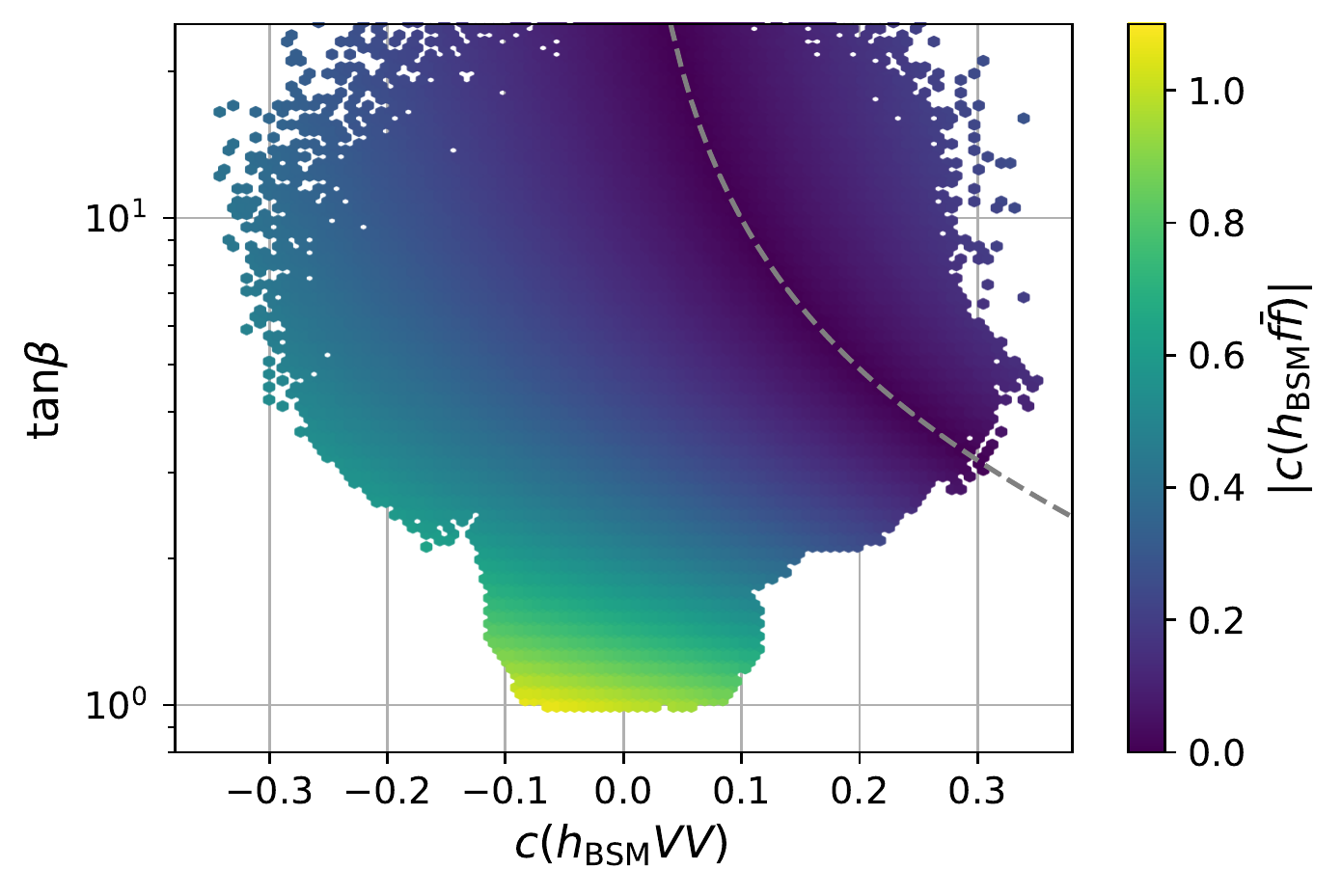}
    \caption{The allowed parameter region for the effective gauge coupling $c(h_\BSM VV)$ of the non-$h_{125}$ \cp-even scalar $h_\BSM$ and $\tan\beta$. The exact alignment limit for $h_{125}$ is realized when $c(h_\BSM VV)=0$. The color code represents the SM-normalized fermion coupling of $h_{\BSM}$, which vanishes in the fermiophobic limit indicated by the dashed line. Parameter points where both $m_{h_{1,2}}\in[120,130]$~GeV are not shown to allow a clear distinction between $h_\BSM$ and $h_{125}$.}
    \label{fig:tbeta}
\end{figure}

\cref{fig:tbeta} shows the scan results in the plane of the two important coupling parameters $c(h_\BSM VV)$ and $\tan\beta$. The Higgs signal rate measurements constrain
\begin{equation}
    |c(h_\BSM VV)| \lesssim 0.3
\end{equation}
for almost all allowed parameter points in this scan, thus the alignment limit for $h_{125}$ --- which would mean $c(h_\BSM VV)=0$ --- is always approximately realized. With \cref{eq:sumrule_HpmWmh} this in turn means that the $H^\pm W^\mp h_\BSM$ coupling reaches always at least $\sqrt{1-0.3^2}\approx \SI{95}{\percent}$ of its maximum possible value. The color map in \cref{fig:tbeta} shows the SM-normalized coupling of $h_\BSM$ to fermions,
\begin{align}
    c(h_\BSM f\bar f) \equiv \frac{g(h_\BSM f\bar f)}{g(h_\SM f\bar f)} =
    \begin{cases}
        \tfrac{c_\alpha}{s_\beta} \; \text{ if } h_1 \equiv h_\BSM \eqcomma \\
        \tfrac{s_\alpha}{s_\beta} \; \text{ if } h_2 \equiv h_\BSM \eqcomma
    \end{cases}
\end{align}
which is identical for all fermions if the Yukawa sector is of type~I.\footnote{For the lepton-specific 2HDM, the coupling of $h_\BSM$ to leptons is $-s_\alpha/c_\beta$ if $h_1 \equiv h_\BSM$ and $c_\alpha/c_\beta$ if $h_2 \equiv h_\BSM$. The couplings to quarks are identical to those in the 2HDM type~I.} It can clearly be seen that $c(h_\BSM f\bar{f})$ becomes small for large $\tan\beta$. Furthermore, in the 2HDM there exists a so-called \emph{fermiophobic} limit for each of the \cp-even neutral scalars (no corresponding limit exists for the $A$ boson), where the couplings of this particle to SM-fermions vanish. This limit is indicated for $h_\BSM$ by the dashed line in \cref{fig:tbeta} and is reached for
\begin{equation}
    \tan\beta = \frac{\sqrt{1 - c(h_\BSM VV)^2}}{c(h_\BSM VV)}\eqdot\label{eq:fermiophobic}
\end{equation}
It is clear from the equation and also visible in \cref{fig:tbeta} that this limit is only reachable if the $h_{125}$ alignment is not exact. Trying to satisfy \cref{eq:fermiophobic} in the case of exact alignment would require $\tan\beta\to\infty$.

The fermiophobic limit is phenomenologically interesting in our study because most of the direct LHC searches for $h_\BSM$ rely on production and decay modes governed by its coupling to fermions, e.g.~gluon fusion, which vanish in the fermiophobic limit. Only the gauge-boson mediated production channels --- such as vector boson fusion (VBF) and $W/Z$-associated production --- are non-zero, but are suppressed by the small $c(h_\BSM VV)$. Similarly, the $h_\BSM$ decay modes to fermions are suppressed (or even vanish in the exact fermiophobic limit). In such a scenario, production of $H^\pm$ followed by $H^\pm\to h_\BSM W^\pm$ could well be the most promising discovery channel for both $H^\pm$ and $h_\BSM$. We come back to the discussion of the fermiophobic Higgs limit in \cref{sec:benchmarks}, where we introduce a dedicated benchmark scenario that features a fermiophobic non-SM-like Higgs boson.

\section{Genuine BSM effects on the \texorpdfstring{$h_{125}$}{h125} properties}
\label{sec:BSMeffects}

As outlined in \cref{sec:2hdm}, the role of the observed Higgs boson $h_{125}$ can be played by either the lighter or the heavier \cp-even neutral Higgs state, $h_1$ or $h_2$. Its tree-level couplings to fermions and gauge bosons become identical to the predictions of the SM in the alignment limit. However, through the interplay with the remaining Higgs states of the model, deviations from the SM Higgs properties can still occur in the exact alignment limit. These can either be loop-induced (effective) coupling modifications, or additional decay modes to BSM Higgs bosons. In this section we will first discuss the charged Higgs boson contribution to the $h_{125} \to \gamma\gamma$ decay, and then elaborate on the possibility of the decays $h_{125}\to h_\BSM h_\BSM$ and $h_{125}\to A A$ in case of very light $h_\BSM$ and $A$ bosons.

\subsection{Charged Higgs boson contribution to \texorpdfstring{$h_{125} \to \gamma \gamma$}{h125->gaga}}
\label{sec:h125_gaga}

\begin{figure}
    \centering
    \includegraphics[scale=1]{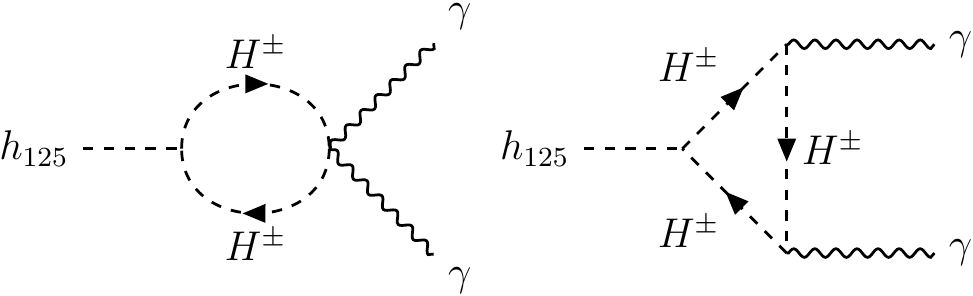}
    \caption{Charged Higgs contribution to $h_{125}\rightarrow\gamma\gamma$ at the one-loop level.}
    \label{fig:Hp_gaga_contr}
\end{figure}

The $h_{125} \to \gamma \gamma$ decay channel is currently measured with an accuracy at the level of $\sim \SI{10}{\%}$~\cite{Sirunyan:2018koj,ATLAS:2020pvn}, which is expected to improve to $\lesssim \SI{3}{\%}$ at the HL-LHC~\cite{Cepeda:2019klc}. The charged Higgs boson induces deviations from the SM prediction at the leading order, as depicted in \cref{fig:Hp_gaga_contr}. These corrections do not vanish in the alignment limit in which all tree-level couplings of $h_{125}$ are exactly equal to the respective SM values. Interestingly, these corrections also do not necessarily vanish if the charged Higgs boson is much heavier than the electroweak scale. This non-decoupling effect (described in detail in~\ccite{Arhrib:2004ak,Bhattacharyya:2013rya,Ferreira:2014naa,Dumont:2014wha,Bhattacharyya:2014oka,Bernon:2015qea,Bernon:2015wef}) opens the possibility to indirectly probe the charged Higgs boson via precision $h_{125}\to \gamma\gamma$ measurement.

The relevant couplings of the \cp-even Higgs bosons $h_i$ ($i=1,2$) to a pair of charged Higgs boson are given by
\begin{equation}
    g_{h_i H^+H^-} = - \frac{1}{v} \Big\{\left[m_{h_1}^2 + 2(m_{H^\pm}^2 - \mbarsq) \right] R_{i1} + 2 (m_{h_1}^2 - \mbarsq ) \frac{R_{i2}}{t_{2\beta}} \Big\}\eqcomma
\end{equation}
where $\mbarsq = m_{12}^2/(s_\beta c_\beta)$ and $R$ is the unitary mixing matrix defined in \cref{eq:mixmat}. In the alignment limit, in which one of the $h_i$ becomes SM-like (implying that $R_{i1} \rightarrow 1$), we obtain
\begin{equation}
    g_{h_{125} H^+H^-} \to -\frac{1}{v}\left[m_{h_{125}}^2 + 2(m_{H^\pm}^2 - \mbarsq)\right] \eqcomma \label{eq:gh125HpHm_align}
\end{equation}
In the $m_{H^\pm} \gg v$ limit, the terms involving $m_{H^\pm}^2$ can compensate the suppression arising through the loop integrals which scale proportional to $v^2/m_{H^\pm}$. Consequently, the charged Higgs boson contribution to the di-photon decay rate can reach a constant value even if $m_{H^\pm}\gg v$.

\begin{figure}
    \centering
    \includegraphics[width= 0.6\textwidth]{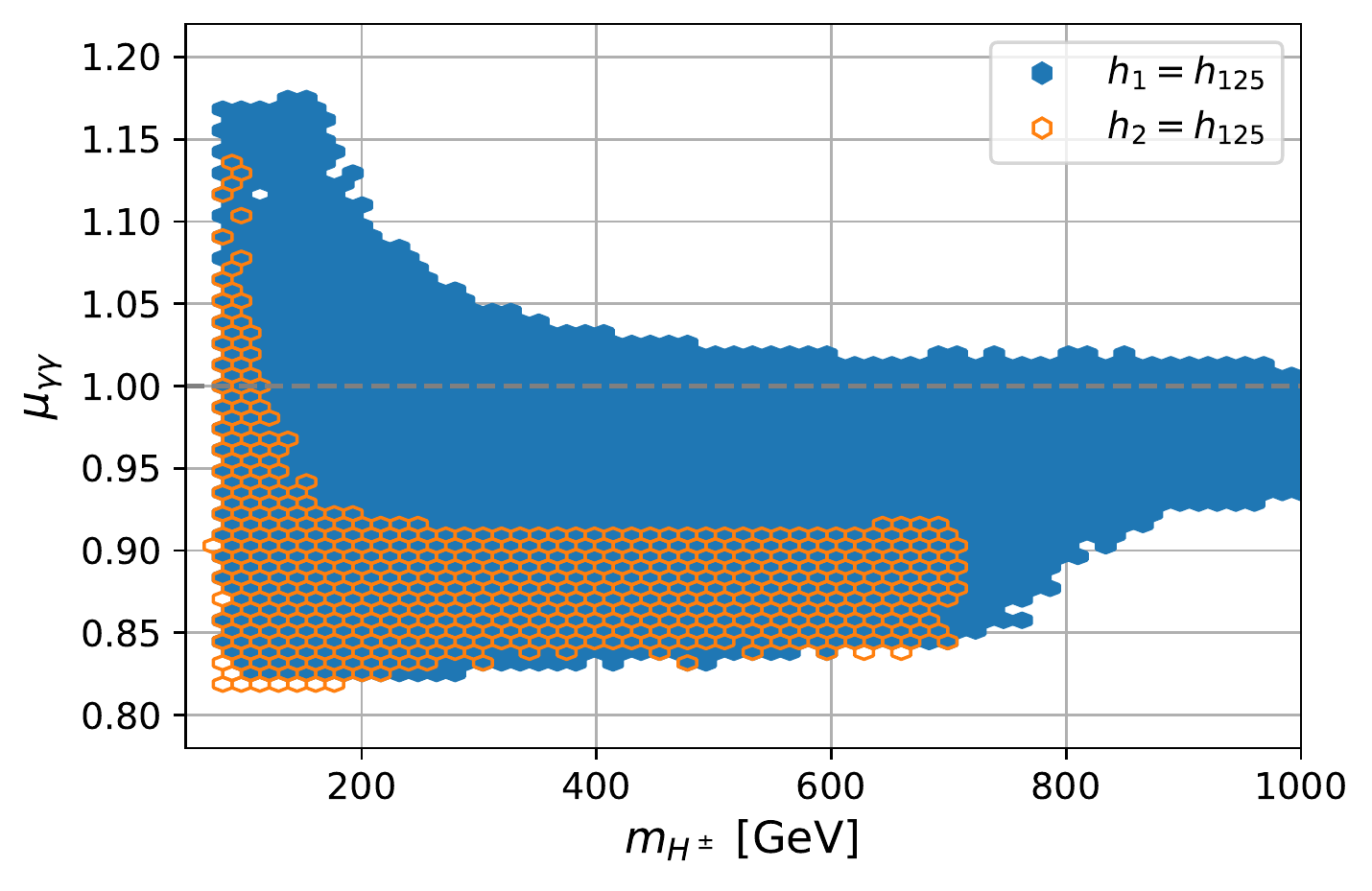}
    \caption{Scatter plot of the signal strength $\mu_{\gamma\gamma}$ of $h_{125}$ as a function of the charged Higgs mass. The dashed line indicates the SM value of $\mu_{\gamma\gamma}=1$. For the blue (orange) points the light (heavy) \cp-even Higgs boson is $h_{125}$. To allow a clear distinction of these cases, parameter points where both $m_{h_1}, m_{h_2}\in[120,130]$~GeV are not shown.}
    \label{fig:mugam}
\end{figure}

While the $h_1 = h_{125}$ and $h_2 = h_{125}$ cases appear to be very similar, they actually have a distinct phenomenology. This is illustrated in \cref{fig:mugam} showing the di-photon signal strength of $h_{125}$ as a function of $m_{H^\pm}$ for the parameter scan described in \cref{sec:2hdm_scan}. For the blue points, the lighter \cp-even Higgs boson is SM-like ($h_1 = h_{125}$); for the orange points, the heavier \cp-even Higgs boson is SM-like ($h_2 = h_{125}$). Scan points for which $m_{h_1}, m_{h_2}\in[120,130]$~GeV are not shown in order to allow for a clear distinction of these cases.

In case of $h_1 = h_{125}$, the di-photon rate can deviate from the SM value by up to $\pm \SI{17}{\%}$ for $m_{H^\pm} \lesssim \SI{200}{\GeV}$. For $m_{H^\pm} \gtrsim \SI{300}{\GeV}$, positive deviations from the SM of up to $\sim \SI{5}{\%}$ and negative deviations of up to $\sim \SI{15}{\%}$ are possible. If the di-photon rate is close to the SM value, the term proportional to $\mbarsq$ compensates the remaining terms in \cref{eq:gh125HpHm_align}. This is exactly what happens in the decoupling limit of the 2HDM~\cite{Gunion:2002zf}, where $m_H^2\approx m_A^2\approx m_{H^\pm}^2 \approx m_{12}^2\gg v^2$ and this contribution vanishes as expected.

This is not possible in the case of $h_2 = h_{125}$. If $m_{h_1}, m_{h_2} \ll m_{H^\pm}$, constraints arising from perturbative unitarity (and perturbativity) force $\mbarsq \sim \mathcal{O}(v^2)$ and the term proportional to $m_{H^\pm}^2\gg\mathcal{O}(v^2)$ in \cref{eq:gh125HpHm_align} can not be compensated (see \ccite{Bernon:2015wef} for more details). Therefore, the charged Higgs boson contribution to the di-photon rate reaches an approximately constant value for $m_{H^\pm} \gtrsim \SI{200}{\GeV}$ with a negative deviation from the SM by $\sim 9-\SI{15}{\%}$. Perturbative unitarity also implies that $m_{H^\pm}$ can not reach values above $\sim \SI{700}{\GeV}$ if $h_2 = h_{125}$. Given these observations, we may optimistically argue that the $h_2 = h_{125}$ scenario in the 2HDM can completely be covered experimentally at the HL-LHC, using two complementary probes: in the heavy $H^\pm$ regime, $m_{H^\pm} \gtrsim \SI{200}{\GeV}$,  the deviations from the SM in the $h_{125}\to\gamma\gamma$ rate will be probed with corresponding precision measurements. In contrast, the light $H^\pm$ regime, $m_{H^\pm} < \SI{200}{\GeV}$, may be probed experimentally with direct $H^\pm$ searches due rather large production cross sections. In this endeavor, though, all possible $H^\pm$ collider signatures have to be searched for, including, in particular, the signatures arising from the $H^\pm \to W^\pm h_\BSM$ decay, as proposed in this work.

The charged Higgs contribution to the $h_{125}$ di-photon rate also affects the selection of the benchmark scenarios to be discussed in \cref{sec:benchmarks} (see also the discussion in \cref{sec:diphoton_rate}).

\subsection{The \texorpdfstring{$h_{125} \to h_\BSM h_\BSM$}{h125-> hBSM hBSM} and \texorpdfstring{$h_{125} \to AA$}{h125 -> A A} decay modes}
\label{sec:hsm_to2hbsm}

The parameter space with a very light non-SM-like neutral Higgs boson $h_\BSM$ or $A$, with a mass $\sim \mathcal{O}(\text{few}~\SI{10}{\GeV})$, is interesting for the charged Higgs boson phenomenology for two reasons: First, it kinematically enables the decay $H^\pm \to W^\pm h_\BSM$ or $H^\pm \to W^\pm A$, respectively, even in case that the charged Higgs boson is very light, e.g., below the top quark mass ($m_{H^\pm} \lesssim m_t$). This, in turn, can lead to a sizable charged Higgs boson production cross section. Second, very light BSM Higgs bosons in the final state can lead to distinct kinematics, e.g., boosted and collimated BSM Higgs decay products, warranting the design of specific analysis techniques in the experimental searches. However, scenarios with a light non-SM-like neutral Higgs boson with mass below $m_{h_{125}}/2 \approx \SI{62.5}{\GeV}$ have to obey stringent constraints, as the additional decay mode $h_{125} \to h_\BSM h_\BSM$ or $h_{125} \to A A$ is kinematically allowed. In the 2HDM, direct searches for such decays, e.g. \ccite{Sirunyan:2018mbx,Sirunyan:2018pzn,Aaboud:2018iil,Aaboud:2018gmx,Sirunyan:2018mot,Aaboud:2018esj,Sirunyan:2020eum}, are currently weaker than the indirect constraints arising from the $\SI{125}{\GeV}$ Higgs boson precision rate measurements (see e.g.~Ref.~\cite{Robens:2019kga}). In this section we therefore investigate these decays in detail and discuss under which circumstances the constraints can be evaded by suppressing the corresponding decay rate.

The triple scalar couplings governing these decay modes are given by
\begin{align}
    4v\cdot g_{h_1 h_1 h_2} & = \frac{\cba}{s_\beta c_\beta}\left[\left(3\mbarsq- m_H^2 - 2 m_h^2\right) s_{2\alpha} - \mbarsq s_{2\beta}\right]\eqcomma \\
    -v\cdot g_{h_i AA}      & = \left(m_{h_i}^2 + 2 m_A^2 - 2\mbarsq\right)R_{i1} + 2 (m_{h_1}^2 - \mbarsq)R_{i2}/t_{2\beta}\eqdot
\end{align}
Even in the alignment limit $h_{125}\to h_\text{SM}$, the resulting couplings
\begin{align}
    g_{h_{125}h_\text{BSM} h_\text{BSM}} & \to \frac{2\mbarsq - m_{h_{125}}^2 - 2 m_{h_\text{BSM}}^2}{2v}\eqcomma \\
    g_{h_{125} AA}                       & \to \frac{2\mbarsq - m_{h_{125}}^2 - 2 m_A^2 }{v}
\end{align}
can lead to decays of $h_{125}\to h_\text{BSM} h_\text{BSM}, AA$ if kinematically allowed. Even for BSM Higgs masses above $m_{h_{125}}/2$, the off-shell branching ratios into these final states can be substantial.

For the purpose of studying charged Higgs phenomenology for $m_{h_\BSM/A} < \SI{62.5}{\GeV}$ it is useful to define scenarios where the $h_{125}\to h_\BSM h_\BSM,AA$ decays are not the most sensitive observables (in order to not be experimentally excluded by these channels). This can be accomplished by choosing $\mbarsq$ and thus $m_{12}^2$ such that the respective triple scalar couplings are zero,
\begin{align}
    g_{h_1 h_1 h_2} & = 0: &  & m_{12}^2 = \frac{(m_{h_2}^2 + 2 m_{h_1}^2)c_\alpha s_\alpha}{3\frac{c_\alpha s_\alpha}{c_\beta s_\beta}-1}\eqcomma \label{eq:zero_gh1h1h2_m12sq}                                      \\
    g_{h_i AA}      & = 0: &  & m_{12}^2 = \frac{1}{2}\left( (2 m_A^2 + m_{h_1}^2) R_{i1} + 2 m_{h_1}^2 R_{i2}/t_{2\beta} \right)\frac{c_\beta s_\beta}{R_{i1} + R_{i2}/t_{2\beta}}\eqdot \label{eq:zero_gh1AA_m12sq}
\end{align}
For the benchmark scenarios to be discussed in \cref{sec:benchmarks}, we only consider the case of a light \cp-even Higgs boson for simplicity (we expect the case of light $A$ boson to be very similar phenomenologically). In this context, we find that $g_{h_1 h_1 h_2} = 0$ is not possible in the exact $h_2=h_{125}\to h_\text{SM}$ alignment limit without violating theoretical constraints and therefore consider scenarios that deviate slightly from alignment (see also \cref{sec:h125hBSMhBSM_suppression}).

\section{Searching for charged Higgs bosons at the LHC}
\label{sec:chargedH_at_LHC}

In this section we will discuss in detail the various LHC production and decay modes of the charged Higgs boson and give an overview of current LHC searches focusing on the 2HDM structure introduced in \cref{sec:2hdm}

\subsection{Charged Higgs boson production at the LHC}
\label{sec:Hpm_prod}

At the LHC there are three main channels for direct charged Higgs boson production (see e.g.~\ccite{Djouadi:2005gj,Akeroyd:2016ymd} for other comprehensive reviews): The charged Higgs boson can be produced in association with a top and a bottom quark, in association with a neutral Higgs boson, and in association with a $W$ boson. These production modes will be discussed in more detail below. In addition to these channels, charged Higgs bosons can also be produced in pairs (see e.g.~\ccite{Djouadi:2005gj}), albeit at a very small rate, or in vector-boson fusion, if the charged Higgs boson originates from a $SU(2)$ Higgs triplet.

All used cross-section values (except of $pp \to H^\pm t b$ production) are available as data tables in the form of ancillary files accompanying the present paper. The cross section calculations in these channels use the MMHT2014 pdf set~\cite{Harland-Lang:2014zoa}.

\subsubsection{\texorpdfstring{$pp \to H^\pm t b$}{pp -> Hpm t b} production}

\begin{figure}
    \centering
    \begin{subfigure}[t]{.45\textwidth}\centering
        \includegraphics[scale=.9]{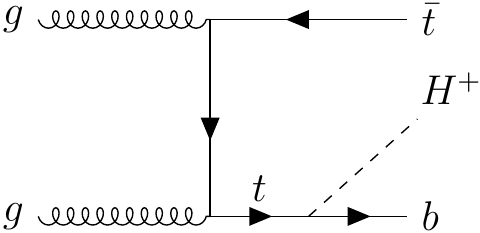}
        \caption{Four-flavor scheme (4FS).}
        \label{fig:Hptb_prod_4FS}
    \end{subfigure}
    \begin{subfigure}[t]{.45\textwidth}\centering
        \includegraphics[scale=.9]{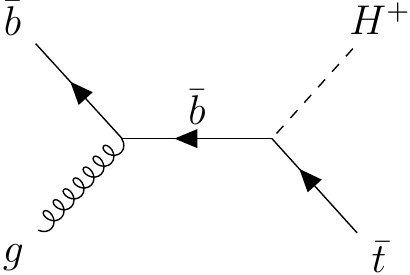}
        \caption{Five-flavor scheme (5FS).}
        \label{fig:Hptb_prod_5FS}
    \end{subfigure}
    \caption{Exemplary Feynman diagrams for $pp\to H^\pm tb$ production.}
    \label{fig:Hptb_prod}
\end{figure}

Charged Higgs boson production in association with a top and a bottom quark can be calculated either in the four-flavor scheme (4FS) as shown in \cref{fig:Hptb_prod_4FS} --- corresponding to the process $gg \rightarrow \bar t b H^+$ --- or in the five-flavor scheme (5FS) as shown in \cref{fig:Hptb_prod_5FS} --- corresponding to the process $g \bar b \rightarrow \bar t H^+$. While the calculation in the 5FS is more precise in the collinear region of phase space in which the transverse momentum of the $b$ quark is small, the 4FS yields more accurate results if the transverse momentum of the $b$ quark is large. To obtain a prediction precise for all phase space regions, both calculations have been matched by identifying terms which would be double-counted if both calculations would be summed naively~\cite{Alwall:2004xw}.

A related issue appears for low charged Higgs boson masses, $m_H^{\pm}$, well below the top-quark mass, $m_t$. In this regime a charged Higgs boson can be produced via the decay of a top quark. This process can be conveniently calculated in the narrow-width approximation. If the mass of the charged Higgs boson is, however, close to the top-quark mass, the narrow-width approximation becomes  invalid and all contributions to the process $pp \to H^\pm t b$ have to be taken into account (see Refs.~\cite{Moretti:2002eu,Assamagan:2004gv,Alwall:2004xw} for LO calculations).

NLO corrections for charged Higgs boson production in association with a top and a bottom quark in the regime of $m_{H^\pm} > m_t$ have been calculated in \ccite{Zhu:2001nt,Gao:2002is,Plehn:2002vy,Berger:2003sm,Kidonakis:2005hc,Dittmaier:2009np,Weydert:2009vr,Klasen:2012wq,Degrande:2015vpa}. In the regime $m_{H^\pm} < m_t$, NLO corrections have been derived in \ccite{Jezabek:1988iv,Li:1990qf,Czarnecki:1992zm,Campbell:2004ch,Czarnecki:1998qc,Chetyrkin:1999ju,Blokland:2004ye,Blokland:2005vq,Czarnecki:2010gb,Gao:2012ja,Brucherseifer:2013iv}. The intermediate mass regime of $m_{H^\pm}\sim m_t$ has been addressed at the NLO level in \ccite{Klasen:2012wq,Berger:2003sm,Degrande:2016hyf}.

\begin{figure}
    \centering
    \includegraphics[width=0.6\textwidth]{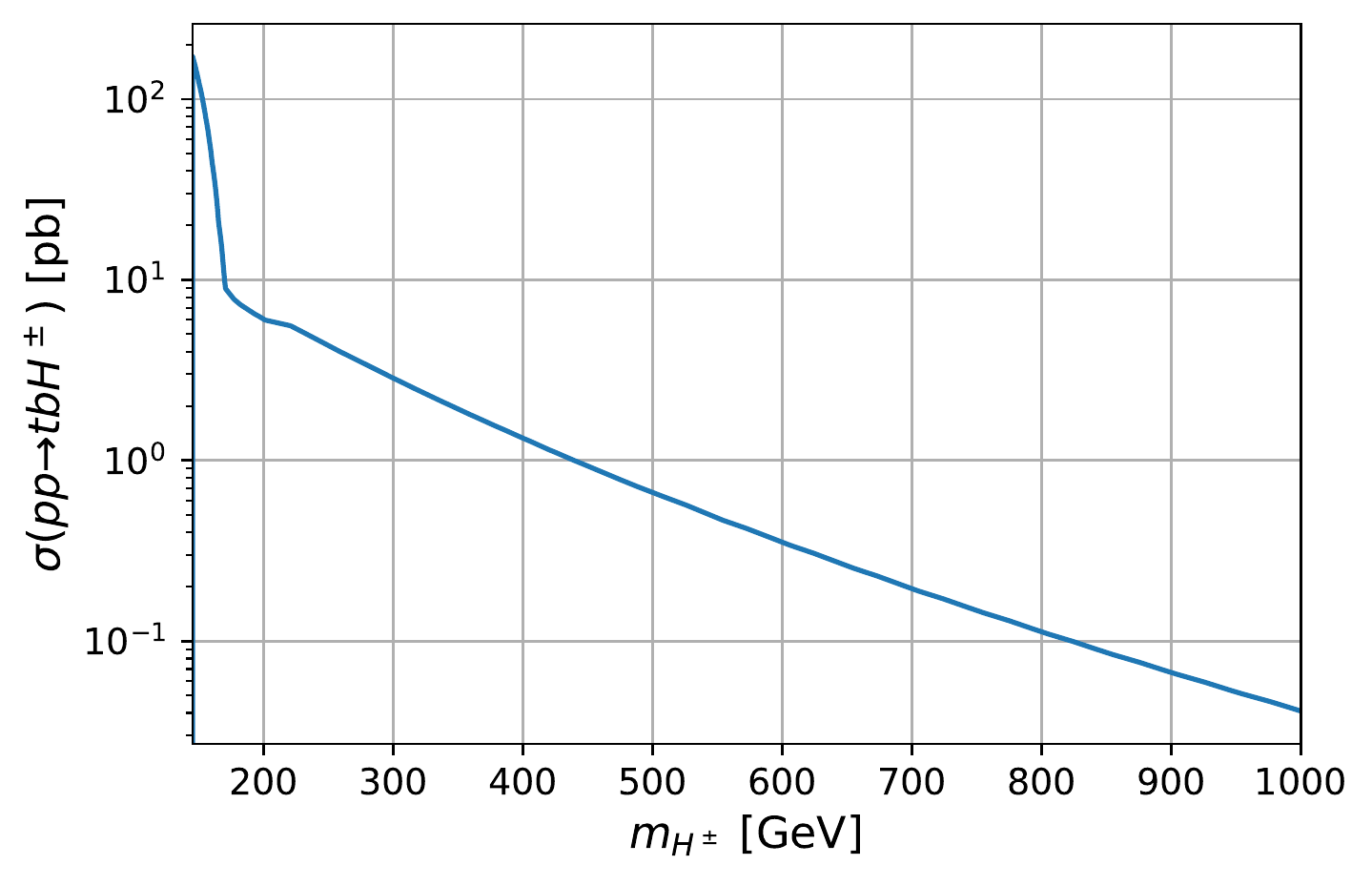}
    \caption{Charged Higgs production cross section in the $pp\to H^\pm tb$ channel at the \SI{13}{\TeV} LHC for $\tan\beta=1$ at NLO~\cite{Degrande:2015vpa,Degrande:2016hyf}.}
    \label{fig:Hptb_cxn}
\end{figure}

For our present study, we use the tabulated NLO results of \ccite{Degrande:2015vpa,Degrande:2016hyf} for $m_{H^\pm} > \SI{145}{\GeV}$. We refer to those references for details on the calculation, the 4FS/5FS matching procedure, and the input parameters. \Cref{fig:Hptb_cxn} shows the corresponding numerical prediction for the 2HDM type-I (and the lepton-specific 2HDM) with $\tan\beta=1$. The cross section for other $\tan\beta$ values can be obtained by dividing by $\tan^2\beta$. For lower charged Higgs masses, we multiply the \SI{13}{\TeV} LHC $\sigma(pp\to t\bar{t})$ cross section of $\sim\SI{803}{\pb}$~\cite{Zyla:2020zbs} by the appropriate branching fraction
\begin{equation}
    2 \,\text{BR}(t\to H^+ b)\left(1-\text{BR}(t\to H^+ b)\right)
\end{equation}
obtained from \texttt{HDECAY}~\cite{Djouadi:1997yw,Djouadi:2018xqq}. While $\sigma(pp\to H^\pm t b)$ can reach values of more than \SI{100}{pb} for $m_{H^\pm}< m_t$, the cross section is substantially smaller above the $t$-threshold with values of \SI{6}{\pb} at $m_{H^\pm}=\SI{200}{\GeV}$ decreasing to $\sim \SI{0.1}{pb}$ at $\SI{800}{\GeV}$.

\subsubsection{\texorpdfstring{$pp\to H^\pm h_i, H^\pm A$}{pp -> Hpm hi, Hpm A} production}

\begin{figure}
    \centering
    \includegraphics[scale=.9]{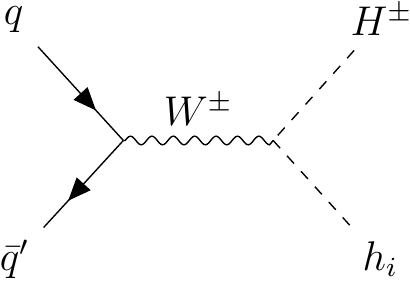}
    \caption{Dominant leading-order $s$-channel Feynman diagram for $pp\to H^\pm h$ production.}
    \label{fig:HpH_prod_feynman}
\end{figure}

Another important production channel is charged Higgs production in association with a neutral Higgs boson~\cite{Kanemura:2001hz,Akeroyd:2003bt,Akeroyd:2003jp,Cao:2003tr,Belyaev:2006rf,Miao:2010rg}, see \cref{fig:HpH_prod_feynman} for the dominant LO $s$-channel diagram. The contribution from this diagram is proportional to the $H^\pm W^\mp h_i$ coupling, which is maximized in the alignment limit for $h_i = h_\BSM$, while it vanishes for $h_i = h_{125}$ (see \cref{sec:chargedH_pheno}). As the experimental data favors an approximate realization of the alignment limit (see discussion in \cref{sec:2hdm_scan}), we only consider the case where the charged Higgs boson is produced in association with the non-SM-like \cp-even Higgs boson $h_\BSM$ or the \cp-odd Higgs boson $A$.

For the $pp\to H^\pm h_\BSM$ process, no dedicated calculation beyond the LO exists. While NLO-QCD corrections could easily be derived using automated NLO tools, we estimate the impact of higher-order corrections to the $pp\to H^+ h_\BSM$ cross section by comparing it to the SM $pp\to W^+ h$ process.
Since the only colored states in both processes are the incoming quarks, QCD corrections arise only as initial state radiation and as virtual corrections to the $W^\pm q\bar{q}$ vertex, which are identical in both processes. Since the different couplings in the second vertex cancel out in the $K$-factor, the only remaining difference between the two process is the different mass of the final state particles, and the resulting differences in phase space and scale. We use this analogy by employing the NNLO-QCD $pp\to W^+ h$ calculation implemented in the code \texttt{vh@nnlo-2.1}~\cite{Brein:2012ne,Harlander:2018yio} adjusted to account for the changed final state masses\footnote{We thank Stefan Liebler for discussions on this possibility and adjusting the code to our needs.} in order to derive a mass-dependent $K$-factor, which is defined as the ratio of the NNLO cross section over the LO cross section. We then use this $K$-factor to rescale the LO $pp\to H^+ h_\BSM$ cross section obtained with \texttt{MadGraph5-2.8.0}~\cite{Alwall:2014hca}. Both calculations are performed in the 4FS, as is appropriate for an $s$-channel $W^\pm$-exchange process, where $b$-quark contributions are negligible.

\begin{figure}
    \centering
    \includegraphics[width=0.7\textwidth]{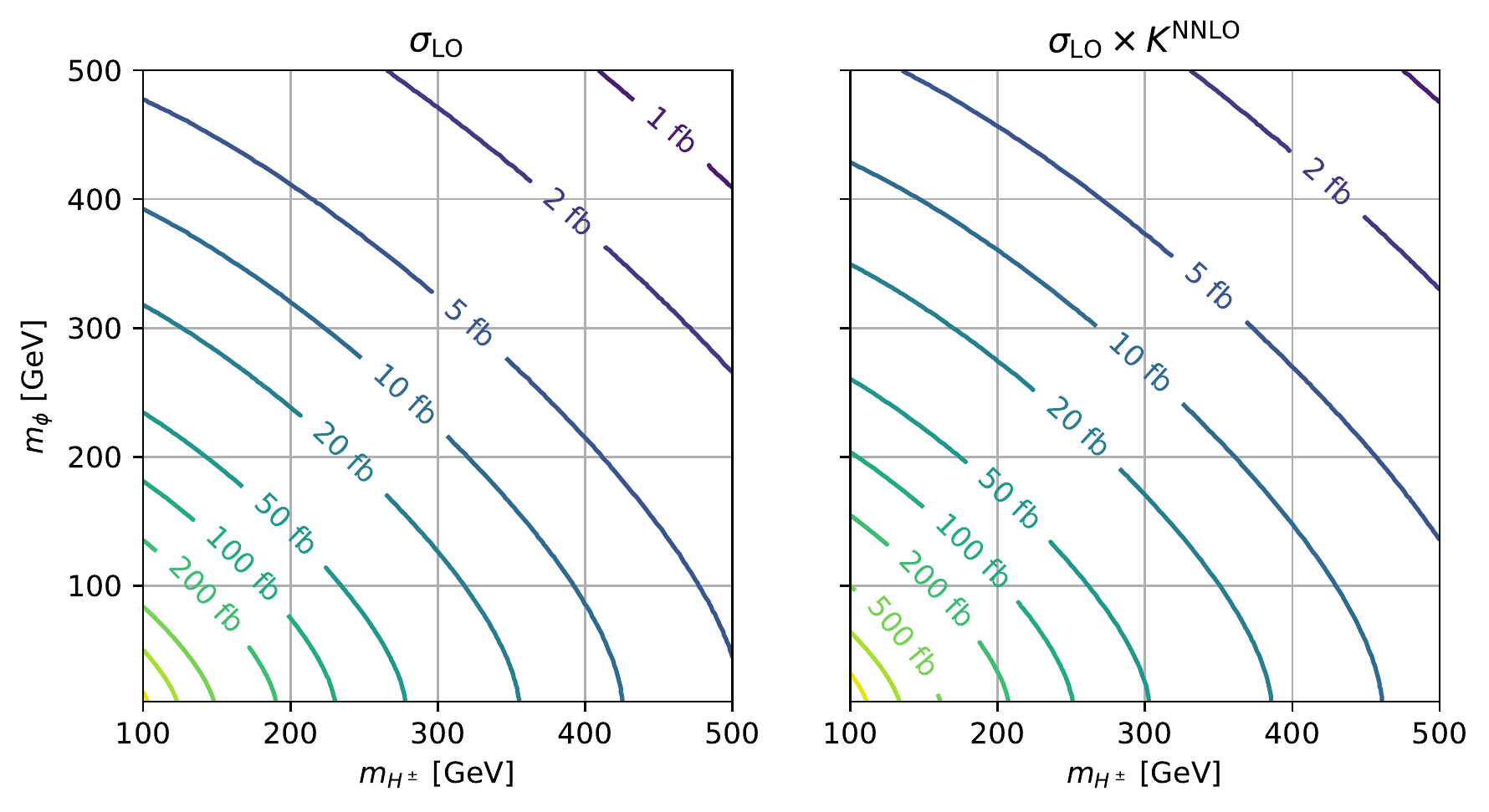}
    \caption{\textit{Left}: LO cross section for $pp\to H^\pm \phi$ production (with $\phi = h_\BSM,A$) as a function of $m_{H^\pm}$ and $m_{\phi}$. \textit{Right}: Same as left but the LO cross section multiplied with NNLO $K$-factor is shown. Both figures assume exact alignment.}
    \label{fig:HcH_cxn}
\end{figure}

The left panel of \cref{fig:HcH_cxn} shows the \SI{13}{\TeV} LHC LO cross section as a function of $m_{H^\pm}$ and $m_{h_\BSM}$, assuming exact alignment of $h_{125}$. The right panel displays the LO cross section multiplied with the NNLO $K$-factor. The $K$-factor only varies slightly within the considered mass plane with a range between \num{1.34} for low masses and \num{1.37} for high masses. Both plots show the cross section summed over both possible charges of $H^\pm$. Cross sections for $p p \to h_\BSM H^\pm$ for different values of $c(h_\BSM VV)$ can be obtained by rescaling with $c(h_\BSM VV)^2$. The $p p \to A H^\pm$ cross sections are independent of all other model parameters.

Let us consider this production mode in conjunction with the subsequent decay of the charged Higgs boson into a $W$ boson and the non-SM-like Higgs boson, $pp\to H^+ h_\BSM \to W^+ h_\BSM h_\BSM$. This final state can also arise from double-Higgsstrahlung, or Higgsstrahlung followed by an $h_\BSM\to h_\BSM h_\BSM$ splitting. These could contribute and may even interfere with our signal process. However, both of these processes involve the coupling $c(h_\BSM VV)$ which is strongly suppressed in the alignment limit. Using \texttt{MadGraph5-2.8.0} we have verified that the total cross sections for the exclusive subprocess $pp\to H^+ h_\BSM \to W^+ h_\BSM h_\BSM$ and the inclusive $pp\to W^+ h_\BSM h_\BSM$ process agree within less than $\SI{3}{\%}$ for all benchmark scenarios defined in \cref{sec:benchmarks}. Thus, the alternative processes can be safely neglected.

In the used approximation, the cross section for the production of the charged Higgs boson in association with an $A$ boson is identical to the $pp\to H^\pm h_\BSM$ cross section.

\subsubsection{\texorpdfstring{$pp \to H^\pm W^\mp$}{pp -> Hpm Wmp} production}

\begin{figure}
    \centering
    \begin{subfigure}[t]{.48\textwidth}\centering
        \includegraphics[scale=.9]{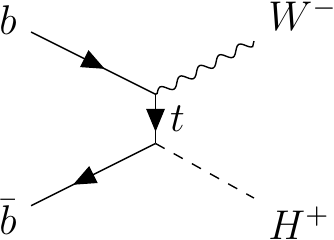}
        \caption{$b\bar b$ initiated (non-resonant)}
        \label{fig:Hptb_prod_bb_nonres}
    \end{subfigure}
    \hfill
    \begin{subfigure}[t]{.48\textwidth}\centering
        \includegraphics[scale=.9]{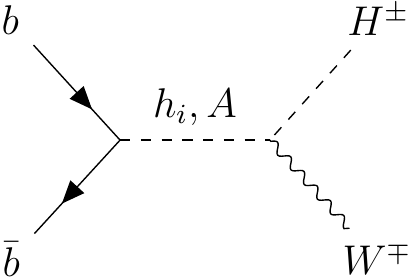}
        \caption{$b\bar b$ initiated (resonant)}
        \label{fig:Hptb_prod_bb_res}
    \end{subfigure}
    \\[1cm]
    \begin{subfigure}[t]{.48\textwidth}\centering
        \includegraphics[scale=.9]{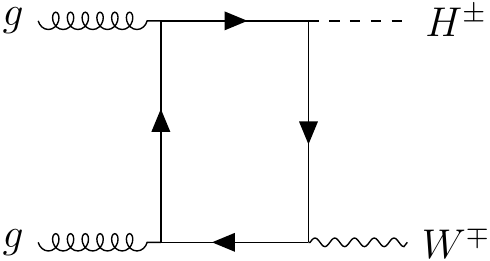}
        \caption{$gg$ initiated (non resonant)}
        \label{fig:Hptb_prod_gg_nonres}
    \end{subfigure}
    \hfill
    \begin{subfigure}[t]{.48\textwidth}\centering
        \includegraphics[scale=.9]{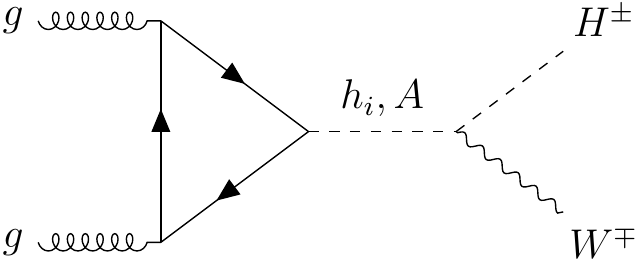}
        \caption{$gg$ initiated (resonant)}
        \label{fig:Hptb_prod_gg_res}
    \end{subfigure}
    \caption{Leading-order Feynman diagrams for $pp\to H^\pm W^\mp$ production.}
    \label{fig:Hptb_prod_feynman}
\end{figure}

At the leading order, four different subprocesses contribute to the production of a charged Higgs boson in association with a $W$ boson: the $b\bar b$-initiated non-resonant channel (see \cref{fig:Hptb_prod_bb_nonres}), the $b\bar b$-initiated resonant channel (see \cref{fig:Hptb_prod_bb_res}), the $gg$-initiated non-resonant channel (see \cref{fig:Hptb_prod_gg_nonres}), and the $gg$-initiated resonant channel mediated by any of the three neutral Higgs bosons (see \cref{fig:Hptb_prod_gg_res}).

The cross section for this process has been calculated at the leading order in Refs.~\cite{Dicus:1989vf,BarrientosBendezu:1998gd,BarrientosBendezu:1999vd,Brein:2000cv,Bao:2010sz}. Higher-order corrections have been derived in Refs.~\cite{Hollik:2001hy,Dao:2010nu,Enberg:2011ae,Kidonakis:2017dmh}. Studies focusing on the collider phenomenology of $H^\pm W^\mp$ production can be found in Refs.~\cite{Moretti:1998xq,Asakawa:2005nx,Eriksson:2006yt,Hashemi:2010ce,Bao:2011sy,Aoki:2011wd}.

\begin{figure}
    \centering
    \includegraphics[width=0.6\textwidth]{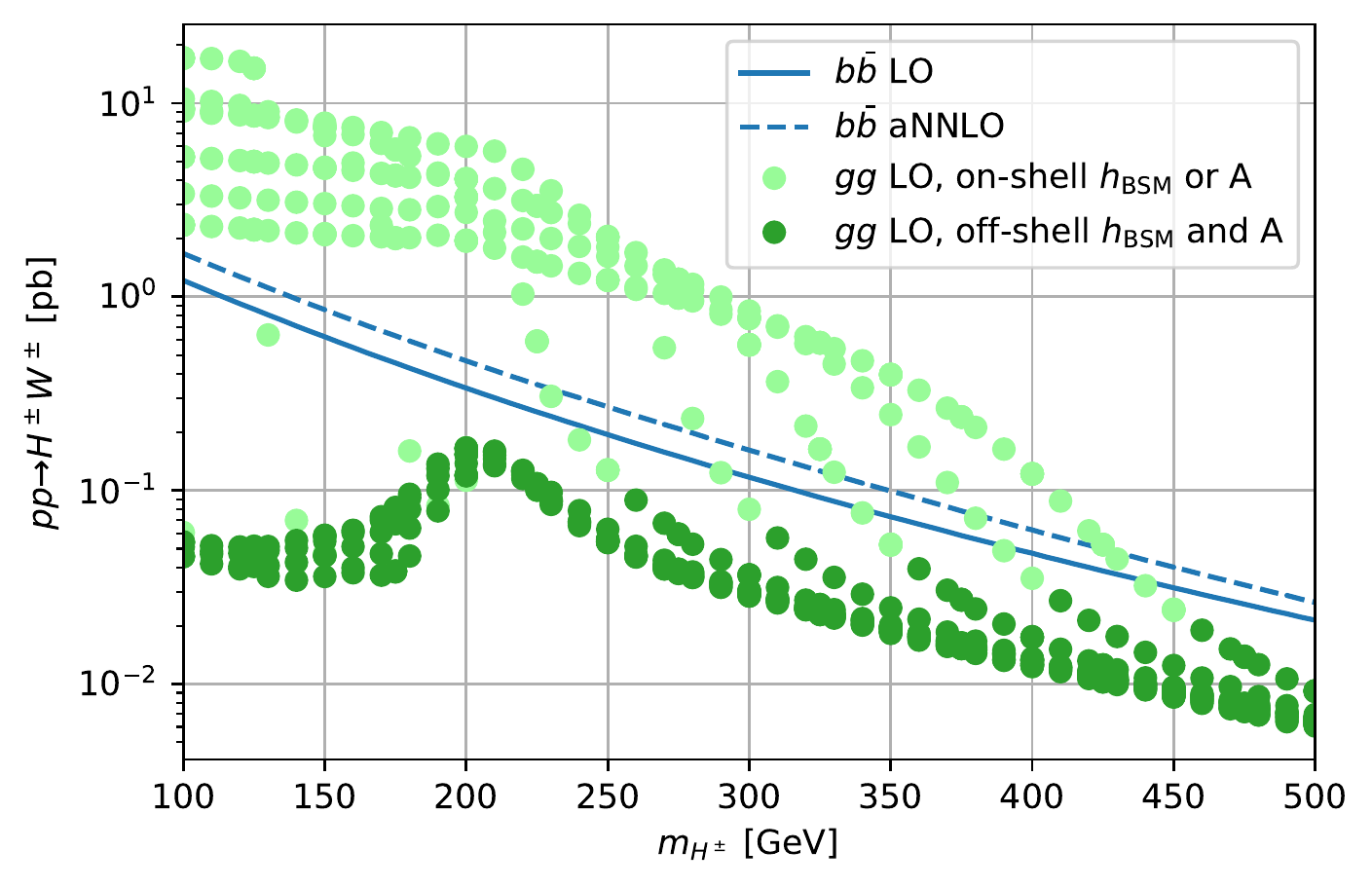}
    \caption{Contributions of the different subprocessess to the overall $pp \to H^\pm W^\mp$ cross section as evaluated in the 2HDM type-I assuming exact alignment and ${\tan\beta = 1}$.}
    \label{fig:HcW_cxn}
\end{figure}

For our present study in the 2HDM type-I, we have derived the \SI{13}{\TeV} LHC cross section for charged Higgs production in association with a $W$ boson using \texttt{MadGraph} assuming exact alignment limit and  $\tan\beta = 1$. We work in the 5FS, as is appropriate for the dominant $b\bar b$-initiated channel, for which we take into account approximate NNLO (aNNLO) corrections by multiplying with the given $K$-factors derived in \ccite{Kidonakis:2017dmh}.
The blue curves in \cref{fig:HcW_cxn} show the cross section of the $b\bar b$-initiated channel at LO (solid curve) as well as at aNNLO (dashed curved).\footnote{We neglect the contribution from the resonant $b\bar b$-initiated channel due to the smallness of the bottom Yukawa couplings in the 2HDM type-I.} While the cross section of the $b\bar b$-initiated channel only depends on $m_{H^\pm}$ in the given approximation and scales with $\tan^{-2}\beta$, the cross section of the $gg$-initiated channels additionally depend on $m_{h_\BSM}$ and $m_A$. We quantify this dependence by varying them in the interval \SIrange{10}{500}{\GeV}, which results in the greenpoints for the $gg$-initiated contributions. As a simple criterion to differentiate the non-resonant and the resonant case, we regard points for which $m_A < m_{H^\pm} + m_W$ and $m_{h_\BSM} < m_{H^\pm} + m_W$ as dominated by the non-resonant process --- since neither $h_\BSM$ nor $A$ can be on-shell\footnote{The contribution of the SM-like Higgs boson is negligible since its coupling to a charged Higgs boson and a $W$ vanishes in the alignment limit.} --- and color them dark-green. All other points are considered to be dominated by resonant production and are shown in light-green.

For $m_{H^\pm}\gtrsim \SI{450}{\GeV}$, the $b\bar{b} \to H^+W^-$ subprocess dominates. For lower charged Higgs boson masses the non-resonant $gg$-initiated contribution remains smaller than the $b\bar b$ initiated cross section by a factor between 2.2 and 30. In this region, the resonant $gg$-initiated channel can, however, yield a significant contribution to the total cross section potentially exceeding the contribution of the $b\bar b$-initiated channel by a factor of $\sim 30$ for $m_{H^\pm} = 100$ GeV.

For our benchmark scenarios defined below, we will always assume that either $m_A = m_{H^\pm}$ or $m_{h_\BSM} = m_{H^\pm}$ in order to satisfy the constraints from electroweak precision observables (see discussion in \cref{sec:2hdm_scan}). Additionally, the second non-SM-like neutral Higgs boson is assumed to be lighter in order to kinematically allow the $H^\pm \rightarrow h_\BSM W^\pm$ or the $H^\pm \rightarrow A W^\pm$ decay processes. For this mass setting, the $gg$-initiated channel is always significantly smaller than the $b\bar b$-initiated channel. Therefore, we approximate the total cross section for charged Higgs boson production in association with a $W$ boson used in our numerical analysis by only taking into account the $b\bar b$-initiated channel, which we rescale by a factor of $\tan^{-2}\beta$ if $\tan\beta \neq 1$.

\subsection{Charged Higgs boson decay modes}
\label{sec:Hpm_decays}

The coupling structure of the charged Higgs boson (see \cref{sec:2hdm}) allows charged Higgs boson decays into SM fermions as well as into a $W$ boson and a neutral Higgs boson. Among the fermionic decay channels, the $H^+ \to t\bar{b}$ as well as the $H^+\to \tau^+\bar{\nu}_\tau$ decays are phenomenologically most important due to the comparatively large respective couplings, followed by the $H^+ \to c\bar{s}$ decay. In the 2HDM of type-I, the partial widths for the fermionic decays are proportional to $\cot^2\beta$, and thus become small for large $\tan\beta$.

\begin{figure}
    \centering
    \includegraphics[width=0.9\textwidth]{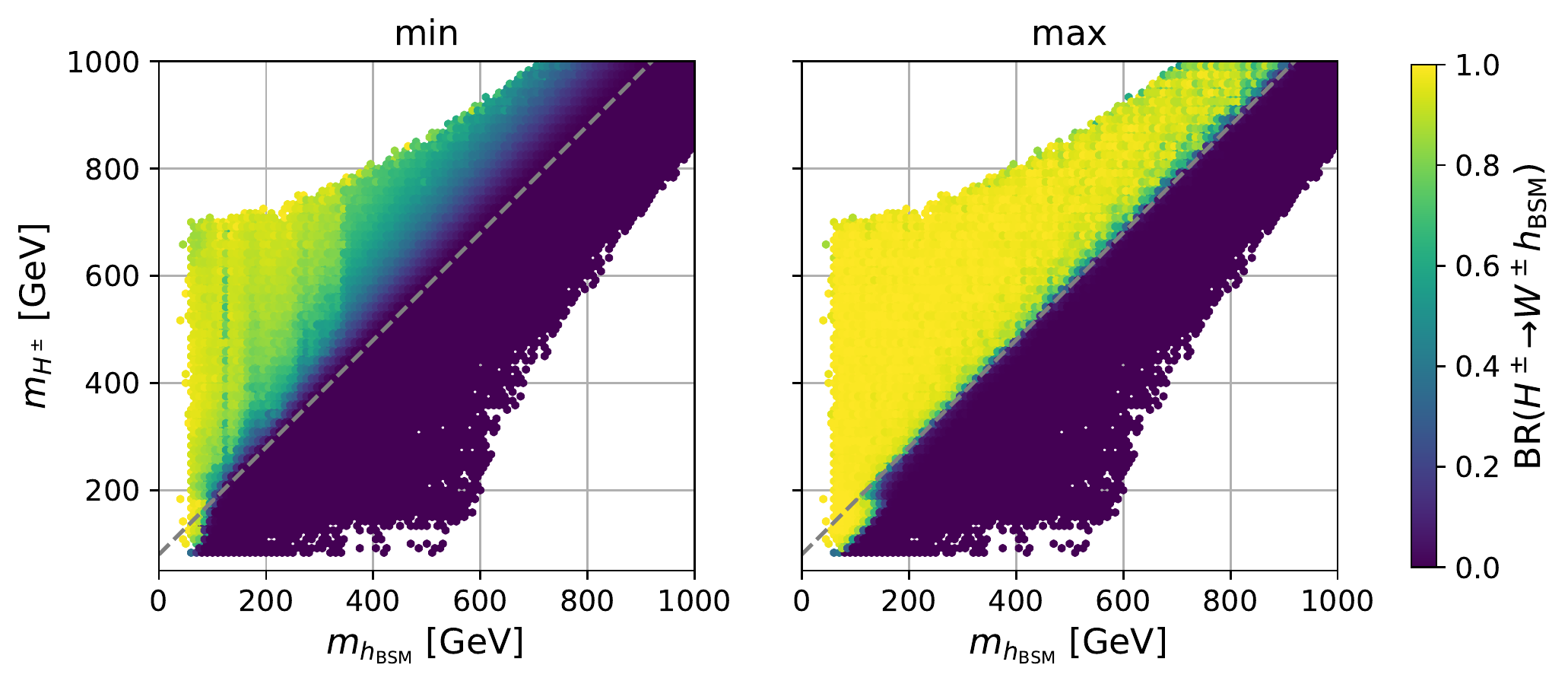}
    \caption{Branching ratio of the charged Higgs $H^\pm$ into $W^\pm h_\BSM$ in the plane of the two Higgs masses, where $h_\BSM$ denotes the non-$h_{125}$
    \cp-even neutral Higgs boson. The left (right) plot shows the minimal
    (maximal) possible BR for each bin. The dashed line indicates where
    $m_{H^\pm}=m_{h_\BSM}+m_W$.}
    \label{fig:b_Hp_Wphi}
\end{figure}

In contrast, as discussed in \cref{sec:nhdm,sec:2hdm}, the charged Higgs coupling to a $W$ boson and a non-SM-like Higgs boson becomes maximal in the alignment limit. Therefore, if kinematically allowed, the decay $H^\pm \to W^\pm h_\BSM$ can easily become the dominant decay mode. In order to assess this statement more quantitatively, we show in \cref{fig:b_Hp_Wphi} the minimal (\emph{left panel}) and maximal (\emph{right panel}) value of the $H^\pm\rightarrow W^\pm h_{\BSM}$ branching ratio in the $(m_{h_\BSM}, m_{H^\pm})$ plane for all allowed scan points. Above the dashed line the decay happens on-shell, $m_{H^\pm} \ge m_W + m_{h_\BSM}$, whereas the $W^\pm$ is off-shell below the dashed line.

In the kinematically allowed region for the on-shell decay, the \emph{minimal} branching ratio often reaches values of above \SI{60}{\%}. As expected, the branching ratio is especially large for low masses of the non-SM-like Higgs boson. The $H^\pm \to W^\pm h_\BSM$ decay mode is, however, also important in the off-shell region if $m_{h_\BSM}$ is below $\sim \SI{150}{\GeV}$ and the $H^\pm \to tb$ decay is suppressed because $m_{H^\pm}<m_t$.
The right panel of \cref{fig:b_Hp_Wphi} shows that in almost the entire parameter region of these two kinematical regimes, i.e.~in the on-shell region and in the off-shell region with $m_{h_\BSM} \lesssim \SI{150}{\GeV}$, the $H^\pm \to W^\pm h_\BSM$ decay can reach branching ratios very close to \SI{100}{\%}. The right panel also shows that the BR can reach \SI{100}{\%} in the off-shell region even for larger $m_{h_\BSM}$, as long as $m_{H^\pm}<m_t$.

\begin{figure}
    \centering
    \includegraphics[width=0.9\textwidth]{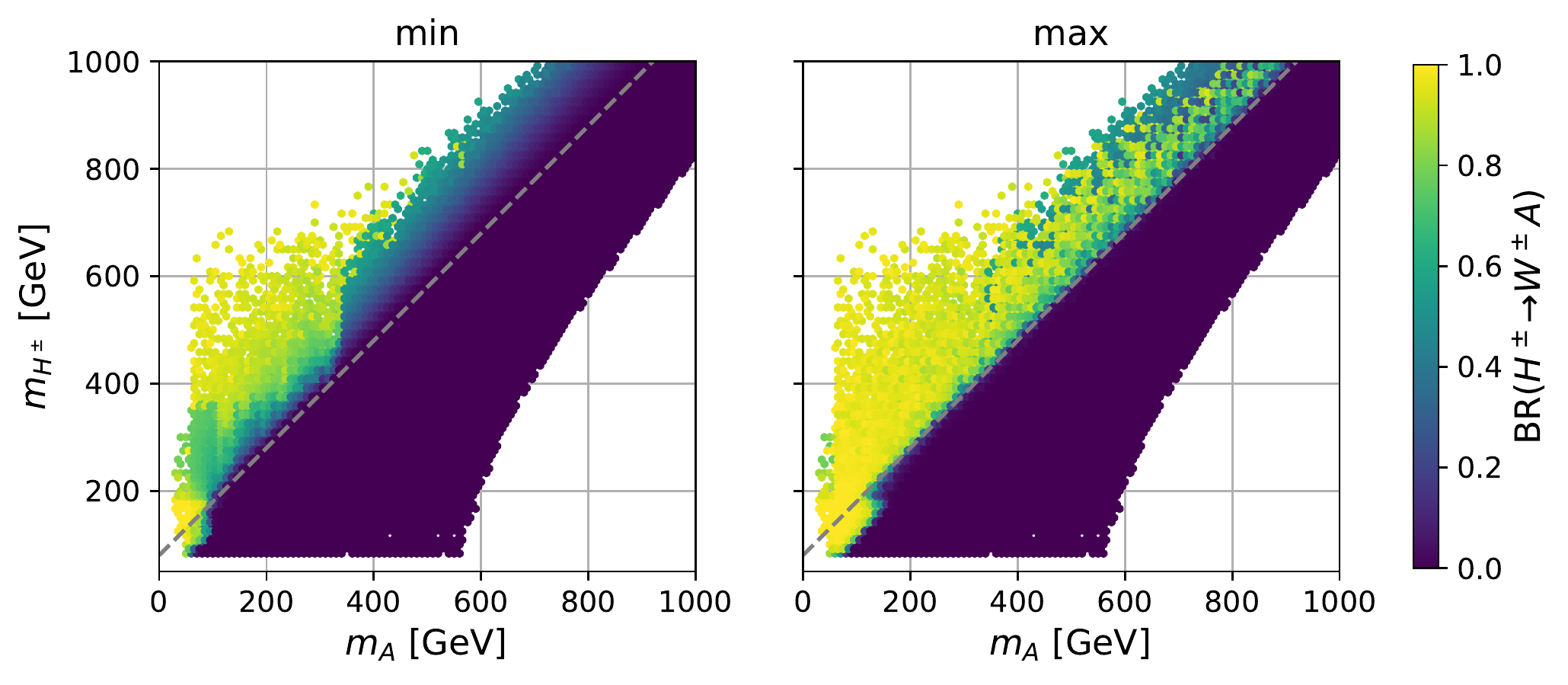}
    \caption{Branching ratio of the charged Higgs $H^\pm$ into $W^\pm A$ in the plane of the two Higgs masses. The left (right) plot shows the minimal (maximal) possible BR for each bin. The dashed line indicates where $m_{H^\pm}=m_A+m_W$.}
    \label{fig:b_Hp_WA}
\end{figure}

Besides the $H^\pm\rightarrow W^\pm h_\BSM$ decay also the $H^\pm\rightarrow W^\pm A$ can be phenomenologically relevant. The minimal and maximal branching ratios for this decay are shown in the left and right panel of \cref{fig:b_Hp_WA} in the $(m_{A}, m_{H^\pm})$ plane, respectively. While the overall behavior is quite similar to the previously discussed $H^\pm\to W^\pm h_\BSM$ decay, the minimal $\text{BR}(H^\pm\rightarrow W^\pm A)$ tends to be smaller than $\text{BR}(H^\pm\rightarrow W^\pm h_\BSM)$.

\begin{figure}
    \centering
    \includegraphics[width=0.9\textwidth]{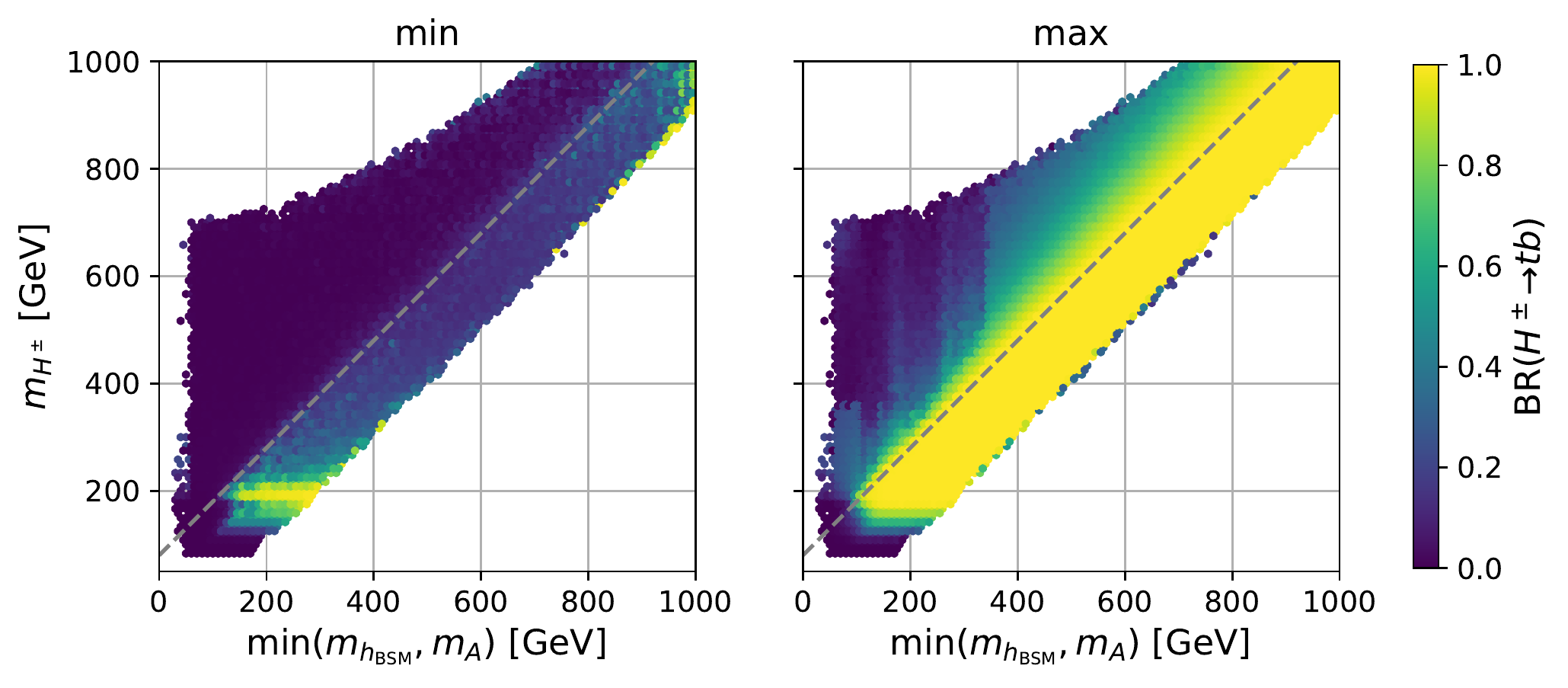}
    \caption{Branching ratio of the charged Higgs $H^\pm$ into $t b$ in the plane of $m_{H^\pm}$ and the smaller of the two non-$h_{125}$ (\cp-even or \cp-odd) neutral Higgs masses. The left (right) plot shows the minimal (maximal) possible BR for each bin. For parameter points above the dashed line, at least one of the on-shell decay modes $H^\pm\to W^\pm h_\BSM / A$ is allowed.}
    \label{fig:b_Hp_tb}
\end{figure}

In the parameter regions in which $\text{BR}(H^\pm\rightarrow W^\pm h_\BSM)$ and $\text{BR}(H^\pm\rightarrow W^\pm A)$ are both small, the charged Higgs boson decays predominantly into a top and bottom quark, if kinematically allowed. This is apparent in the left and right panels of \cref{fig:b_Hp_tb} which display the minimal and maximal value, respectively, of $\text{BR}(H^\pm\rightarrow tb)$ in the $(\text{min}(m_{h_\BSM},m_A), m_{H^\pm})$ plane. If the decays $H^\pm\rightarrow W^\pm h_\BSM/A$ are kinematically suppressed (i.e., below the dashed line), the charged Higgs boson decays almost always to $\sim \SI{100}{\%}$ to a top and a bottom quark.\footnote{The clear exception is when $H^\pm\lesssim m_t$, where the dominant fermionic decay modes are $H^+\to\tau^+\bar{\nu}_\tau$ and $H^+\to c\bar{s}$.}

We observe a steep rise of the maximal $\text{BR}(H^\pm\rightarrow tb)$ for $\text{min}(m_{h_\BSM},m_A)$ greater than $\sim \SI{350}{\GeV}$, which corresponds to the decrease in the minimal value of $\text{BR}(H^\pm \to W^\pm h_\BSM)$ and $\text{BR}(H^\pm \to W^\pm A)$, seen in the left panels of \cref{fig:b_Hp_Wphi,fig:b_Hp_WA}, respectively. At this mass value --- corresponding to $\sim 2 m_t$ --- the decays of the $h_\BSM$ and $A$ bosons to a pair of top quarks become kinematically accessible suppressing the branching ratio of all other decays. Since the di-top final state is experimentally challenging, only rather weak bounds exist for scalars decaying dominantly to $t\bar{t}$. Consequently, lower values of $\tan\beta$ --- and thereby higher values of $\text{BR}(H^\pm\to tb)$ --- remain allowed by current direct searches.

It is interesting to note that $\text{BR}(H^\pm\to tb)$ can be small even if the decay channel is open and neither of the competing $H^\pm\to W^\pm h_\BSM/A$ decays is kinematically allowed. This is visible in the left plot of \cref{fig:b_Hp_tb} in which the minimal $\text{BR}(H^\pm\to tb)$ is only $\sim \SI{30}{\%}$ for most of region below the dashed line as long as $m_{H^\pm}\gtrsim\SI{220}{\GeV}$. For these points, the alignment limit is only approximately realized resulting in a non-zero $H^\pm W^\mp h_{125}$ coupling (see \cref{sec:2hdm}). In addition, $\tan\beta$ is comparably high and therefore $\text{BR}(H^\pm\to tb)$ is rather small. As a consequence, for these points the charged Higgs boson decays dominantly into a the SM-like Higgs boson and a $W$ boson. This parameter region can therefore also be probed by searches for a bosonically decaying charged Higgs boson. The exception is the mass region $m_{H^\pm}\approx m_t + m_b$, where the $H^\pm\to tb$ decay is resonant and always dominates the decay width if the $H^\pm\to W^\pm A/h_\BSM$ decay modes are not accessible. This is the origin of the yellow region in \cref{fig:b_Hp_tb} (left).

\subsection{Current LHC searches for a charged Higgs boson}
\label{sec:current_searches}

\begin{table}
    \centering
    \small
    \begin{tabular}{c c c l}
        \toprule
        Production process        & Higgs decay processes       & Final state particles & Exp.~searches                                                                          \\
        \midrule
        $pp \to H^\pm t b$        & $H^\pm \to \tau \nu_{\tau}$ & $t b (\tau \nu_\tau)$ & \ccite{Aad:2014kga,Khachatryan:2015qxa,Aaboud:2016dig,Aaboud:2018gjj,Sirunyan:2019hkq} \\
        $pp \to H^\pm t b$        & $H^\pm \to t b$             & $t b  t b$            & \ccite{Khachatryan:2015qxa,Aaboud:2018cwk,ATLAS:2020jqj,Aad:2021xzu}                   \\
        $pp \to tt, t\to H^\pm b$ & $H^\pm \to c b$             & $t b  c b$            & \ccite{Sirunyan:2018dvm}                                                               \\
        $pp \to tt, t\to H^\pm b$ & $H^\pm \to c s$             & $t b  c s$            & \ccite{Aad:2013hla,Khachatryan:2015uua}                                                \\
        $pp \to H^\pm qq'$ (VBF)  & $H^\pm \to W^\pm Z$         & $W^\pm Z qq'$         & \ccite{Sirunyan:2017sbn,Aaboud:2018ohp,Aad:2021lzu}                                    \\
        $pp \to tt, t\to H^\pm b$ & $H^\pm \to W^\pm A$         & $W^\pm\mu^+\mu^-$     & \ccite{Sirunyan:2019zdq}                                                               \\
        \bottomrule
    \end{tabular}
    \caption{Experimentally covered LHC signatures for a (singly) charged Higgs boson, $H^\pm$.}
    \label{tab:existing_searches}
\end{table}

Existing LHC searches for a charged Higgs boson (as listed in \cref{tab:existing_searches}) have mainly focused on the charged Higgs boson decays into SM fermions. One exception are searches for a charged Higgs boson produced in a vector-boson-fusion process and decaying to a $W$ and a $Z$ boson, but these production and decay channels are only possible in higher-multiplet extensions of the SM (e.g.~Higgs triplet models). Another exception is the search for a light charged Higgs boson decaying to a $W$ boson and an $A$ boson, with the $A$ boson decaying to a $\mu^+\mu^-$ pair~\cite{Sirunyan:2019zdq}. While this search is the first experimental attempt to target the decay signatures discussed in this paper at the LHC, its results are only of very limited use, as the experimental limit has not been released for the two-dimensional mass plane ($m_{H^\pm}, m_A$), but only for one-dimensional slices of the two-dimensional parameter space, assuming either $m_{H^\pm} = m_A + \SI{85}{\GeV}$ or $m_{H^\pm} = \SI{160}{\GeV}$.

\begin{figure}[tb]
    \centering
    \includegraphics[height=0.45\textwidth]{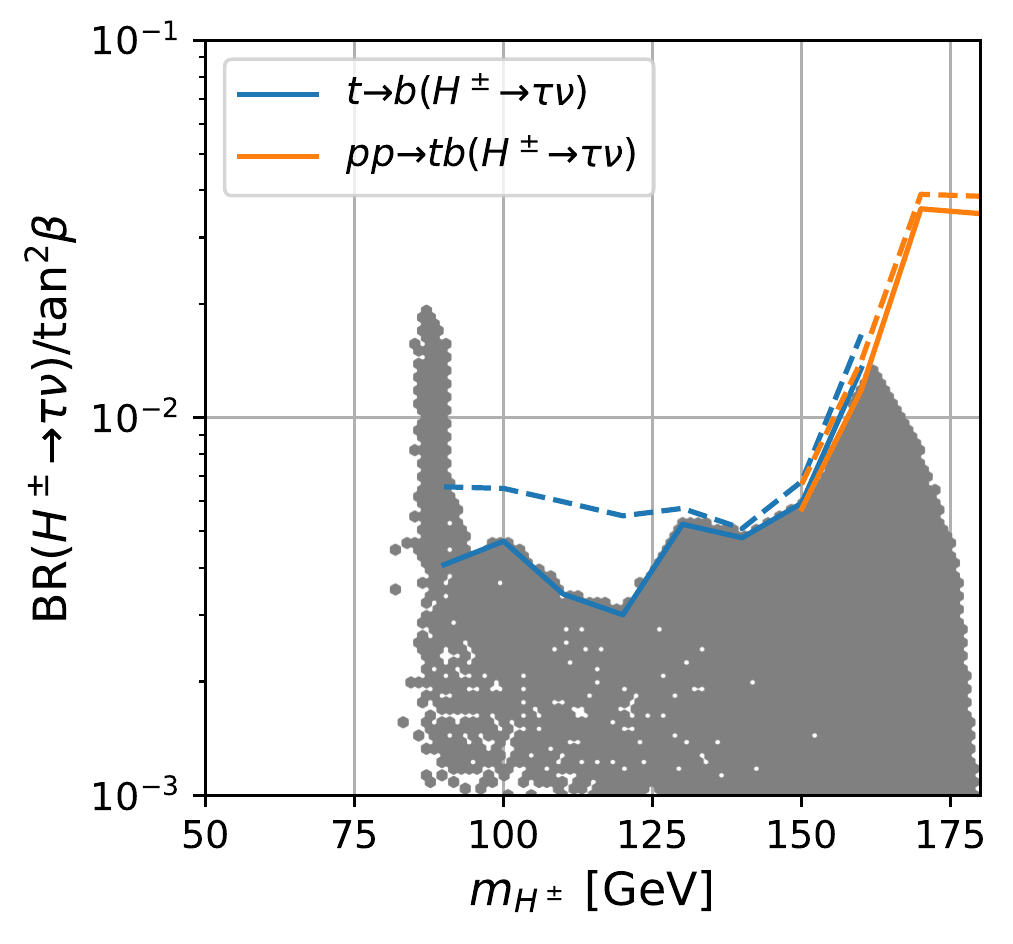}
    \includegraphics[height=0.45\textwidth]{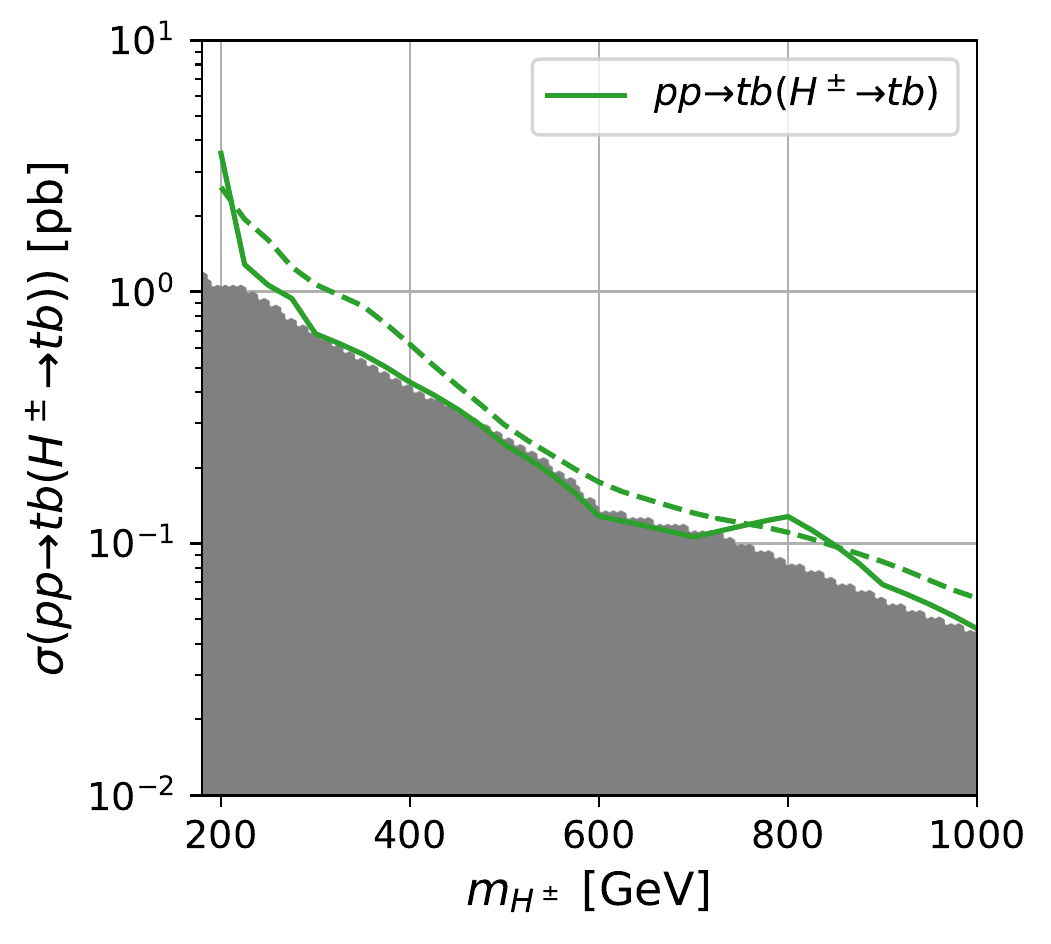}
    \caption{Branching ratio of the charged Higgs boson into $\tau\nu_\tau$ (\textit{left panel}, normalized by $\tan^2\beta$), and the \SI{13}{\TeV} LHC signal cross section in the  $pp\to tb H^\pm, H^\pm\to tb$ channel (\textit{right panel}) as a function of the charged Higgs boson mass, $m_{H^\pm}$. The \textit{gray points} pass all relevant constraints. In the \textit{left} plot, we include the observed (\textit{solid line}) and expected (\textit{dashed line}) limits from the most sensitive $t\rightarrow H^\pm b, H^\pm \rightarrow \tau\nu_\tau$ (\textit{blue}) and $pp\rightarrow H^\pm tb, H^\pm \rightarrow \tau\nu_\tau$ (\textit{orange}) searches~\cite{Aaboud:2018gjj}. In the \textit{right panel}, the limit from the latest $pp\rightarrow H^\pm t b, H^\pm \rightarrow t b$ search~\cite{ATLAS:2020jqj} is shown (\textit{green}).}
    \label{fig:searches}
\end{figure}

We illustrate the impact of the current LHC searches for a charged Higgs boson on our phenomenological 2HDM scan in \cref{fig:searches}. In the left panel we show the branching ratio $\text{BR}(H^\pm\rightarrow\tau\nu_\tau)$ normalized by $t_\beta^2$ as a function of $m_{H^\pm}$. This normalization is chosen as it allows us to show limits from the different LHC charged Higgs production modes that are relevant in this mass range on the same scale. The gray points are allowed by all constraints listed in \cref{sec:2hdm_scan}. For charged Higgs boson masses below $\sim \SI{80}{\GeV}$, no points in our scan pass all constraints. The most important constraints in this region are set by combined LEP searches for charged Higgs pair production~\cite{Abbiendi:2013hk}.\footnote{This LEP search limit cannot be shown explicitly in the left panel of \cref{fig:searches} as it does not scale with $1/t_\beta^2$.}

Above the kinematic reach of the LEP searches, $m_{H^\pm} \gtrsim \SI{90}{\GeV}$, the LHC searches for a charged Higgs boson decaying into $\tau\nu$ and produced either from top quark decays, $t\to H^\pm b$, or through $pp \to tbH^\pm$~\cite{Aaboud:2018gjj} become relevant (blue and orange curves in \cref{fig:searches} (\emph{left})).\footnote{Gray points located \emph{above} the LHC search limits at around \SI{90}{\GeV} in \cref{fig:searches} (\emph{left}) are formally \emph{not} excluded since \HB selects a different search --- in this case the aforementioned combination of LEP searches~\cite{Abbiendi:2013hk} --- to be the most sensitive channel for the given parameter point. This selection happens on the basis of the expected limit of the search, even if the corresponding observed limit has a different exclusion power (see \ccite{Bechtle:2020pkv} for a detailed discussion).} They limit $\text{BR}(H^\pm\to\tau\nu_\tau)/t_\beta^2$ to be below $\sim 10^{-2}$. These searches loose sensitivity above the kinematic threshold for the $H^\pm \to tb$ decay at $m_{H^\pm} \sim m_t + m_b$, where $\text{BR}(H^\pm\to\tau\nu_\tau)$ becomes small due to the large partial width of the $H^\pm \to tb$ decay, as can be seen from the distribution of gray points in this mass range in the left panel of \cref{fig:searches}.

The impact of LHC searches for a charged Higgs boson decaying into a top and a bottom quark is shown in the right panel of \cref{fig:searches}, which displays the signal cross section $pp\to tb H^\pm \to tb tb$ as a function of $m_{H^\pm}$. While this channel has a comparatively large cross section of up to $\sim 1~\si{pb}$, the $tb$ final state is experimentally hard to disentangle from SM background processes. Therefore, the most recent experimental search~\cite{ATLAS:2020jqj} constrains the parameter space only slightly within the region $m_{H^\pm} \sim 500-\SI{700}{\GeV}$.

From the discussion of the charged Higgs boson decay rates in \cref{sec:Hpm_decays} we know that many of our scan points in fact exhibit a large $H^\pm \to W^\pm h_\BSM$ and/or $H^\pm \to W^\pm A$ decay rate, which in turn suppresses the decay modes with fermionic final states. This is also evident from the large swath of gray points in \cref{fig:searches} that are far below the current limits from the discussed LHC searches. These observations strongly motivate experimental efforts to complement the existing searches by probing the $H^\pm \to W^\pm h_\BSM$ and $H^\pm \to W^\pm A$ decay modes directly. In the upcoming section we focus on the collider signatures arising from these decays and present suitable benchmark scenarios for the design of such dedicated experimental searches.

\section{Unexplored LHC signatures and benchmark models}
\label{sec:benchmarks}

As discussed above, vast parts of the phenomenologically viable parameter space of the 2HDM of type-I feature a charged Higgs boson that dominantly decays into a $W$ boson and a non-SM like neutral Higgs boson ($h_\BSM$ or $A$). However, the collider signatures arising from these decays are to a large extent not covered by LHC searches.

\begin{table}[t]
    \centering
    \small
    \begin{tabular}{ccc}
        \toprule
        Production process   & Higgs decay processes                                            & Final state particles                                                   \\
        \midrule
        $pp \to H^\pm t b$   & $H^\pm \to W^\pm \phi$ and $\phi\to \begin{cases} bb \\ \tau\tau \\ WW \\ ZZ\\ \gamma\gamma \end{cases} $ & $t b W^\pm + \begin{bmatrix}bb \\ \tau\tau \\ WW\\ ZZ\\ \gamma\gamma \end{bmatrix}$                                \\
        \midrule
        $pp \to H^\pm \phi$  & $H^\pm \to W^\pm \phi$ and $\phi\to \begin{cases} bb \\ \tau\tau \\ WW \\ ZZ\\ \gamma\gamma \end{cases} $ & $W^\pm + \begin{bmatrix}bb \\ \tau\tau \\ WW\\ ZZ\\ \gamma\gamma \end{bmatrix} \otimes \begin{bmatrix}bb \\ \tau\tau \\ WW\\ ZZ\\ \gamma\gamma \end{bmatrix}$ \\
        \midrule
        $pp \to H^\pm W^\mp$ & $H^\pm \to W^\pm \phi$ and $\phi\to \begin{cases} bb \\ \tau\tau \\ WW \\ ZZ\\ \gamma\gamma \end{cases} $ & $W^\pm W^\mp + \begin{bmatrix}bb \\ \tau\tau \\ WW\\ ZZ\\ \gamma\gamma \end{bmatrix}$                              \\
        \bottomrule
    \end{tabular}
    \caption{LHC signatures arising from the charged Higgs boson decay $H^\pm \to W^\pm \phi$, with a neutral non-SM-like Higgs boson, $\phi = h_\BSM,A$, for the most relevant production (\emph{first column})  and decay (\emph{second column}) processes. The resulting final state particles are given in the \emph{third column} (further decays of the SM particles are not explicitly shown). The ``$\otimes$'' symbol in the center right cell indicates that any combination of the final states within the square brackets can occur due to the independent decays of the two neutral Higgs bosons.}
    \label{tab:uncovered_signatures}
\end{table}

We summarize the most relevant LHC signatures in \cref{tab:uncovered_signatures}. These arise from charged Higgs production via one of the three main LHC production modes discussed in \cref{sec:Hpm_prod}, and the successive decay of the charged Higgs boson into a $W$ boson and a neutral Higgs boson $\phi$, which can either be the $h_\BSM$ boson or the $A$ boson. For the neutral Higgs boson decays we include the $bb$, $\tau\tau$, $WW$, $ZZ$ and $\gamma\gamma$ final states, as these are typically the most frequent ($bb$, $\tau\tau$) or experimentally cleanest ($WW$, $ZZ$, $\gamma\gamma$) channels. The corresponding rates are model-dependent and will be discussed in more detail below in the context of benchmark scenarios. Besides the final states included in \cref{tab:uncovered_signatures}, the experimentally more challenging decays $\phi\to cc$ and $\phi \to gg$ may also become relevant in certain scenarios (see e.g.~\ccite{Aaboud:2018fhh,Sirunyan:2019qia,Alves:2019ppy} for related searches and studies for the $h_{125}$). We shall therefore also include their rates in the following discussions if relevant.

While a detailed analysis of the LHC discovery potential of the various collider signatures must be postponed to future work, we briefly want to comment on a few features that may be exploited in collider searches, and the most important SM backgrounds. For the $pp\to tbH^\pm$ production process --- which is typically the dominant charged Higgs boson production mode, see \cref{sec:Hpm_prod} --- inclusive SM processes with pairs of top quarks, $t\bar{t} (+X)$ and $t\bar{t}h_{125}$, are inevitably a major background, almost irrespective of how the neutral non-SM-like Higgs boson $\phi$ of the signal process decays (see also discussion in \ccite{Alves:2017snd}). For the experimentally clean signature arising from the decay $\phi\to \gamma\gamma$, we expect that a good signal-background separation can be achieved by using similar techniques as in the $t\bar{t}h_{125}, h_{125}\to \gamma\gamma$ analyses~\cite{Aad:2020ivc,Sirunyan:2020sum}. However, as the signal rate is typically very low --- except for specific scenarios with a very light and/or fermiophobic $\phi$ (see below) --- sensitivity may only be reached with a very large amount of data. Hence, we expect that for the more conventional 2HDM scenarios near the alignment limit the signal processes from $\phi \to bb$ and $\phi \to \tau\tau$ provide a more promising avenue (see also \ccite{Coleppa:2014cca,Kling:2015uba}). Moreover, under the additional model-assumption on the relative size of $\text{BR}(\phi \to bb)$ and $\text{BR}(\phi \to \tau\tau)$ as predicted in the 2HDM, search results from the two signatures may be combined to maximize the sensitivity to the 2HDM parameter space.

For the $pp\to H^\pm \phi \to W^\pm \phi \phi$ process a multitude of signatures arises because the two $\phi$ bosons  decay independently. As the production rate is already comparatively small, we also expect the typically more frequent decay modes $\phi\to bb$ and $\phi\to \tau\tau$ to exhibit the highest sensitivity in most of the 2HDM parameter space. In this case, a leptonically decaying $W$ boson provides a triggerable isolated lepton. Moreover, dedicated signal regions for resolved, semi-boosted and fully-boosted $bb$/$\tau\tau$ pairs can be defined to enhance the sensitivity. Again, the main SM background to these signatures arises from (semi-leptonic) top quark pair production.

For the third process, $pp \to H^\pm W^\mp \to W^\pm W^\mp \phi$, we again expect the semi-leptonic analysis to be the most sensitive selection (see also \ccite{Basso:2012st}). SM backgrounds arise from inclusive electroweak boson production, $W^+W^- + X$ and top quark pair production. Note, however, that the signal cross section scales with $\cot^2\beta$, and is therefore typically smaller than the cross section of the previous processes. We therefore expect this  production channel to be not as sensitive as the $pp\to tbH^\pm$ and $pp\to H^\pm \phi$ channels.

2HDM scenarios with large decay rates for $H^\pm \to W^\pm \phi$ typically feature a concomitant signature in the neutral Higgs sector --- the $A\to Zh_\BSM$ (if $\phi = h_\BSM$) or $h_\BSM \to Z A$ (if $\phi = A$) decay --- with a sizable rate. This can be ascribed to the EW precision measurements discussed in \cref{sec:2hdm_scan} which constrain the mass of one of the neutral Higgs bosons to be close to the charged Higgs boson mass. Consequently, these scenarios are simultaneously probed by searches for $pp\to A\to Zh_\BSM$ or $pp\to h_\BSM \to Z A$, and current limits from these searches constrain parts of the relevant parameter space~\cite{Aaboud:2018eoy,Sirunyan:2019xls} (see also discussion of benchmark scenarios below). Yet, direct charged Higgs boson searches for the $H^\pm \to W^\pm \phi$ decay signatures are highly warranted, as they are able to directly probe the charged Higgs sector independently of the (model-dependent) correlation between neutral and charged Higgs boson masses enforced by the EW precision constraints. In particular, in the optimistic case of a discovery in either of these channels, the model correlations would strongly suggest to look for a corresponding signal in the complementary channel as well.

Another related channel is of course the inverse decay $\phi\to H^\pm W^\pm$ if the mass hierarchy of $H^\pm$ and $\phi$ is reversed. This decay is governed by the same coupling as the $H^\pm \to W^\pm \phi$ decay, and is therefore also maximized in the alignment limit of $h_{125}$. However, through the mass correlation imposed by the EW precision measurements, this channel is even more strongly related to the $h_\BSM/A \to Z A/h_\BSM$ channels, since it also shares the initial production mode of the neutral scalar. We expect the $W^\pm H^\mp$ final state to be experimentally more challenging than the $Z \phi$ final state, regardless of the successive $H^\pm$ and $\phi$ decay. As a result, the 2HDM parameter region where searches for $pp\to \phi \to H^\pm W^\mp$ are competitive with $pp\to A/h_\BSM \to Z h_\BSM/A$ searches is limited to the small mass region where the $Z\phi$ decay is kinematically suppressed, while the $H^\pm W^\mp$ decay is not. We stress that this decay channel is nevertheless very interesting to search for, since the mass correlations imposed by the EW precision constraints are model-dependent. We will further comment on this channel below, whenever our benchmark scenarios include parameter regions where it can appear.

In order to facilitate future searches for the $H^\pm \to W^\pm \phi$ signatures, we present in the following five benchmark scenarios, which are parametrized in the ($m_{H^\pm}, m_{\phi}$) plane:
\begin{itemize}

    \item \BPh scenario with large $\text{BR}(H^\pm \to W^\pm h_{\BSM})$:
          We choose $m_{H^\pm} = m_A$, and $m_H^\pm > m_{h_\BSM}$ in most of the parameter plane. We assume the exact alignment limit and take a very small value $\tan\beta =3$, which maximizes the $H^\pm \to W^\pm h_{\BSM}$ decay rate and the $pp\to tbH^\pm$ production rate, respectively. The light non-SM-like Higgs boson $h_\BSM$ mainly decays to SM fermions ($bb$, $\tau\tau$) and gluons ($gg$);

    \item \BPA scenario with large $\text{BR}(H^\pm \to W^\pm A)$:
          We choose $m_{H^\pm} = m_{h_\BSM}$, and $m_H^\pm > m_A$ in most of the parameter plane. Analogous to the previous scenario, we assume the exact alignment limit and take a very small value $\tan\beta =3$ to obtain a  large $pp\to tbH^\pm$ production rate. The $A$ boson predominantly decays to $bb$, $gg$ and $\tau\tau$.

    \item \BPff scenario with fermiophobic $h_\BSM$:
          The charged Higgs boson and the $A$ boson are chosen to be mass degenerate, $m_{H^\pm} = m_A$, and $h_\BSM$ is lighter in most of the parameter space. We depart slightly from the exact alignment limit, $c(h_\BSM VV) = 0.2$, and chose $\tan\beta$ to fulfill the fermiophobic Higgs condition, \cref{eq:fermiophobic}, i.e.~$\tan\beta \approx 4.9$. The fermiophobic Higgs boson $h_\BSM$ decays dominantly to di-photons (for $m_{h_\BSM} < \SI{90}{\GeV}$) or massive SM vector bosons (for $m_{h_\BSM} \ge \SI{90}{\GeV}$).

    \item \BPlight scenario with light $h_\BSM$ ($2 m_{h_\BSM} \le m_{h_{125}}$):
          We choose $m_{H^\pm} = m_A$. In order to avoid constraints from LHC Higgs rate measurements we suppress the decay rate of $h_{125}\to h_\BSM h_\BSM$ by choosing $m_{12}^2$ according to \cref{eq:zero_gh1h1h2_m12sq}. The light $h_\BSM$ boson decays predominantly to $bb$, $\tau\tau$ and $\gamma\gamma$. The dominant charged Higgs boson production mode is $pp \to H^\pm h_\BSM$ with a $\SI{13}{\TeV}$ LHC cross section $\sim \order{\SI{100}{\fb} - \SI{1}{\pb}}$.

    \item \BPlightLS scenario with leptophilic $h_\BSM$:
          Defined analogously to the \BPlight scenario but in the lepton-specific 2HDM (instead of the 2HDM type-I). Consequently, $h_\BSM$ decays almost exclusively to tau leptons.

\end{itemize}
Each of these scenarios features a distinct collider phenomenology and shows the importance of the $H^\pm \rightarrow W^\pm h_\BSM$ or $W^\pm A$ decay mode. While the first two ``standard'' scenarios, \BPh and \BPA, aim to maximize the signal rate in the $pp\to tbH^\pm\to tbW^\pm \phi$ ($\phi = h_\BSM, A$, respectively) channel, the latter three ``specialized'' scenarios highlight exceptional phenomena (fermiophobic, leptophilic, or very light $h_\BSM$) that may occur for specific parameter choices, and lead to very different collider signatures. A significant part of the parameter planes evade all current constraints evaluated as specified in \cref{sec:2hdm_scan}.

\begin{table}
    \centering
    \begin{tabular}{lc c c c c c c}
        \toprule
                   & $m_{h_{125}}$                                                        & $m_{H^\pm}$               & $m_{h_\text{BSM}}$ & $m_A$       & \multirow{2}{*}{$c(h_\text{BSM}VV)$} & \multirow{2}{*}{$\tan\beta$} & $m_{12}^2$                   \\
                   & [\si{\GeV}]                                                          & [\si{\GeV}]               & [\si{\GeV}]        & [\si{\GeV}] &                                      &                              & [$\si{\GeV}^2$]              \\
        \midrule
        \BPh       & \multirow{2}{*}{125.09}                                              & \multirow{2}{*}{150--300} & 65--200            & $m_{H^\pm}$ & \multirow{2}{*}{0}                   & \multirow{2}{*}{3}           & 500                          \\
        \BPA       &                                                                      &                           & $m_{H^\pm}$        & 65--200     &                                      &                              & 5000                         \\
        \midrule
        \BPff      & \multirow{2}{*}{125.09}                                              & 150--300                  & 65--200            & $m_{H^\pm}$ & 0.2                                  & \cref{eq:fermiophobic}       & 1200                         \\
        \BPlight   &                                                                      & 100--300                  & 10--62.5           & $m_{H^\pm}$ & -0.062                               & 16.6                         & \cref{eq:zero_gh1h1h2_m12sq} \\
        \BPlightLS & \multicolumn{7}{c}{same as \BPlight but in the lepton-specific 2HDM}                                                                                                                                                                     \\
        \bottomrule
    \end{tabular}
    \caption{Parameter choices in the five benchmark scenarios for the $H^\pm \to W^\pm \phi$ ($\phi = h_\BSM, A$) decay the 2HDM\@. All scenarios except \BPlightLS are defined in the type I 2HDM.}
    \label{tab:BPpars}
\end{table}

In order to evade constraints from electroweak precision observables (see \cref{sec:2hdm_scan}), we set the mass of the other Higgs boson, $A$ or $h_\BSM$, respectively, equal to the charged Higgs boson mass, as described above. The other neutral \cp-even Higgs boson is considered to be the discovered Higgs boson $h_{125}$ with a mass fixed to $\SI{125.09}{\GeV}$. The values for the remaining 2HDM input parameters are given in \cref{tab:BPpars}. Their choices are explained in the description of the respective benchmark scenarios. We provide the full data tables for all benchmark scenarios as ancillary files.

While we aimed to pick typical and illustrative scenarios, different choices of the fixed parameters could have led to different phenomenology and different parameter regions excluded by existing constraints. When designing experimental searches for these signatures, the search ranges should therefore never be constrained to the allowed region in the targeted benchmark scenario, but chosen as large as possible for the experimental analysis.

It is also possible to define similar scenarios in the type-II or flipped 2HDM\@. As discussed in \cref{sec:2hdm}, the light $H^\pm$ region in these models is tightly constrained by flavor constraints. As a consequence, the charged Higgs boson mass would have to be significantly higher --- $m_{H^\pm}\gtrsim \SI{580}{\GeV}$~\cite{Misiak:2017bgg} --- resulting in smaller signal cross sections for charged Higgs bosons decaying to a $W$ and a non-SM like neutral Higgs boson.

\subsection{\BPh scenario with large \texorpdfstring{BR($H^\pm\rightarrow W^\pm h_\BSM$)}{BR(Hpm -> Wpm hBSM)}}

With our first benchmark scenario, the \BPh scenario, we aim to provide a reference model that maximizes the rates of the $H^\pm \to W^\pm h_\BSM$ decay and $pp \to t b H^\pm$ production mode. The $h_\BSM$ boson decays predominantly to $bb$, $\tau\tau$ and $gg$. This ``standard'' scenario exhibits a collider phenomenology that is found in large parts of the viable parameter space, without the need of tuning specific parameters.

We assume the exact alignment limit, $c(h_\BSM VV) = 0$, and set $m_A =
m_{H^\pm}$. The scenario is parametrized in the ($m_{H^\pm}$, $m_{h_\BSM}$)
plane, with $m_{h_\BSM} < m_{H^\pm}$ in most of the parameter plane. As a
consequence, the decay $H^\pm\rightarrow W^\pm h_\BSM$ is one of the most
important decay channels of the charged Higgs boson. The parameter
$\tan\beta$ is chosen as low as possible without violating flavor constraints
for light $H^\pm$, which maximizes the rates in the $pp \to tbH^\pm$
production mode. The charged Higgs boson phenomenology is not significantly
affected by the choice of the parameter $m_{12}^2$ (see
\cref{tab:BPpars}). With the $m_{12}^2$ value chosen here theoretical constraints are successfully evaded.

Similar scenarios in the exact aligment limit can easily be defined for larger values of $\tan\beta$. In that case all fermionic production and decay modes would be suppressed --- since the fermionic couplings scale with $1/\tan\beta$ --- while the $pp\to H^\pm h_\BSM$ production cross section would be unaffected, and the $H^\pm \to W^\pm h_\BSM$ branching ratio would be enhanced --- due to the suppressed fermion decays. Therefore, this low $\tan\beta$ scenario is intentionally chosen as a \emph{worst case} scenario for searches targeting $pp\to H^\pm h_\BSM \to W^\pm h_\BSM h_\BSM$, while showing the complementarity with searches in fermionic channels.

\begin{figure}[!tb]
    \centering
    \includegraphics[width=0.9\textwidth]{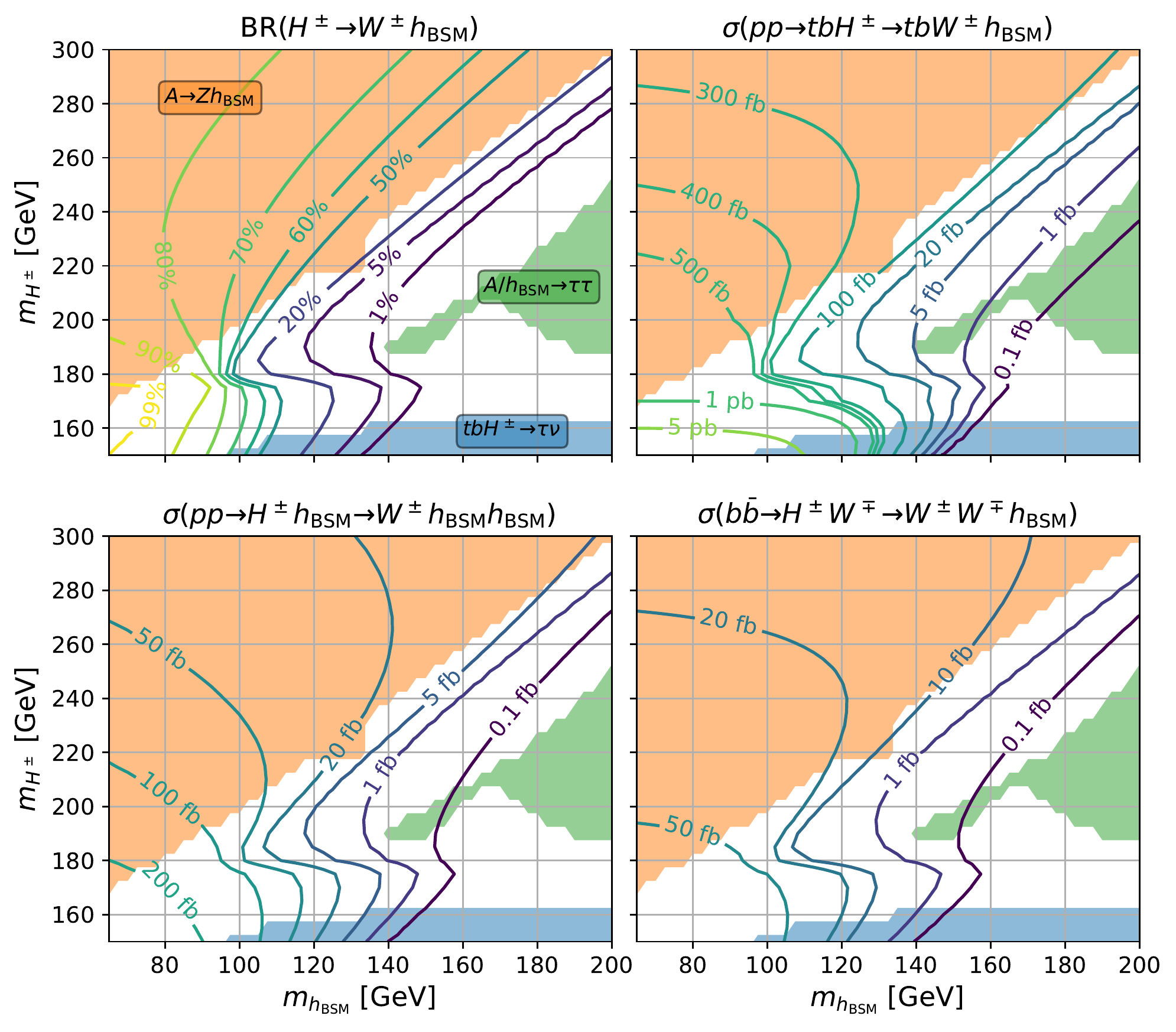}
    \caption{Benchmark scenario \BPh in the ($m_{H^\pm}, m_{h_\text{BSM}}$) parameter plane.
    The \emph{colored contour lines} indicate $\text{BR}(H^\pm\to W^\pm h_\text{BSM})$ in the \emph{top-left} panel and the \SI{13}{\TeV} LHC signal cross sections in the $pp\to tb H^\pm \to tb W^\pm h_\text{BSM}$ (\emph{top-right}), $pp\to H^\pm h_\text{BSM} \to W^\pm h_\text{BSM}h_\text{BSM}$ (\emph{bottom-left}), and $pp\to W^\mp H^\pm \to W^\pm W^\mp h_\text{BSM}$ (\emph{bottom-right}) channels in the remaining panels. The \textit{colored regions} of parameter space are excluded by current constraints from searches in the $pp\to A\to Z h_\text{BSM}$~\cite{Aaboud:2017cxo,Aaboud:2018eoy,Sirunyan:2019xls}, $pp\to A/h_\BSM\to\tau^+\tau^-$~\cite{Aad:2020zxo} and $pp\to tb H^\pm, H^\pm \to \tau^\pm \nu_\tau$~\cite{Aaboud:2018gjj} channels, as denoted by the labels.}\label{fig:hBSM_aligned_rates}
    \label{fig:BPh}
\end{figure}

The branching ratio $\text{BR}(H^\pm \to W^\pm h_\BSM)$ is shown in the upper left panel of \cref{fig:BPh} in the ($m_{H^\pm}, m_{h_\text{BSM}}$) parameter plane (colored contour lines). The colored regions are excluded by the following LHC searches: experimental searches for $pp\to A\to Z h_\BSM$~\cite{Aaboud:2017cxo,Aaboud:2018eoy,Sirunyan:2019xls} (orange region) exclude the upper left part of the parameter plane, $pp\to A/h_\BSM\to\tau^+\tau^-$ searches~\cite{Aad:2020zxo} (green region) constrain the central right part of the parameter plane, and searches for $pp\to tb H^\pm, H^\pm \to \tau^\pm \nu_\tau$~\cite{Aaboud:2018gjj} (blue region) exclude the lower right part of the parameter plane. In the remaining unconstrained parameter region, $\text{BR}(H^\pm\rightarrow W^\pm h_\BSM)$ can reach values above $\SI{99}{\%}$ (for $m_{H^\pm} \sim \SI{170}{\GeV}$ and $m_{h_\BSM} \sim \SI{70}{\GeV}$) rendering the $H^\pm\rightarrow W^\pm h_\BSM$ decay a prime target for future searches in this part of the parameter space. The branching ratio decreases for increasing $m_{h_\BSM}$ due to the decreasing phase space. All of the LHC searches that exclude parts of the scenario rely on fermionic production modes. Therefore, at larger $\tan\beta$ the existing experimental constraints would become significantly weaker.

The total signal cross sections for the three main charged Higgs production modes (as discussed in \cref{sec:Hpm_prod}) with the subsequent $H^\pm$ decay into a $W$ boson and an $h_\BSM$ boson are shown in the remaining three panels of \cref{fig:BPh}. In the unconstrained parameter region, the signal cross section can maximally reach about \SI{13}{\pb} for $tb$-associated charged Higgs production, \SI{340}{\fb} for $h_\BSM$-associated charged Higgs boson production, and \HB{\SI{95}{\fb}} for $W$-associated charged Higgs boson production. All of these maximal cross sections are reached for the lowest considered masses.

\begin{figure}
    \centering
    \includegraphics[width=0.6\textwidth]{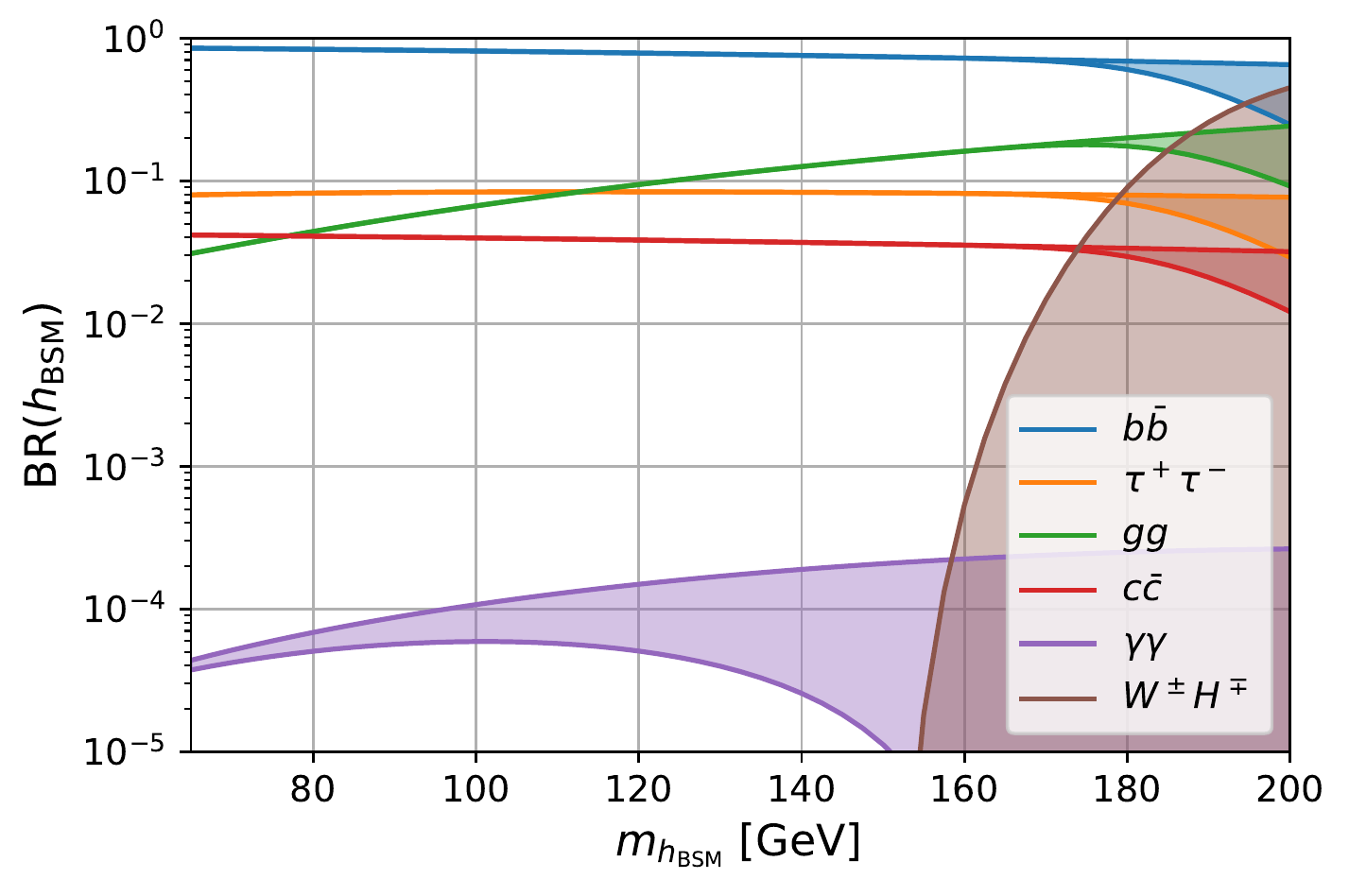}
    \caption{Branching ratios of $h_\text{BSM}$ in the \BPh benchmark scenario as a function of $m_{h_\text{BSM}}$. The decays $h_\text{BSM}\to\gamma\gamma$ and $h_\text{BSM}\to W^\pm H^\mp$ additionally depend on $m_{H^\pm}$ and induce a dependence in the remaining BRs through the total decay width. The bands for each BR indicate the impact of this dependence for $\SI{150}{\GeV} < m_{H^\pm} < \SI{300}{\GeV}$.}
    \label{fig:BPh_hBSM_BR}
\end{figure}

\Cref{fig:BPh_hBSM_BR} shows the branching ratios of $h_\BSM$ for the most important decay modes as a function of $m_{h_\BSM}$. This completes the rate information required for designing experimental searches to probe this benchmark scenario. Since the decay modes depend on the mass of $H^\pm$ directly or indirectly through the total width, the ranges of BRs within the $m_{H^\pm}$ range of the scenario are shown. The \BPh scenario is defined in the exact alignment limit, therefore $h_\BSM$ does not couple to massive vector bosons. Accordingly, $h_\BSM$ decays dominantly to two bottom quarks ($b\bar{b}$, blue curve) with branching ratios of up to \SI{80}{\%} (\SI{70}{\%}) for low (high) masses. The decay into gluons ($gg$, green curve) becomes increasingly important for rising $m_{h_\BSM}$, reaching decay rates of up to $\sim \SI{25}{\%}$. The branching ratios for $h_\BSM$ decays to a pair of tau leptons ($\tau^+\tau^-$, orange curve) and a pair of charm quarks ($c\bar{c}$, red curve) are approximately constant reaching values of $\sim \SI{8}{\%}$ and \SI{4}{\%}, respectively. The partial width of $h_\BSM$ decaying into a pair of photons depends on the charged Higgs boson mass (see \cref{sec:h125_gaga}). The corresponding BR can reach values of up to \SI{0.03}{\%}. For large $m_{h_\BSM} \gtrsim \SI{155}{\GeV}$ and low values of $m_{H^\pm}$, the (off-shell) decay $h_\BSM \to W^\pm H^\mp$ becomes possible. The BR in this channel, shown as the brown filled region in \cref{fig:BPh_hBSM_BR}, can reach values up to \SI{45}{\%} at $m_{h_\BSM} = \SI{200}{\GeV}$ and very low $m_{H^\pm}$ values. This decay rate is anti-correlated with all other decay rates, leading to the filled regions and declining slopes of the minimal decay rates for other decay modes. As discussed above, this illustrates that the $h_\BSM \to W^\pm H^\mp$ decay can become equally large as the $H^\pm \to W^\pm h_\BSM$ decay.

Finally, the charged Higgs boson contribution to the di-photon decay mode (see \cref{sec:h125_gaga}) also induces deviations in the $h_{125}$ di-photon rate of up to \SI{13}{\%} from the SM (see \cref{sec:diphoton_rate} for more details).

\subsection{\BPA scenario with large \texorpdfstring{BR($H^\pm\rightarrow W^\pm A$)}{BR(Hpm -> Wpm A)}}

Our second benchmark model, the \BPA scenario, is designed to feature a maximal rate for the $H^\pm\rightarrow W^\pm A$ decay and a dominant production through the $pp \to t b H^\pm$ process. In analogy with the previous scenario, we choose $m_{h_\BSM} = m_{H^\pm}$, and $m_A$ is allowed to vary. A small value of $\tan\beta = 3$ is chosen to obtain a large $pp \to t b H^\pm$ production cross section. The choice of $m_{12}^2$ (see \cref{tab:BPpars}) has no significant impact on the charged Higgs boson phenomenology but is chosen differently from the \BPh scenario in order to satisfy theoretical constraints that depend differently on the \cp-even and \cp-odd Higgs masses. As in the \BPh scenario, the low value for $\tan\beta$ is an intentionally chosen \emph{worst case} scenario for the bosonic production and decay channels of $H^\pm$ compared to the fermionic channels.

\begin{figure}[!t]
    \centering
    \includegraphics[width=0.9\textwidth]{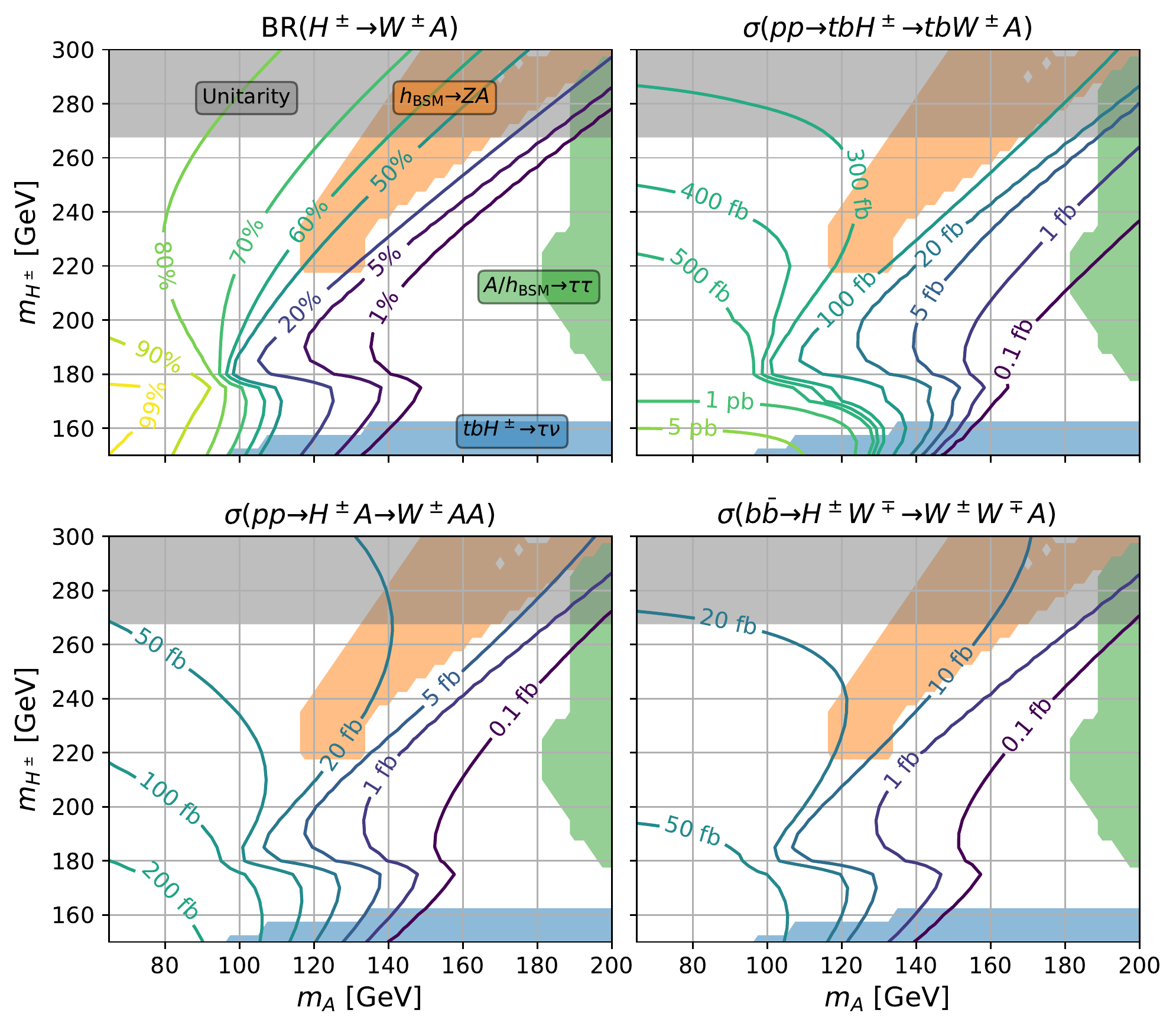}
    \caption{Benchmark scenario \BPA in the ($m_{H^\pm}, m_A$) parameter plane. The \emph{colored contour lines} indicate $\text{BR}(H^\pm\to W^\pm A)$ in the \emph{top-left} panel and the \SI{13}{\TeV} LHC signal cross sections in the $pp\to tbH^\pm \to tb W^\pm A$ (\emph{top-right}), $pp\to H^\pm A \to W^\pm A A $ (\emph{bottom-left}), and $pp\to W^\pm \to H^\mp \to W^\pm W^\mp A$ (\emph{bottom-right}) channels in the remaining panels. The \textit{colored regions} of parameter space are excluded by current constraints from searches in the $pp\to h_\BSM\to Z A$~\cite{Aaboud:2017cxo,Aaboud:2018eoy,Sirunyan:2019xls}, $pp\to A/h_\BSM\to\tau^+\tau^-$~\cite{Aad:2020zxo} and $pp\to tb H^\pm, H^\pm \to \tau^\pm \nu_\tau$~\cite{Aaboud:2018gjj} channel. Theoretical constraints from perturbative unitarity (\emph{gray region}) also impact \BPA.
    }
    \label{fig:BPA}
\end{figure}

We display the branching ratio of the $H^\pm \to W^\pm A$ decay in the upper left plot of \cref{fig:BPA}. The BR values are identical to those found in the \BPh scenario (shown in the upper left plot of \cref{fig:BPh}) under the exchange $A \leftrightarrow h_\BSM$. The $pp\to h_\BSM \to Z A$ searches~\cite{Aaboud:2017cxo,Aaboud:2018eoy,Sirunyan:2019xls} (orange region) cover a significantly smaller parameter region than the $pp\to A\to Z h_\BSM$ searches in the \BPh scenario due to the lower $gg\to h_\BSM$ production cross section. The $pp\to A/h_\BSM \to \tau^+\tau^-$ search~\cite{Aad:2020zxo} excludes parts of the parameter region with $m_A \gtrsim \SI{180}{\GeV}$. The region excluded by $pp\to tb H^\pm, H^\pm \to \tau^\pm \nu_\tau$ searches~\cite{Aaboud:2018gjj} is identical to the corresponding region in the \BPh scenario. As an additional constraint in the \BPA scenario, charged Higgs boson masses above $\sim \SI{270}{\GeV}$ are excluded by perturbative unitarity. As for the \BPh scenario, the maximal $\text{BR}(H^\pm\to W^\pm A)$ reaches values above \SI{99}{\%}. In the \BPA scenario, however, larger parts of the parameter space with $\text{BR}(H^\pm\to W^\pm A) \gtrsim \SI{65}{\%}$ at large $m_{H^\pm}$ values are still unconstrained making this scenario a very interesting target scenario for future $H^\pm\to W^\pm A$ searches. As in the \BPh scenario, the existing searches all rely on fermionic production modes and quickly loose sensitivity for larger values of $\tan\beta$.

The future potential of this channel becomes even more apparent from the $\SI{13}{\TeV}$ LHC signal cross section values, as shown in the remaining plots of \cref{fig:BPA}. These are the same as in the \BPh scenario under the exchange $A\leftrightarrow h_\BSM$, however, the experimental and theoretical constraints differ. Large cross section are possible in particular in the mass region of $m_A\lesssim \SI{120}{\GeV}$, which is now unexcluded compared to the \BPh scenario. Through most of this region the cross section for $tb$-associated charged Higgs boson production, $\sigma(pp\to tb H^\pm,H^\pm\to W^\pm A)$ lies above $\sim \SI{200}{\fb}$ and the cross section for charged Higgs boson production in association with an $A$ ($W$) boson lies above \SI{20}{\fb} \HB{(\SI{10}{\fb})}.

\begin{figure}
    \centering
    \includegraphics[width=0.6\textwidth]{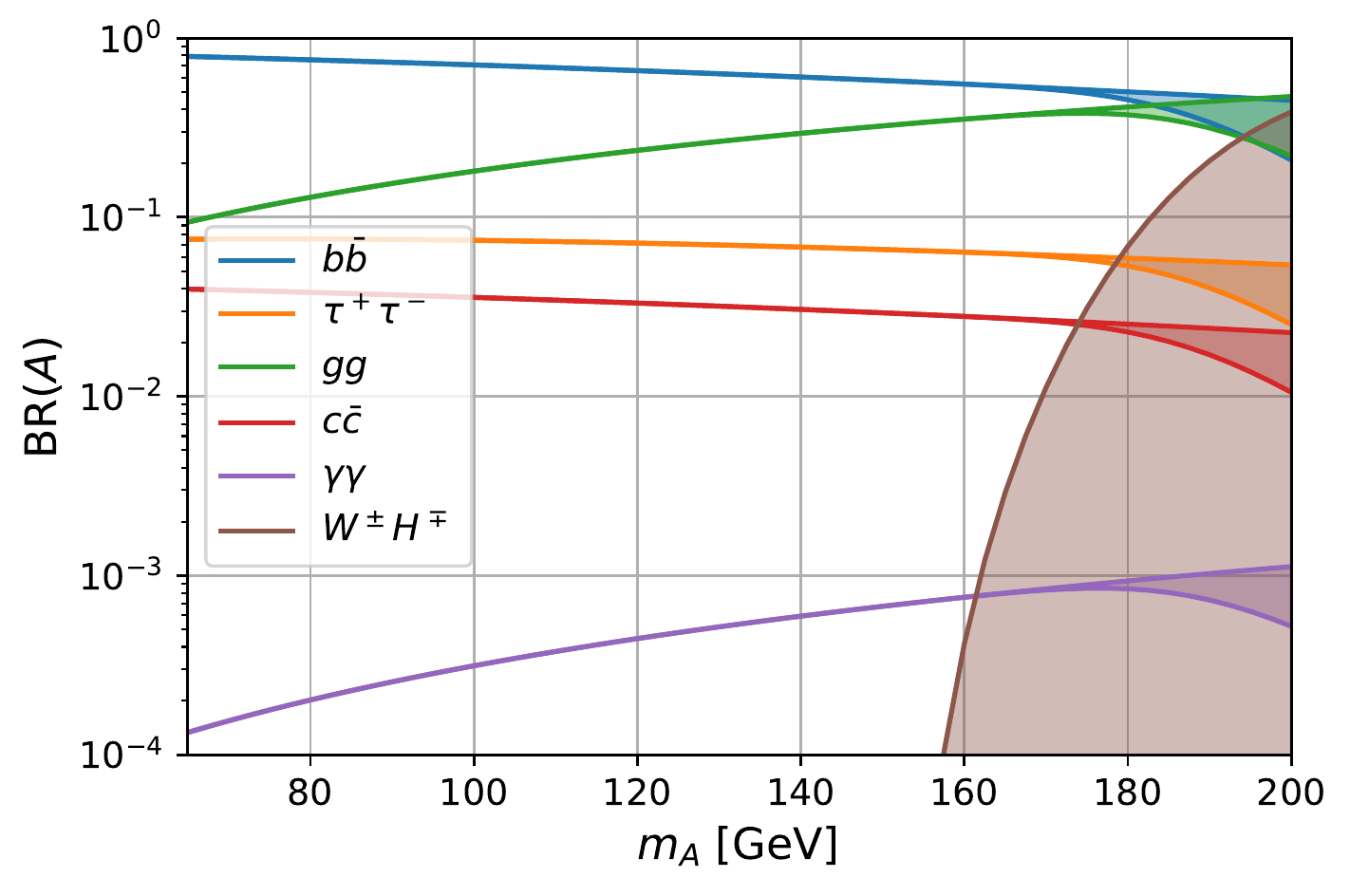}
    \caption{Branching ratios of $A$ in the \BPA benchmark scenario as a function of $m_{A}$. The decays $A\to\gamma\gamma$ and $A\to W^\pm W^\mp$ additionally depend on $m_{H^\pm}$ and induce a dependence in the remaining BRs through the total width. The bands for each BR indicate the impact of this dependence for $\SI{150}{\GeV} < m_{H^\pm} < \SI{300}{\GeV}$.}
    \label{fig:BPA_A_BR}
\end{figure}

Analogous to \cref{fig:BPh_hBSM_BR} for the previous scenario, we show the branching ratios of the $A$ boson in the \BPA scenario in \cref{fig:BPA_A_BR}. The overall behavior of the branching ratios is similar to those of the $h_\BSM$ boson in the previous scenario (see \cref{fig:BPh_hBSM_BR}). As a consequence of the $A$ boson being a \cp-odd scalar, the decays of the $A$ boson into two gluons and two photons are enhanced in comparison to the decays of the $h_\BSM$ boson in the \BPh scenario. For $m_A \sim \SI{200}{\GeV}$, $\text{BR}(A\to gg)$ and $\text{BR}(A\to \gamma\gamma)$ reach values of \SI{47}{\%} and \SI{0.1}{\%}, respectively.

In the \BPA scenario, the charged Higgs boson contribution to the di-photon decay width (see \cref{sec:h125_gaga}) induces deviations of the $h_{125}$ di-photon rate with respect to the SM of up to \SI{9}{\%} (see \cref{sec:diphoton_rate} for more details).

\subsection{\BPff scenario with fermiphobic \texorpdfstring{$h_\text{BSM}$}{hBSM}}

As a first benchmark model specialized on a rather exceptional parameter region with a very distinct collider phenomenology, we discuss the \BPff scenario. We again choose $m_A = m_{H^\pm}$, and $m_{h_\BSM}$ is allowed to vary. In this scenario we depart from the exact alignment limit by setting the coupling of $h_{\BSM}$ to the massive vector bosons to one fifth of the respective SM Higgs couplings. By choosing $\tan\beta$ according to \cref{eq:fermiophobic} (taking a value of $\sim 4.9$), we realize the fermiophobic limit for $h_\BSM$, which implies that $h_\BSM$ does not couple to fermions. Note that the realization of the fermiophobic limit very sensitively depends on the chosen $\tan\beta$ value. Already small deviations from \cref{eq:fermiophobic} result in substantial couplings of $h_\BSM$ to fermions. This is discussed in more detail in \cref{sec:ff_limit}. In contrast, the choice of $m_{12}^2$ (see \cref{tab:BPpars}) only has a minor impact on the phenomenology of the \BPff scenario but is important to satisfy the theoretical constraints. Earlier studies of scenarios with a fermiophobic Higgs boson can be found in Refs.~\cite{Akeroyd:1998ui,Akeroyd:2003xi,Akeroyd:2003bt,Akeroyd:2005pr,Haisch:2017gql,Arhrib:2017wmo}. This choice of parameters can be considered very fine-tuned from a theoretical perspective. Nevertheless, the collider signatures of this scenario are strikingly different from the scenarios discussed above and should not remain unexplored.

\begin{figure}[!t]
    \centering
    \includegraphics[width=.8\textwidth]{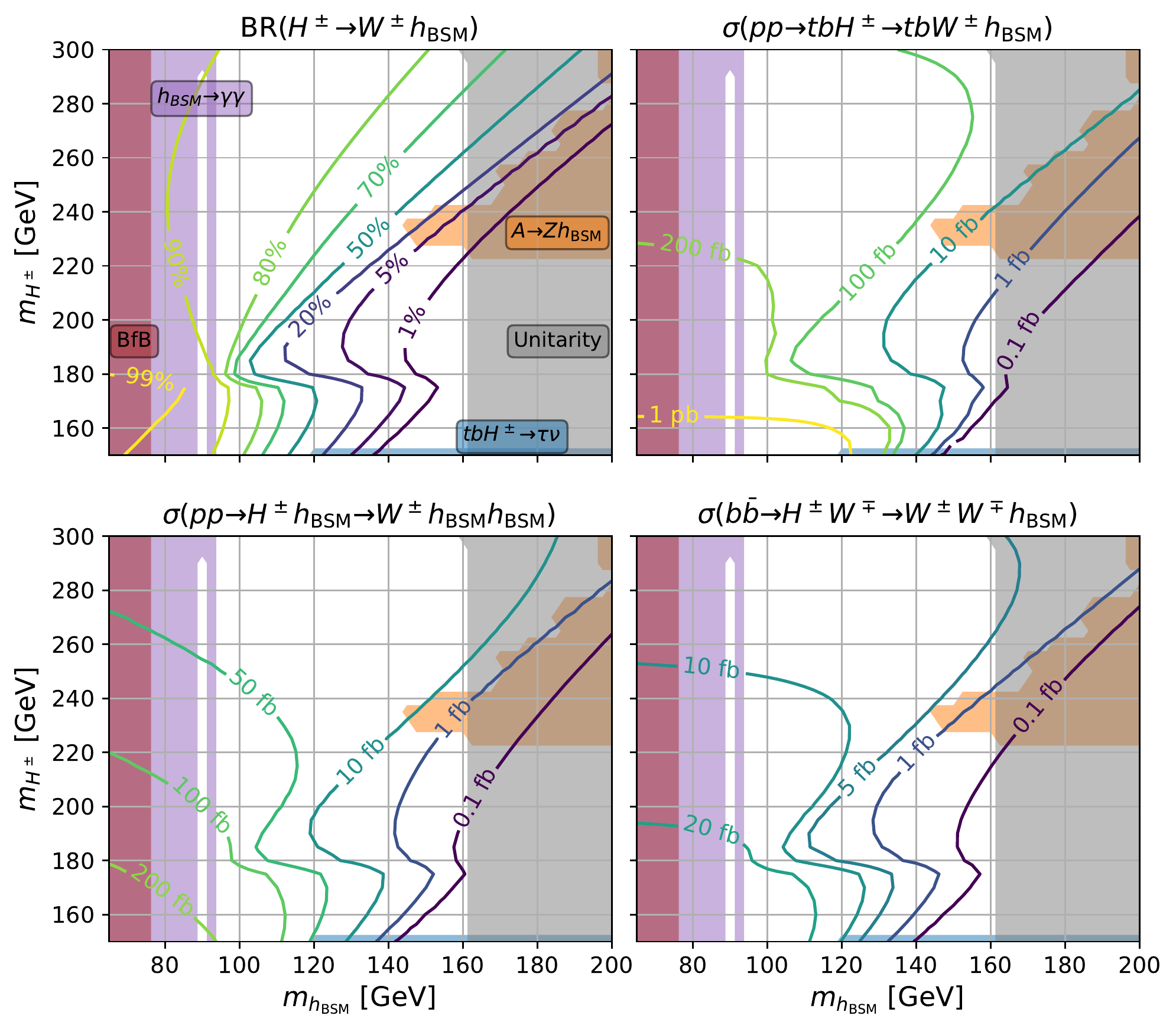}
    \caption{Benchmark scenario \BPff in the ($m_{H^\pm}, m_{h_\text{BSM}}$) parameter plane. The \emph{colored contour lines} indicate $\text{BR}(H^\pm\to W^\pm h_\text{BSM})$ in the \emph{top-left} panel and the \SI{13}{\TeV} LHC signal cross sections in the $pp\to tb H^\pm \to tb W^\pm h_\text{BSM}$ (\emph{top-right}), $pp\to H^\pm h_\text{BSM} \to W^\pm h_\text{BSM}h_\text{BSM}$ (\emph{bottom-left}), and $pp\to W^\mp H^\pm \to W^\pm W^\mp h_\text{BSM}$ (\emph{bottom-right}) channels in the remaining panels. The \emph{colored regions} of parameter space are excluded by current constraints from searches in the $pp\to A\to Z h_\text{BSM}$~\cite{Sirunyan:2019xls}, $pp\to h_\BSM\to\gamma\gamma$~\cite{Aad:2014ioa,CMS:2015ocq,ATLAS:2018xad} and $pp\to tb H^\pm, H^\pm \to \tau^\pm \nu_\tau$~\cite{Aaboud:2018gjj} channels. Theoretical constraints from perturbative unitarity and boundedness from below (BfB) are also relevant.}
    \label{fig:BPff}
\end{figure}

The branching ratio $\text{BR}(H^\pm\to W^\pm h_\BSM)$ for the \BPff scenario is shown in the top-left panel of \cref{fig:BPff}. As before, the results are displayed in the ($m_{H^\pm}, m_{h_\text{BSM}}$) parameter plane with current experimental and theoretical constraints shown as colored areas. Theoretical constraints in the \BPff scenario are the boundedness from below (BfB) requirement on the Higgs potential (dark magenta) --- excluding $m_{H^\pm} \lesssim \SI{76}{\GeV}$ --- and perturbative unitarity (gray) --- excluding $m_{H^\pm} \gtrsim \SI{160}{\GeV}$.
Experimental constraints arise from searches for $pp\to h_\BSM\to\gamma\gamma$~\cite{Aad:2014ioa,CMS:2015ocq,ATLAS:2018xad} (purple) --- excluding $m_{h_\BSM} \lesssim \SI{95}{\GeV}$ except for a narrow region around the $Z$-boson mass ---, searches for $pp\to A\to Z h_\text{BSM}$~\cite{Sirunyan:2019xls} (orange) --- excluding an otherwise unexcluded small patch around $m_{H^\pm} \sim \SI{230}{\GeV}$ and $m_{h_\BSM}\sim \SI{155}{\GeV}$ ---, as well as searches for $pp\to tb H^\pm, H^\pm \to \tau^\pm \nu_\tau$~\cite{Aaboud:2018gjj} (blue) --- excluding a small region around $m_{H^\pm} \sim \SI{150}{\GeV}$ and $m_{h_\BSM}\gtrsim \SI{120}{\GeV}$. In the remaining allowed parameter region, $\text{BR}(H^\pm\to W^\pm h_\BSM)$ values of up to $\sim \SI{90}{\%}$ are possible. Especially the region of low $m_{h_\BSM}$ and high $m_{H^\pm}$ features large branching fractions.

The corresponding signal cross sections, shown in the remaining panels of \cref{fig:BPff}, have a slightly different behavior. Since the production cross sections for a charged Higgs boson decrease with rising $m_{H^\pm}$, also the cross sections for charged Higgs boson production with the subsequent $H^\pm \to W h_\BSM$ decay tend to decrease for rising $m_{H^\pm}$.
The cross section for $pp\to tb H^\pm, H^\pm \to W^\pm h_\BSM$, shown in the top-right panel of \cref{fig:BPff}, reaches values almost \SI{5}{\pb}. For $pp\to H^\pm h_\BSM \to W^\pm h_\BSM h_\BSM$ production, shown in the bottom-left panel of \cref{fig:BPff}, the maximal cross section is $\sim \SI{200}{\fb}$. The $pp\to H^\pm W^\mp \to W^\pm W^\mp h_\BSM$ production cross section, displayed in the bottom-right panel of \cref{fig:BPff}, reaches values $\gtrsim \HB{\SI{35}{\fb}}$.

\begin{figure}
    \centering
    \includegraphics[width=0.6\textwidth]{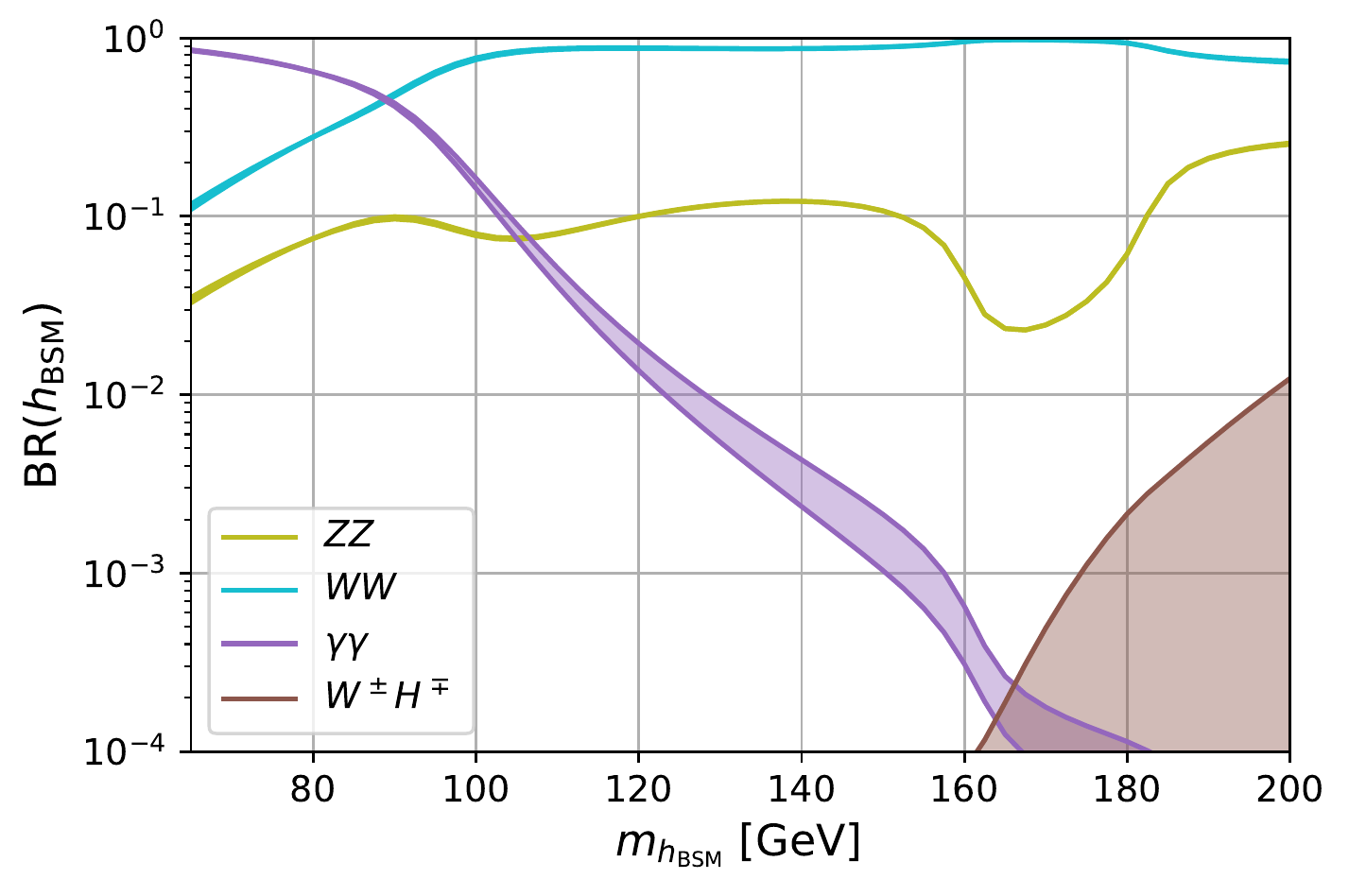}
    \caption{Branching ratios of $h_\text{BSM}$ in the \BPff benchmark scenario as a function of $m_{h_\text{BSM}}$. The decays $h_\text{BSM}\to\gamma\gamma$ and $h_\text{BSM}\to W^\pm H^\mp$ additionally depend on $m_{H^\pm}$ and induce a small dependence in the remaining BRs through the total width. The bands for each BR indicate the impact of this dependence for $\SI{150}{\GeV} < m_{H^\pm} < \SI{300}{\GeV}$.}
    \label{fig:BPff_hBSM_BR}
\end{figure}

Due to fermiophobic character of the $h_\BSM$ boson, its decays shown in \cref{fig:BPff_hBSM_BR} are very different from the ones in the \BPh scenario. For $m_{h_\BSM}\lesssim \SI{90}{\GeV}$, the $h_\BSM$ boson decays dominantly to a pair of photons reaching a maximal branching ratio of $\sim \SI{90}{\%}$.\footnote{Due to $h_\BSM$ being fermiophobic its decay to a pair of photons is only mediated by the $W$ boson and the charged Higgs boson in the loop (see also Ref.~\cite{Akeroyd:2007yh}).} For $m_{h_\BSM}\gtrsim \SI{90}{\GeV}$, the (off-shell) decays into a pair of $W$ or $Z$ bosons become increasingly important with branching ratios of $\sim \SI{90}{\%}$ and $\sim \SI{10}{\%}$, respectively. Note that the decay modes into massive vector bosons are possible as the alignment of $h_{125}$ is not exact in this scenario.

This departure from alignment, together with the charged Higgs boson effects, leads to deviations of the $h_{125}$ di-photon decay rate from the SM by up to \SI{14}{\%} (see \cref{sec:diphoton_rate} for more details) in the \BPff scenario.

\subsection{\BPlight scenario with light \texorpdfstring{$h_\BSM$}{hBSM}}

The second rather specialized benchmark model is the \BPlight scenario which features a light $h_\BSM$ with $m_{h_\BSM} \le \SI{62.5}{\GeV}$. Similar to the other scenarios we choose again $m_A = m_{H^\pm}$. In order to suppress the $h_{125}\rightarrow h_\BSM h_\BSM$ decay channel we choose $m_{12}^2$ according to \cref{eq:zero_gh1h1h2_m12sq}. As already mentioned in \cref{sec:hsm_to2hbsm}, unitarity and absolute vacuum stability requirements enforce a strong correlation between $c(h_\BSM VV)$ and $\tan\beta$. For $\tan\beta$ values consistent with flavor constraints, $\tan\beta \gtrsim 3$, we need to slightly depart from the exact alignment limit. We therefore choose $c(h_\BSM VV) = -0.062$ in this scenario, and we set $\tan\beta$ to 16.6 in order to fulfill the unitarity and vacuum stability requirements. See \cref{sec:h125hBSMhBSM_suppression} for more details on these parameter choices showing that this kind of scenario in the 2HDM clearly requires significant tuning of the parameters. However, similar phenomenology may be far easier to achieve in more complex models, for which this scenario can serve as a simple benchmark.

\begin{figure}
    \centering
    \includegraphics[width=.9\textwidth]{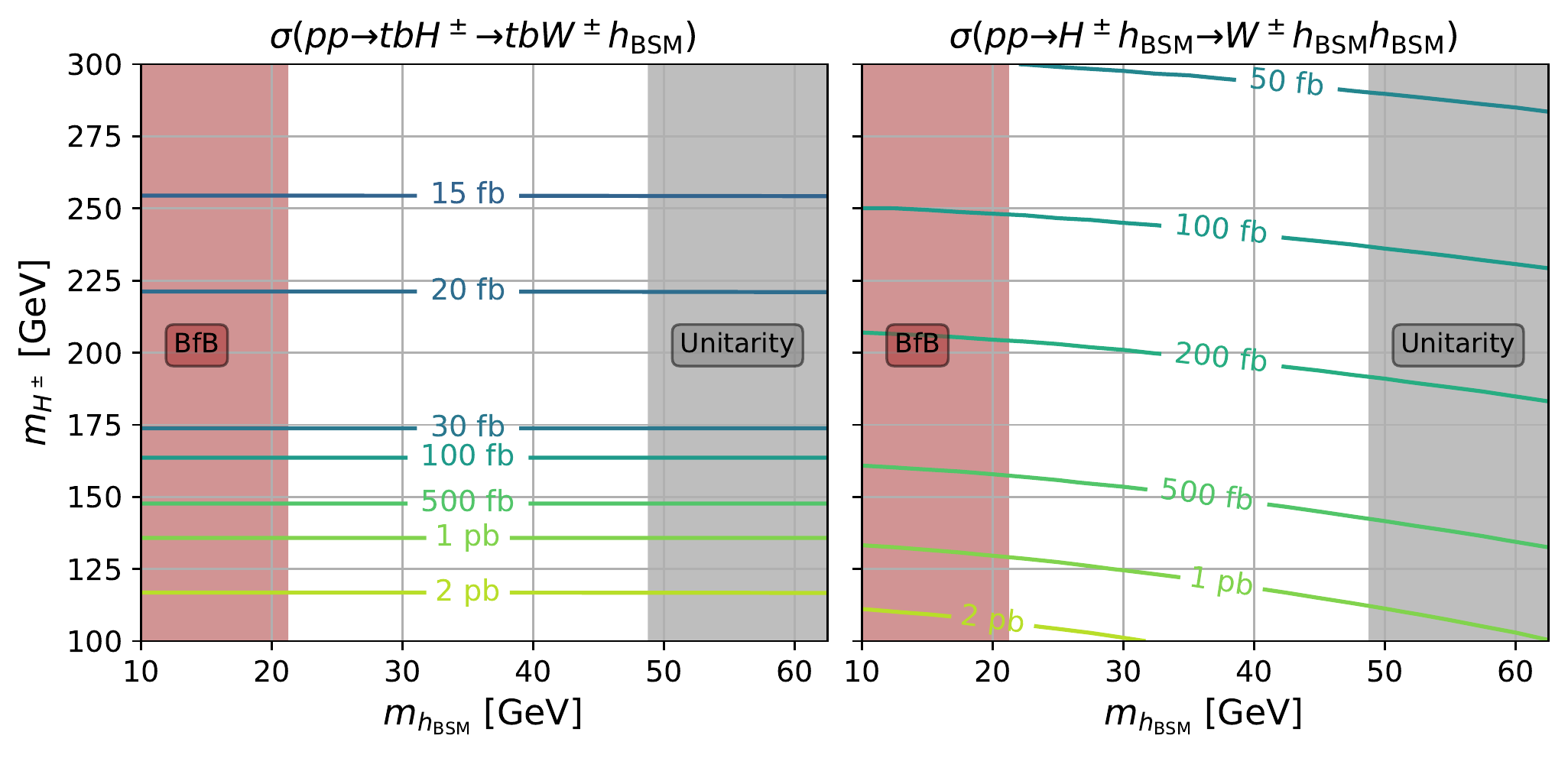}
    \caption{Signal cross sections in the ($m_{H^\pm}, m_{h_\text{BSM}}$) parameter plane of benchmark scenario \BPlight. The contour lines indicate the \SI{13}{\TeV} LHC cross sections for the processes $pp\to tb H^\pm, H^\pm \to W^\pm h_\BSM$ (\emph{left}) and $pp\to H^\pm h_\BSM \to W^\pm h_\BSM h_\BSM$ (\emph{right}). The shaded regions are excluded by perturbative unitarity and boundedness from below (BfB).}
    \label{fig:BPlight}
\end{figure}

Due to the low $m_{h_\BSM}$ values and the relatively high $\tan\beta$, $\text{BR}(H^\pm \to W^\pm h_\BSM)$ is larger than \SI{98}{\%} in the entire benchmark plane. The signal cross section for $tb$-associated charged Higgs boson production with a subsequent $H^\pm \to W h_\BSM$ decay is shown in the left panel of \cref{fig:BPlight} in the ($m_{H^\pm}, m_{h_\BSM}$) parameter plane. The colored areas are excluded by theoretical constraints: the region of $m_{H^\pm} \lesssim \SI{21}{\GeV}$ is  excluded by requiring boundedness from below (BfB) of the scalar potential (red), and the region of $m_{H^\pm} \gtrsim \SI{49}{\GeV}$ is excluded by perturbative unitarity (gray). The signal cross section $\sigma(p\to tb H^\pm, H^\pm \to W^\pm h_\BSM)$ strongly depends on $m_{H^\pm}$ but it is nearly independent of $m_{h_\BSM}$. It ranges from around \SI{10}{\fb} at $m_{H^\pm} = \SI{300}{\GeV}$ to above \SI{2}{\pb} at $m_{H^\pm} \sim \SI{100}{\GeV}$. We find similar cross-section values for charged Higgs boson production in association with a $h_\BSM$ boson and the subsequent $H^\pm \to W h_\BSM$ decay, shown in the right panel of \cref{fig:BPlight}. Also for this production mode, cross-section values above \SI{1}{\pb} can be reached for $m_{H^\pm} \sim \SI{100}{\GeV}$. At $H^\pm$ masses above $\sim \SI{150}{\GeV}$ the cross section for the $pp\to H^\pm h_\BSM$ channel even surpasses the one for the $pp\to tbH^\pm$ channel. The cross section for charged Higgs boson production in association with a $W$ boson (not shown) is negligible in the \BPlight scenario, reaching values of only \HB{$\lesssim \SI{6}{\fb}$}.

\begin{figure}
    \centering
    \includegraphics[height=5.5cm]{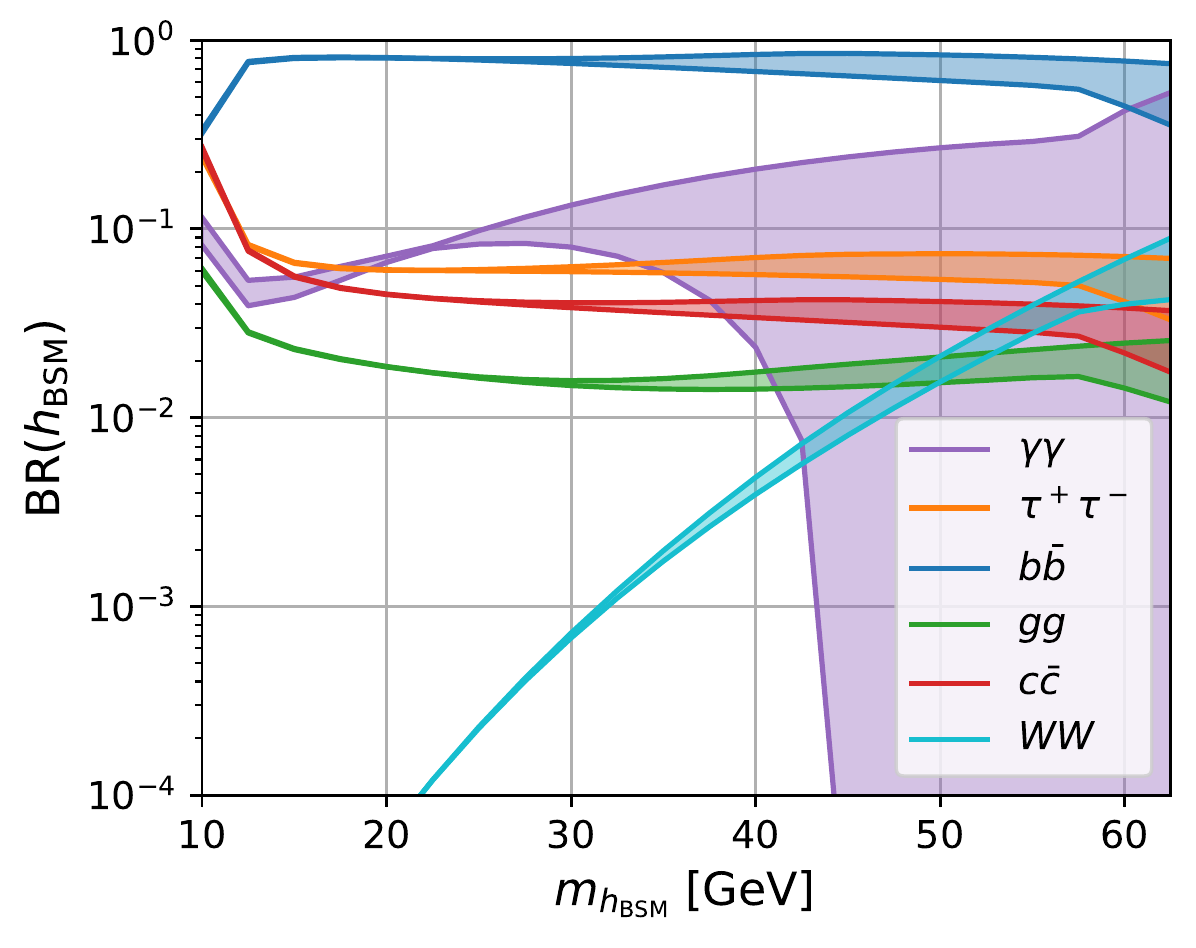}
    \includegraphics[height=5.5cm]{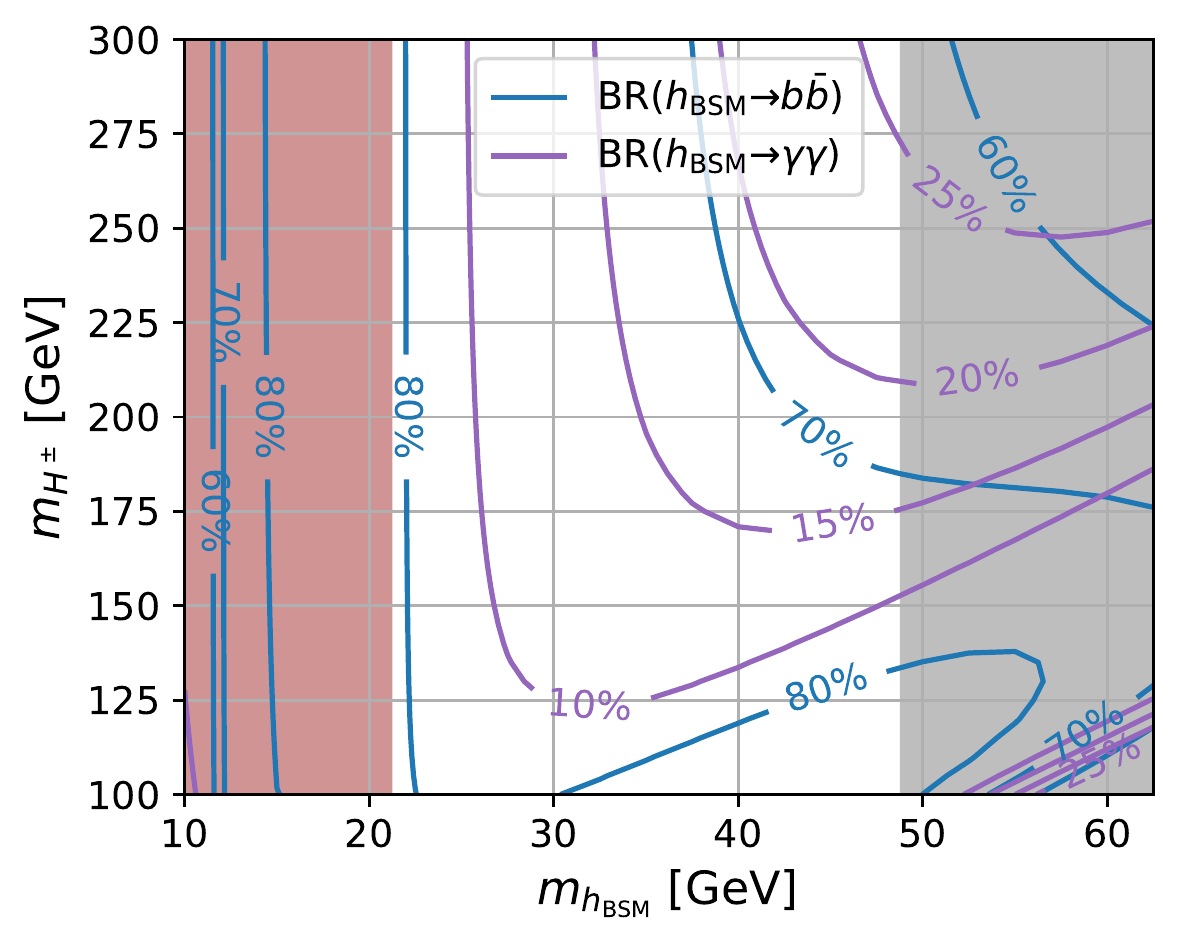}
    \caption{Branching ratios of $h_\text{BSM}$ in the \BPlight benchmark scenario. The \emph{left panel} shows all important BRs as a function of $m_{h_\text{BSM}}$. The decay $h_\text{BSM}\to\gamma\gamma$ additionally depends on $m_{H^\pm}$ and induces a dependence in the remaining BRs through the total width. The bands for each BR indicate the impact of this dependence within the parameter ranges of the scenario. The \emph{right panel} shows the interplay of the two dominant BRs in the ($m_{H^\pm}, m_{h_\text{BSM}}$) parameter plane of \BPlight.}
    \label{fig:BPlight_hBSM_BR}
\end{figure}

The branching ratios of the $h_\BSM$ decays are displayed in the left panel of \cref{fig:BPlight_hBSM_BR}. For very low masses, $m_{h_\BSM}\sim \SI{10}{\GeV}$, $h_\BSM$ decays with almost equal probabilities to $b$ quarks, $c$ quarks, and $\tau$ leptons. For masses in the intermediate range, $\SI{10}{\GeV} < m_{h_\BSM} \lesssim \SI{40}{\GeV}$, the $h_\BSM \to b\bar{b}$ decay dominates with a branching ratio of $\sim 80-\SI{90}{\%}$. For higher mass values, it is possible that $h_\BSM$ decays to two photons with a branching ratio of up to \SI{25}{\%} in the allowed region. In the same region, the branching ratio of $h_\BSM$ decaying to two photons can, however, also be very close to zero.

This large variation originates from the varying charged Higgs mass. This dependence is shown in more detail in the right panel of \cref{fig:BPlight_hBSM_BR}, which displays $\text{BR}(h_\BSM\to b\bar b)$ (blue contours) and $\text{BR}(h_\BSM\to \gamma\gamma)$ (purple contours) in the $(m_{h_\BSM},m_{H^\pm})$ parameter plane. While the decay of $h_\BSM\to b\bar{b}$ dominates for low $m_{h_\BSM}$ values irrespectively of the $H^\pm$ mass, $\text{BR}(h_\BSM\to \gamma\gamma)$ can reach values of up to $\sim \SI{25}{\%}$ in the allowed region for high values of $m_{h_\BSM}$ and $m_{H^\pm}$. The dependence of $\text{BR}(h_\BSM\to \gamma\gamma)$ on $H^\pm$ is larger than in the other benchmark scenarios, since the top-quark and the $W$ boson contribution to $h_\BSM\to\gamma\gamma$ are suppressed by the high value of than $\tan\beta$ and the approximately realized alignment limit, respectively.

In the \BPlight scenario, the $h_{125}$ di-photon decay rate deviates from the SM by up to \SI{14}{\%} (see \cref{sec:diphoton_rate} for more details).

\subsection{\BPlightLS scenario with light leptophilic \texorpdfstring{$h_\BSM$}{hBSM}}

As the last benchmark scenario we present the \BPlightLS scenario. This scenario is defined with the same parameter values as the \BPlight scenario (see \cref{tab:BPpars}), however, instead of type-I, the \BPlightLS scenario is defined in the lepton-specific 2HDM (see \cref{sec:2hdm}). As a direct consequence, $h_\BSM$ decays almost exclusively to tau leptons (with a branching ratio above $\SI{99}{\%}$). Therefore, in contrast to the \BPlight scenario, LHC searches need to focus on the $h_\BSM \to \tau^+\tau^-$ decay (following the charged Higgs boson decay $H^\pm \to W^\pm h_\BSM$) in order to probe this scenario. The HL-LHC sensitivity to a phenomenologically similar scenario has previously been studied in \ccite{Chun:2018vsn}.

As for the \BPlight scenario, $m_{h_\BSM} \le \SI{62.5}{\GeV}$ is required. While it is in principle possible to define benchmark scenario within the lepton-specific 2HDM analogously to the \BPh or \BPA scenarios, we found that such scenarios are completely excluded experimentally by searches for a neutral Higgs decaying to tau leptons. These searches do, however, not cover the region $m_{h_\BSM} \le \SI{62.5}{\GeV}$.

\begin{figure}[tb]
    \centering
    \includegraphics[width=\textwidth]{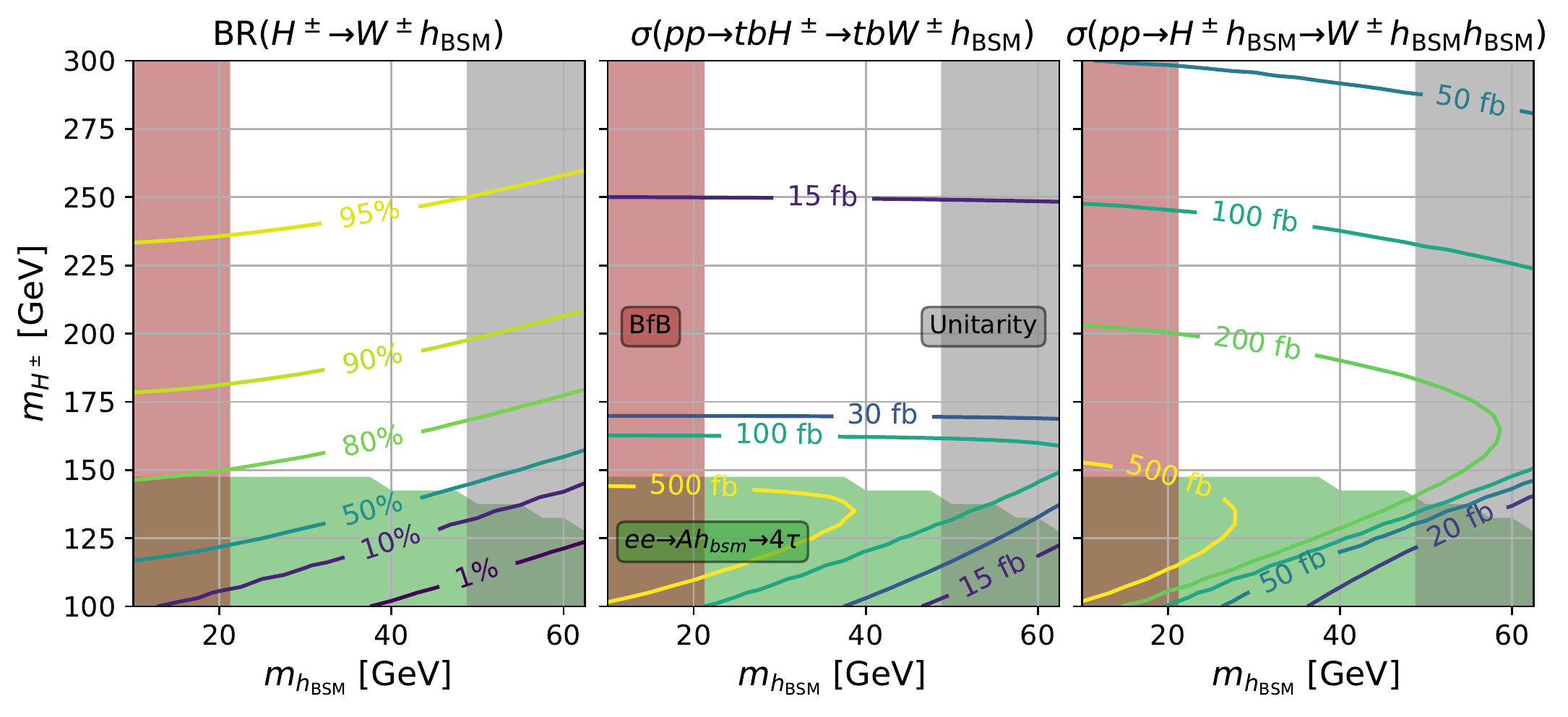}
    \caption{Signal cross sections in the ($m_{H^\pm}, m_{h_\text{BSM}}$) parameter plane of benchmark scenario \BPlightLS. All model parameters except for the Yukawa type are identical to the \BPlight scenario as given in \cref{tab:BPpars}. The \emph{colored contours} indicate $\text{BR}(H^\pm\to W^\pm h_\text{BSM})$ in the \emph{left} panel and the \SI{13}{\TeV} LHC cross sections for the search processes $pp\to tb H^\pm, H^\pm \to W^\pm h_\BSM$ (\emph{center}) and $pp\to H^\pm h_\BSM \to W^\pm h_\BSM h_\BSM$ (\emph{right}). The shaded regions are excluded by perturbative unitarity, boundedness from below, and LEP searches for di-Higgs production in the $4\tau$ final state~\cite{Schael:2006cr}.}
    \label{fig:BPlightLS}
\end{figure}

The branching ratio $\text{BR}(H^\pm\to W^\pm h_\BSM)$ for the \BPlightLS scenario is shown in the left panel of \cref{fig:BPlightLS} in the ($m_{H^\pm}, m_{h_\BSM}$) parameter plane. The colored areas are excluded by the following constraints: as for the \BPlight scenario, the region of $m_{h_\BSM}\lesssim \SI{21}{\GeV}$ is excluded by requiring boundedness from below (red) and the region of $m_{h_\BSM}\gtrsim \SI{50}{\GeV}$ by unitarity constraints (gray). While no experimental search constrains the \BPlight scenario, the region of $m_{H^\pm}\lesssim \SI{140}{\GeV}$ is excluded in the \BPlightLS scenario by LEP searches for $e^+e^-\to Ah_\BSM\to 4\tau$ (green). In the remaining unexcluded parameter space, $\text{BR}(H^\pm\to W^\pm h_\BSM)$ is always above \SI{50}{\%}, and reaches almost \SI{100}{\%} for $m_{H^\pm}\sim \SI{300}{\GeV}$. $\text{BR}(H^\pm\to W^\pm h_\BSM)$ is smaller than in the \BPlight scenario due to the fact that the competing $\text{BR}(H^\pm\to\tau^\pm\bar{\nu}_\tau)$ decay is enhanced by the large $\tan\beta$ value in the lepton-specific 2HDM.

The signal cross section for charged Higgs boson production in association with a top and a bottom quark and its subsequent decay to a $W$ boson and a $h_\BSM$ boson is shown in the center panel of \cref{fig:BPlightLS}. In the allowed parameter space, $\sigma(p\to tb H^\pm, H^\pm \to W^\pm h_\BSM)$ can reach almost \SI{500}{\fb} for $m_{H^\pm}\sim \SI{150}{\GeV}$. For higher values of $m_{H^\pm}$ the cross section quickly drops below \SI{100}{\fb}. In this region, $\sigma(p\to H^\pm h_\BSM, H^\pm \to W^\pm h_\BSM h_\BSM)$ --- shown in the right panel of \cref{fig:BPlightLS} --- exceeds $\sigma(p\to tb H^\pm, H^\pm \to W^\pm h_\BSM)$, still reaching e.g.\ \SI{200}{\fb} for $m_{H^\pm} \sim \SI{200}{\GeV}$ and $m_{h_\BSM}\sim \SI{30}{\GeV}$. Note in particular, that --- since $\text{BR}(h_\BSM\to\tau^+\tau^-)\approx1$ --- the signal cross sections for the $t b W^\pm \tau^+\tau^-$ and $W^\pm\tau^+ \tau^-\tau^+\tau^-$ final states are almost as large as the corresponding cross sections without the $h_\BSM$ decays. Charged Higgs production in association with a $W$ boson is negligible in the \BPlightLS scenario.

In the \BPlightLS scenario, the $h_{125}$ di-photon decay rate deviates from the SM by up to \SI{7}{\%} (see \cref{sec:diphoton_rate} for more details).

\section{Conclusions}
\label{sec:conclusions}

The presence of charged Higgs bosons is a generic prediction of multiplet extensions of the SM Higgs sector. Collider searches for charged Higgs bosons are, therefore, an important puzzle piece in the search for new physics in the Higgs sector. While many LHC searches for charged Higgs bosons decaying to SM fermions exist, the bosonic decay modes of charged Higgs bosons have so far received much less attention.

Focussing on the 2HDM, we discussed the charged Higgs boson phenomenology taking into account all applicable constraints of theoretical as well as experimental nature. We considered type I and lepton specific Yukawa sectors, where light charged Higgs bosons are not excluded by flavor observables. We revisited two genuine BSM effects on the discovered SM-like Higgs boson --- the non-decoupling charged Higgs boson contribution to the SM-like Higgs boson decay into two photons, as well as the decay of the SM-like Higgs boson into two non-SM-like Higgs bosons --- and their impact on the charged Higgs boson phenomenology. These effects even appear in the alignment limit, in which the SM-like Higgs boson has exactly SM-like couplings.

We then investigated in detail the production and decay modes of the charged Higgs bosons. We demonstrated that the charged Higgs boson decays predominantly to a $W$ boson and a neutral non-SM-like Higgs state (which could be either \cp-even or \cp-odd) in large parts of the allowed parameter space. This decay mode is especially large close to the observationally-favored alignment limit, in which the charged Higgs boson coupling to a non-SM-like Higgs boson and a $W$ boson is maximized.

We discussed current experimental searches for charged Higgs bosons at the LHC and pointed out the so-far unexplored decay signatures that arise from the above-mentioned decay to a $W$ boson and a neutral non-SM-like Higgs state. In order to facilitate future searches for these signatures, and as the main result of this work, we introduced five benchmark scenarios, each featuring a distinct phenomenology: the \BPh scenario, the \BPA scenario, the \BPff scenario, the \BPlight scenario, and the \BPlightLS scenario. All scenarios exhibit a large decay rate of the charged Higgs boson to a $W$ boson and a neutral non-SM-like Higgs boson. The scenarios are defined in a 2HDM of type I, except for \BPlightLS where the Yukawa sector is chosen to be lepton specific.

In the \BPh scenario, the charged Higgs boson decays dominantly to a non-SM-like \cp-even Higgs boson and a $W$ boson. The signal cross section for the production of the charged Higgs boson in association with a top and a bottom quark and its subsequent decay reaches up to \SI{5}{\pb} in significant parts of the yet-unconstrained parameter space. The experimentally most interesting decay modes of the non-SM-like Higgs boson in this scenario are the decay into bottom quarks and the decay into tau leptons.

In contrast to the \BPh scenario, the charged Higgs boson decays dominantly to a \cp-odd $A$ boson and a $W$ boson in the \BPA scenario, while the $h_\BSM$ boson has the same mass as the charged Higgs boson. Also in this scenario, signal cross sections of up to \SI{5}{\pb} can be reached in large parts of unconstrained parameter space. Due to the \cp-odd nature of the $A$ boson, its decay width into two photons is enhanced with respect to the corresponding decay width of the $h_\BSM$ boson in the \BPh scenario.

As in the \BPh scenario, the charged Higgs boson decays dominantly to a non-SM-like \cp-even Higgs boson and a $W$ boson in the \BPff scenario reaching signal cross sections of up to \SI{1}{\pb}. In contrast to the previous scenarios, the non-SM-like \cp-even Higgs boson is fermiophobic in the \BPff scenario, which results in large branching ratios of the $h_\BSM$ decays into massive vector bosons and photons.

In the \BPlight scenario the non-SM-like \cp-even Higgs boson is much lighter than the SM-like Higgs boson, allowing in principle for decays of the SM-like Higgs boson into two light non-SM-like \cp-even Higgs bosons. We, however, showed that this decay mode can be suppressed by suitable parameter choices. In that case, strong constraints on additional decay modes of the SM-like Higgs boson arising from precision rate measurements can be avoided. LHC searches for the charged Higgs boson decay into a light, non-SM-like \cp-even Higgs boson and a $W$ boson therefore would provide an important complementary probe of such scenarios. In this scenario, the branching ratio of $h_\BSM$ to two photons can become quite large, i.e.~up to \SI{25}{\%} within the allowed parameter region.

For the \BPlightLS scenario, we use the same parameters as in the \BPlight scenario. However, in contrast to all other benchmark scenarios, which are defined in the 2HDM type-I, the \BPlightLS scenario is defined in the lepton-specific 2HDM. Consequently, the $h_\BSM$ boson almost exclusively decays to tau leptons, while the remaining phenomenology is very similar to the \BPlight scenario.

We hope that the presented work and in particular the five benchmark scenarios serve as an encouragement as well as an useful set of tools for future experimental searches for bosonic charged Higgs boson decays at the LHC.

\section*{Acknowledgments}

We are grateful to Liron Barak, David Brunner, Dirk Kr\"ucker and Isabell Melzer-Pellmann for useful discussions of experimental questions, and thank Hass AbouZeid and Mike Hance for collaboration at an initial stage of the project. We thank Stefan Liebler for discussions and assistance with \texttt{vh@nnlo-2}. H.B.~and T.S.~acknowledge support by the Deutsche Forschungsgemeinschaft (DFG, German Research Foundation) under Germany’s Excellence Strategy — EXC 2121 ``Quantum Universe'' — 390833306. The work of J.W. is supported by the Swedish Research Council, contract number 2016-05996 and was in part funded by the European Research Council (ERC) under the European Union’s Horizon 2020 research and innovation programme, grant agreement No 668679.

\clearpage
\appendix

\section{Benchmark scenarios: di-photon rate of \texorpdfstring{$h_{125}$}{h125}}
\label{sec:diphoton_rate}

\begin{figure}[t]
    \centering
    \includegraphics[width=.6\textwidth]{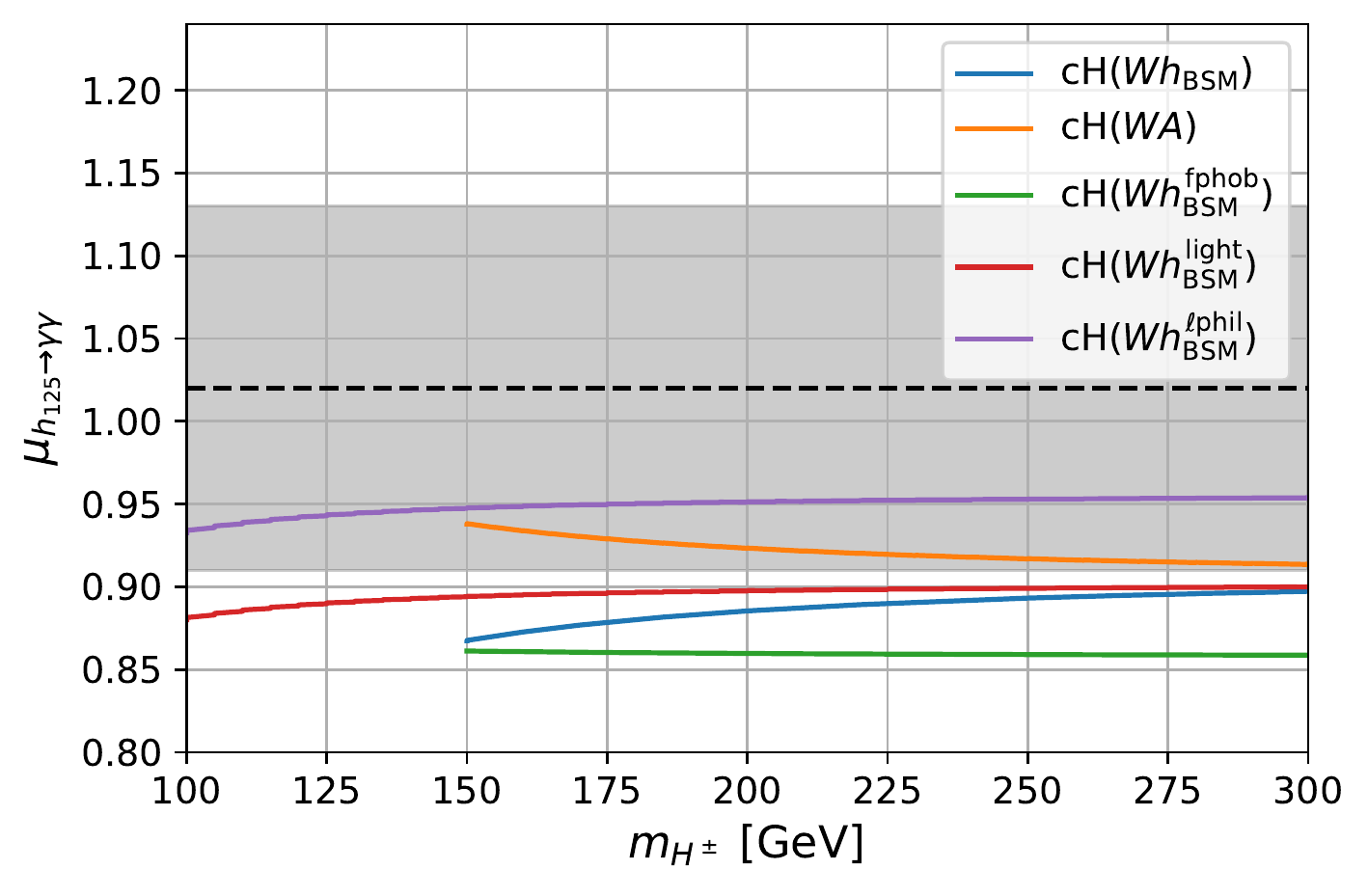}
    \caption{Di-photon rate of $h_{125}$ for the fermionic production modes (ggF+bbH) in dependence of $m_{H^\pm}$ for the various benchmark scenarios defined in \cref{sec:benchmarks}. The dashed line indicates the central value of the latest ATLAS measurement~\cite{ATLAS:2020pvn}, the shaded region is the corresponding $1\sigma$ uncertainty and the displayed $\mu_{h_{125}\to \gamma\gamma}$ range corresponds to the $2\sigma$ uncertainty region (assuming a Gaussian uncertainty).}
    \label{fig:mu_gamgam}
\end{figure}

As discussed in \cref{sec:h125_gaga}, the presence of the charged Higgs boson can result in deviations of the $h_{125}$ di-photon rate even in the exact alignment limit. \cref{fig:mu_gamgam} shows these deviations for the various benchmark scenarios defined in \cref{sec:benchmarks}: the \BPh scenario (blue), the \BPA scenario (orange), the \BPff scenario (green), the \BPlight scenario (dark red), and the \BPlightLS scenario (purple). For all these scenarios the di-photon rate of $h_{125}$ is suppressed with respect to the SM with the deviations ranging from $\sim \SI{5}{\%}$ (for the \BPlightLS scenario) to $\sim \SI{14}{\%}$ (for the \BPff scenario). While the deviation is entirely caused by the charged Higgs effects in the \BPh and \BPA scenarios, which are defined in the exact alignment limit, the deviations in the other scenarios are also slightly affected by the departure from exact alignment. In the \BPff scenario, $c(h_{125}t\bar{t})\approx 1.02$ and $c(h_{125}VV)\approx 0.98$, while the misalignment is even smaller in the \BPlight and \BPlightLS scenarios with $c(h_{125}t\bar{t})\approx 1.002$ and $c(h_{125}VV)\approx 0.998$.

This suggests that large parts of the considered parameter space can be probed by future Higgs precision measurements (e.g.\ at the HL-LHC~\cite{Cepeda:2019klc}). The deviation of the di-photon rate, however, strongly depends on the value of $m_{12}^2$. Correspondingly, it is possible to define phenomenologically similar benchmark scenarios with smaller (or even positive) deviations of the di-photon rate with respect to the SM.

\section{Realizing the fermiophobic limit}
\label{sec:ff_limit}

\begin{figure}[t]
    \centering
    \includegraphics[width=.6\textwidth]{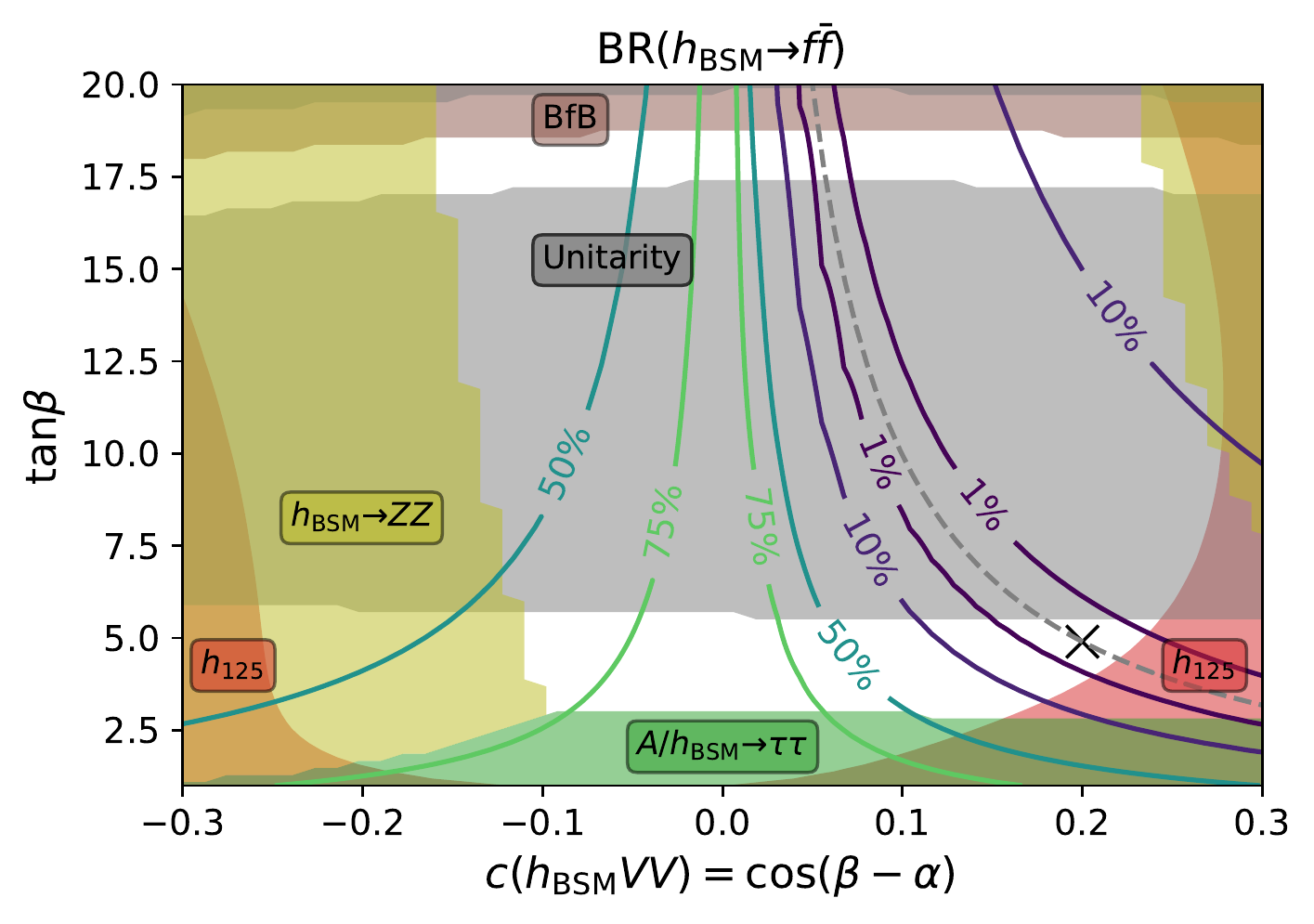}
    \caption{The ($c(h_\BSM VV)$, $\tan\beta$) parameter plane orthogonal to the \BPff benchmark scenario. The masses are fixed to $m_{h_\text{BSM}}=\SI{150}{\GeV}$ and $m_A = m_{H^\pm}=\SI{200}{\GeV}$ and all remaining parameters are as given in \cref{tab:BPpars}. The contours indicate $\text{BR}(h_\text{BSM}\to f\bar{f})$ and the $\times$ marks where this plane touches the mass plane of the \BPff scenario. The dashed line denotes the fermiophobic limit. The shaded regions are excluded by existing theoretical constraints --- boundedness from below and perturbative unitarity ---  experimental searches --- $pp\to A/h_\text{BSM}\to\tau^+\tau^-$~\cite{CMS:2015mca,Aad:2020zxo} and $pp\to h_\text{BSM}\to ZZ$~\cite{Sirunyan:2018qlb} --- as well as $h_{125}$ measurements.}
    \label{fig:ff_tuning}
\end{figure}

The \BPff scenario is defined as a ($m_{h_\BSM}$, $m_{H^\pm}$) parameter plane with $\tan\beta \sim 4.9$ and $\cos(\beta-\alpha) = 0.2$ taking fixed values (see \cref{sec:benchmarks}). In order to motivate the choice of $\tan\beta$ and $\cos(\beta-\alpha)$, we show $\text{BR}(h_\BSM\to\bar f f)$ as a function of $\tan\beta$ as well as $\cos(\beta-\alpha)$ for $m_{h_\BSM} = \SI{150}{\GeV}$ and $m_{H^\pm} = \SI{200}{\GeV}$ in \cref{fig:ff_tuning}. All other parameters are chosen as in the \BPff scenario. The shaded regions are excluded by boundedness from below (brown) --- excluding the region of $\tan\beta\gtrsim 18$ ---, perturbative unitarity (gray) --- excluding the region of $6\lesssim\tan\beta\lesssim 17$ ---, searches for $pp\to A/h_\text{BSM}\to\tau^+\tau^-$~\cite{CMS:2015mca,Aad:2020zxo} (green) --- excluding the region of $\tan\beta\lesssim 2.5$ ---, searches for $pp\to h_\text{BSM}\to ZZ$~\cite{Sirunyan:2018qlb} --- excluding the region of $\cos(\beta-\alpha)\lesssim -0.15$ and $\cos(\beta-\alpha)\gtrsim 0.25$ ---  as well as $h_{125}$ measurements --- excluding the region of large $|\cos(\beta-\alpha)|$ and low $\tan\beta$.

In the still unconstrained parameter region, the fermiophobic limit (see \cref{sec:2hdm_scan}) can be realized either for $\tan\beta \sim 5$ or $\tan\beta\sim 17.5$. While choosing $\tan\beta\sim 17.5$ would allow to approach the alignment more closely, the high $\tan\beta$ value would result in a relatively low charged Higgs production cross section (see \cref{sec:Hpm_prod}). Therefore, we choose $\tan\beta \sim 4.9$ for the definition of the \BPff scenario (as marked by the cross in \cref{fig:ff_tuning}).

\section{Suppressing \texorpdfstring{$h_{125}\to h_{\BSM}h_{\BSM}$}{h125 -> hBSM hBSM} in the \BPlight scenario}
\label{sec:h125hBSMhBSM_suppression}

\begin{figure}
    \centering
    \includegraphics[width=.6\textwidth]{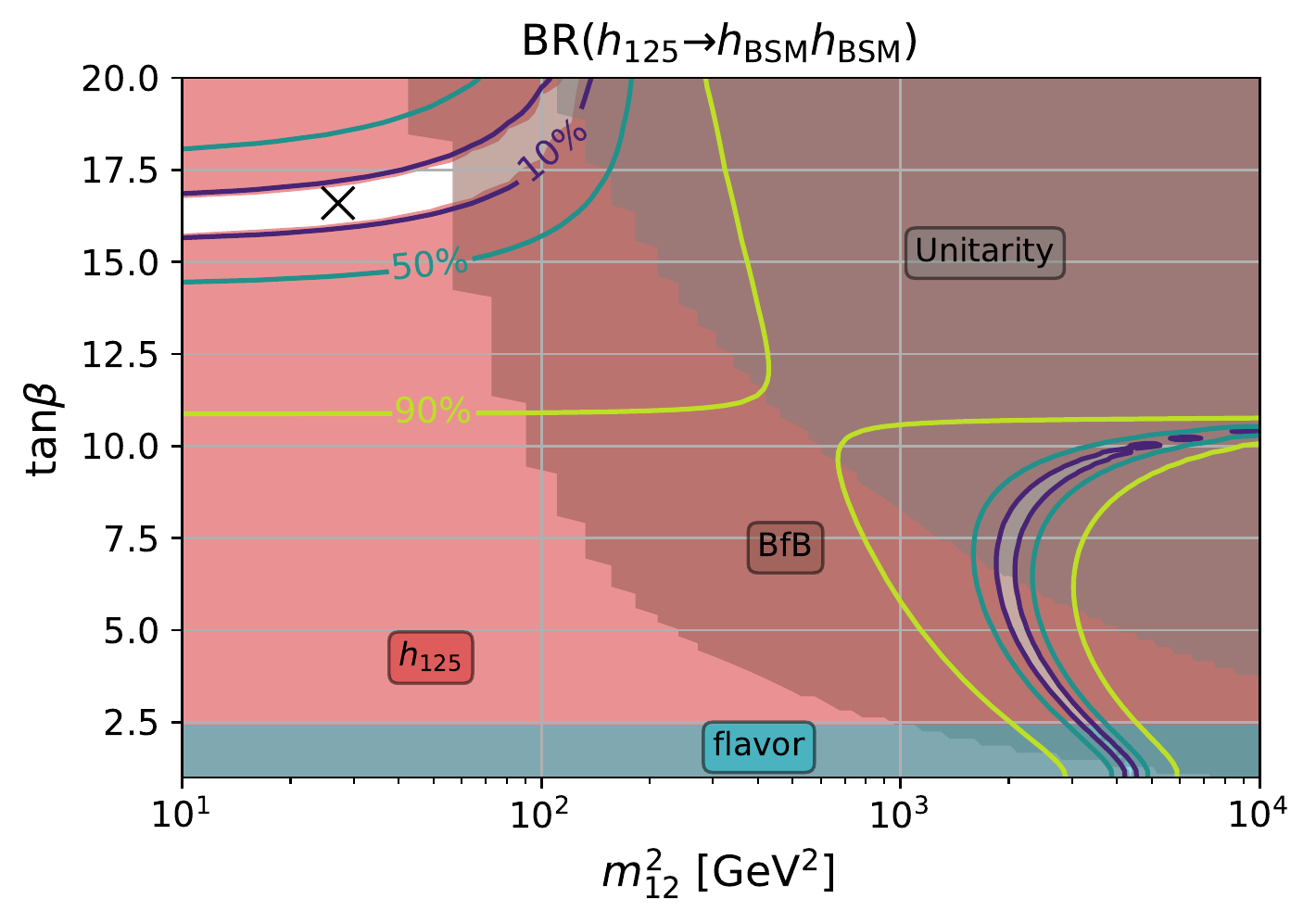}
    \caption{The ($\tan\beta$, $m_{12}^2$) parameter plane orthogonal to  the \BPlight benchmark scenario. The masses are fixed to $m_{h_\text{BSM}}=\SI{30}{\GeV}$ and $m_A = m_{H^\pm}=\SI{200}{\GeV}$ and all remaining parameters are as given in \cref{tab:BPpars}. The contours indicate $\text{BR}(h_{125}\to h_\text{BSM}h_\text{BSM})$ and the $\times$ marks where this plane touches the mass plane of the \BPlight scenario. The shaded regions are excluded by existing theoretical constraints --- boundedness from below and perturbative unitarity --- as well as $h_{125}$ measurements and flavor constraints.}
    \label{fig:vlighth_tuning}
\end{figure}

As discussed in \cref{sec:hsm_to2hbsm}, the $h_{125}\to h_{\BSM}h_{\BSM}$ decay can be suppressed by an appropriate choice of $m_{12}^2$. This choice of $m_{12}^2$ for the \BPlight scenario is motivated in \cref{fig:vlighth_tuning} showing $\text{BR}(h_{125}\to h_\BSM h_\BSM)$. All parameters are chosen as for the \BPlight scenarios except of $m_{12}^2$ and $\tan\beta$ which are treated as free parameters while $m_{h_\BSM} = \SI{30}{\GeV}$ and $m_{H^\pm}= \SI{200}{\GeV}$ are fixed. The shaded regions are excluded by boundedness from below (brown) and perturbative unitarity (gray) --- excluding the region of high $m_{12}^2$ and $\tan\beta$ ---, flavor constraints --- excluding the region of $\tan\beta\lesssim 2.5$ ---, as well as $h_{125}$ measurements --- excluding most of the remaining region, except for a band of $15.5\lesssim \tan\beta\lesssim 17$.

While $\text{BR}(h_{125}\to h_\BSM h_\BSM)$ can be close to zero in the region of $\tan\beta\lesssim 10$ and $m_{12}^2>\SI{1000}{\GeV^2}$, this region is completely excluded by existing constraints. The only still unconstrained region with $\text{BR}(h_{125}\to h_\BSM h_\BSM) \sim 0$ lies in the region of $15.5\lesssim\tan\beta\lesssim 17$ and $m_{12}^2 \lesssim \SI{60}{\GeV}^2$. The \BPlight scenario, marked by the cross, is chosen in the center of this region.

\clearpage
\printbibliography

\end{document}